\documentclass[structabstract]{aa}  
\usepackage{graphicx,lscape}
\usepackage{txfonts}
\begin{document}
   \title{Spectroscopic atlas of H$\alpha$ and H$\beta$ in a sample of northern Be stars\thanks{Table 
           B.1 and all observed spectra are available in electronic form at the CDS via anonymous ftp to cdsarc.u-strasbg.fr (130.79.128.5)
           or via http://cdsweb.u-strasbg.fr/cgi-bin/qcat?J/A+A/}}

   \author{G. Catanzaro\thanks{I wish to dedicate this paper to my child never born, to keep track of his passage through my life.}}

   \institute{INAF-Catania Astrophysical Observatory, via S. Sofia 78, I--95123, Catania, Italy\\
              \email{giovanni.catanzaro@oact.inaf.it}
             }

   \date{Received 10 September 2012 / Accepted 7 December 2012 }

 
  \abstract
   {Be stars are fast-rotating early-type emission line stars. It is generally assumed that observed emission is generated in a rotating
    disk-like envelope, as supported by the observed correlation between the stellar projected rotational velocity $v \sin i$ and the width
    of the emission lines. Then, high-resolution spectroscopic observations of Balmer lines profiles play an important role in putting 
    constraints on Be stars modeling.}
   {We present Balmer lines spectroscopy for a sample of 48 Be stars. For most of them, H$\alpha$ and H$\beta$ have been 
    observed more than two times, in a total period spanning almost two years between 2008 and 2009.}
   {Spectral synthesis of the H$\alpha$ profile was performed following two steps: photospheric contribution was 
    computed by using Kurucz's code ATLAS9 and SYNTHE, and disk emission was derived by the approach of Hummel \& 
    Vrancken (2000, A\&A, 302, 751).}
   {For 26 out of 48 stars, a modeling of the total H$\alpha$ emission, i.e. photospheric absorption plus disk net emission, 
    has been attempted. By this modeling we derived an estimation of the disk radius, as well as the inclination angle between the 
    rotational axis with line of sight and the base density at the stellar equator. For the stars observed more 
    than once, we also discuss the variability of H$\alpha$ and H$\beta$ for what concerns both the equivalent width and the spectral profile. 
    We found 16 stars with variable equivalent width and 7 stars with clear signs of profile variations.}
   {For all the stars in our sample, we derive all the fundamental astrophysical quantities, such as, effective temperature, gravity, and
    projected rotational velocity.  We found 13 stars whose equivalent width is variable with a confidence level greater than 80$\%$ and 
    7 object for which spectral profiles show change with time. According to the commonly used classification scheme, 
    we classified 16 stars as belonging to class 1, 13 to class 2, 11 are shell stars, 6 objects do not show net emission, and 2 stars display 
    transitions from class 1 and 2. For the class 1 stars, we confirm the correlation between $v \sin i$ and peak separation.
    Concerning the geometry of the disk, we derived the base density at the stellar equator, the radius, and the inclination angle between 
    rotational axis and line of sight. The maximum concentration of stars occurs for disk dimensions ranging in the interval of 6 to 8 
    stellar radii and for inclination angles going from 23$^\circ$ to $35^\circ$.}

   \keywords{Stars: emission-line, Be --
                stars: fundamental parameters
               }

   \maketitle
%

\section{Introduction}
Classical Be stars are early-type B-type stars whose spectra have one or more emission lines in the
Balmer series. In particular, the H$\alpha$ emission line is typically the dominant feature in the spectra
of such stars, and many authors have modeled H$\alpha$ line profiles to understand the Be star phenomenon better.

The emission lines observed in the spectra are explained in terms of the recombination that occurs in a flattened 
circumstellar disk, according to the widely accepted model first proposed by Struve (\cite{struve31}). The
disk is a decretion disk; i.e., the source of the disk material is the central star, generated by the equatorial flow
of stellar material. One of the key factors in creating the disk is supposed to be the very high value of rotational velocity.
In fact, Be stars are known to have higher rotational velocities than a sample of normal B-type stars. From statistical 
considerations on the $v \sin i$ distribution among Be stars, Porter (\cite{porter96}) estimated that Be stars rotate at
a equatorial velocity equal to 80 $\%$ of the critical rotation velocity:

\begin{displaymath}
 v_{\rm eq}\,=\,0.8\,\sqrt{\frac{GM_*}{R_*}} .
\end{displaymath}

The observed emission lines take a variety of shapes, which following the scheme proposed by Hanuschik (\cite{hanu96}),
range from wine bottle profiles, singly or double-peaked profiles, to shell spectra, when the central absorption must
extend below the stellar continuum flux. The various shapes are explained as a dependency of $i$, the inclination angle
of the star's rotation axis to the observer's line of sight. In particular, shell profiles occur only when the disk is
viewed equator-on ($i\,=\,90^\circ$), while the single peak and wine bottle occur only for near pole-on 
($i\,=\,0^\circ$) viewings, and double-peaked profiles occur at mid-inclination angles.

Double-peaked profiles have been observed both symmetric, which are the two peaks have the same intensity, and asymmetric,
the peaks have different height over the continuum level. The current theory is that asymmetry arises from one-armed
density waves in the circumstellar disk, which is also known as the global disk oscillation model. In this model, a 
one-armed oscillation mode is superposed on an unperturbed, axisymmetric disk (Okazaki, \cite{okazaki97}). Another aspect 
of Be stars emission is their variability. For example, about one third of all double-peaked profiles exhibit
changing asymmetry, with the so-called violet-to-red ratio (V/R) being cyclically variable on timescales of years to 
decades.

Observations in different spectral regions help astronomers probe different regions of the stellar disk and then
put constraints on modeling these stars. For example, recently Meilland et al. (\cite{meilland07}) have used
VLTI/AMBER to observe $\alpha$ Arae in the Br$_\gamma$ line, which was constrained very strongly the rotational property 
of its disk, concluded that its rotation is purely Keplerian. 

In this paper we present an atlas of observed H$\alpha$ and H$\beta$ spectral lines in a sample of bright Be stars. We payed 
particular attention to modeling the H$\alpha$ profile and to the time variability of Balmer lines profiles.

\begin{table}
\caption{Spectral types, luminosity classes, Str\"omgren photometry, and derived effective temperatures for our 
program stars. Spectral types and luminosity classes are taken from the SIMBAD database}
\label{param}
\centering
\begin{tabular}{rlrccccrr} 
\hline\hline             
HD~~ & ~~Spec. & b-y~~ & m$_1$ & c$_1$ & H$\beta$ & T$_{\rm eff}$ (K) \\
\hline
  6811 & B7\,Ve      &    0.006 & 0.082 & 0.697 & 2.676 & 12600 \\
 10516 & B2\,Vep     &      --  &   --  &  --   &  --   & 25410 \\
 11415 & B3\,III     & $-$0.059 & 0.094 & 0.419 & 2.665 & 15680 \\
 37202 & B2\,IVp     &  --~~~~  &  --   &  --   &  --   & 21480 \\
 37490 & B3\,III     & $-$0.016 & 0.060 & 0.184 & 2.576 & 20200 \\
 41335 & B2\,Vne     &    0.031 & 0.020 & 0.002 & 2.463 & 20230 \\
 43285 & B6\,Ve      & $-$0.055 & 0.104 & 0.481 & 2.670 & 14840 \\
 44458 & B1\,Vpe     &   --     &   --  & --    &  --   & 26600 \\
 45542 & B6\,III     & $-$0.049 & 0.106 & 0.544 & 2.657 & 14080 \\
 47054 & B8\,Ve      & $-$0.022 & 0.096 & 0.698 & 2.677 & 12500 \\
 50658 & B8\,IIIe    & $-$0.015 & 0.086 & 0.596 & 2.626 & 13600 \\
 50820 & B3\,IVe+    &  --~~~~  &  --   &  --   &  --   & 17900 \\
 52918 & B1\,V       & $-$0.072 & 0.063 & 0.021 & 2.591 & 24720 \\
 53416 & B8          &  --~~~~  &  --   &  --   &  --   & 12000 \\
 58050 & B2\,Ve      & $-$0.087 & 0.081 & 0.169 & 2.495 & 20260 \\
 58343 & B2\,Vne     &    0.011 & 0.073 & 0.264 & 2.566 & 18600 \\
 58715 & B8\,Ve      & $-$0.037 & 0.124 & 0.791 & 2.729 & 11700 \\
 60855 & B2/B3\,V    & $-$0.017 & 0.059 & 0.194 & 2.609 & 19984 \\
 61224 & B8/9 IV     &    0.041 & 0.061 & 0.773 & 2.656 & 12000 \\
 65875 & B2.5Ve      &    0.004 & 0.040 & 0.102 & 2.451 & 22540 \\
 71072 & B4\,IV      &  --~~~~  &  --   &  --   &  --   & 16550 \\
 91120 & B8/9\,IV/V  &    0.005 & 0.095 & 0.956 & 2.742 & 10600 \\ 
109387 & B6III\,pe   & $-$0.023 & 0.072 & 0.351 & 2.566 & 16850 \\ 
138749 & B6\,Vnne    & $-$0.040 & 0.083 & 0.479 & 2.683 & 14910 \\
142926 & B9\,pec     & $-$0.029 & 0.109 & 0.764 & 2.677 & 12000 \\
142983 & B8\,Ia/Iab  & $-$0.022 & 0.081 & 0.664 & 2.590 & 12860 \\
143275 & B0.2\,IVe   & $-$0.016 & 0.036 & $-$0.020 & 2.605 & 26700 \\
162428 & A0          &  --~~~~  &  --   &   --  &   --  & 12200 \\
162732 & Bp\,she     & $-$0.045 & 0.096 & 0.786 & 2.669 & 11740 \\
164284 & B2\,Ve      &    0.057 & 0.022 & 0.001 & 2.495 & 21650 \\
164447 & B8\,Vne     & $-$0.013 & 0.095 & 0.660 & 2.721 & 12900 \\
183362 & B3\,Ve      & $-$0.009 & 0.056 & 0.184 & 2.528 & 20260 \\
183656 & B6\,sh      &    0.050 & 0.037 & 0.809 & 2.641 & 11700 \\
183914 & B8\,Ve      & $-$0.030 & 0.109 & 0.725 & 2.737 & 12270 \\
187567 & B2.5\,IVe   &    0.052 & 0.007 & 0.093 & 2.531 & 23110 \\
189687 & B3\,IVe     & $-$0.057 & 0.081 & 0.281 & 2.635 & 17980 \\
191610 & B2.5\,Ve    & $-$0.039 & 0.062 & 0.162 & 2.553 & 20600 \\
192044 & B7\,Ve      & $-$0.032 & 0.086 & 0.603 & 2.632 & 13460 \\
193911 & B8\,IIIne   & $-$0.015 & 0.073 & 0.644 & 2.627 & 13100 \\
210129 & B7\,Vne     & $-$0.024 & 0.101 & 0.476 & 2.643 & 15000 \\
212571 & B1\,Ve      & --~~~~   &  --   &  --   &   --  & 23300 \\
214168 & B2\,Ve      & $-$0.045 & 0.055 & 0.024 & 2.609 & 24790 \\
216057 & B5\,Vne     & $-$0.013 & 0.088 & 0.470 & 2.671 & 15090 \\
216200 & B3\,IVne    &    0.100 & 0.040 & 0.331 & 2.634 & 17600 \\
217050 & B4\,IIIep   & $-$0.006 & 0.062 & 0.472 & 2.501 & 15090 \\
217543 & B3\,Vpe     & $-$0.034 & 0.079 & 0.241 & 2.650 & 18860 \\
217675 & B6\,Ve      &    0.002 & 0.042 & 0.477 & 2.650 & 15050 \\
217891 & B6\,Ve      & $-$0.050 & 0.109 & 0.476 & 2.630 & 14910 \\
\hline
\end{tabular}
\end{table}

\begin{table*}
\caption{Adopted astrophysical quantities for our stars and final best-fit parameters. For each star we report, the number of 
spectra collected, their classification (abs means absorption lines only), the projected equatorial rotational velocity as measured 
by us in our spectra, mass and radius as estimated from Drilling \& Landolt (\cite{drill99}), peaks separation, critical velocity, 
equatorial velocity, inclination angle, base density at the stellar equator, and the estimated disk radius. In the last column
we also reported the references of other similar catalogs in which spectra of the star have been published: 1) Hanuschik (\cite{hanu88}),
2) Hanuschik et al. (\cite{hanu96}), 3) Hanuschik (\cite{hanu96a}), 4) Slettebak et al. (\cite{slettebak92}), 5) Saad et al. (\cite{saad06}),
6) Silaj et al. (\cite{silay10}).}
\label{summary}
\centering
\begin{tabular}{rrrrrrcccccccrr} 
\hline\hline             
HD & HR & Name & N & Cl. & v$\sin i$ & M$_*$ & R$_*$ & $\Delta v_{\rm peak}$ & v$_c$ & v$_{\rm eq}$ & $i$ & $\rho _0$ & R$_d$/R$_*$ & rem\\
   &    &  &   &    & (km/s) & (M$_\odot$) & (R$_\odot$) & (km s$^{-1}$) & (km s$^{-1}$) & (km s$^{-1}$) & ($^\circ$) & (g cm$^{-3}$) & & \\
\hline
  6811&  335 & $\phi$ And &  3 & 2   &   80 & 4.0 & 3.1 &  -- & 500 & 400 & 12 & 8.50$\cdot$10$^{-14}$ & 6.4 & 3,5 \\
 10516&  496 & $\phi$ Per &  1 & 2   &  240 & 17.0& 10.0& 180 & 570 & 450 & 22 & 4.20$\cdot$10$^{-12}$ & 4.3 & 3,4,5\\
 11415&  542 & $\epsilon$ Cas &  1 & abs &   40 & 6.5 & 4.2 & --  & --  &  -- & -- &           --          & --  & --\\
 37202& 1910 & $\zeta$ Tau &  2 & 2   &  120 &10.5 & 6.1 & 290 &  -- &  -- &  --&    --   &  --         & 1,2,3,4,5 \\
 37490& 1934 & $\omega$ Ori&  2 & 1   &  180 & 8.9 & 5.1 & 190 & 580 & 460 & 23 & 8.00$\cdot$10$^{-12}$ & 3.1 & 1,2,3,4\\
 41335& 2142 & V696 Mon &  2 & 2   &  310 & 8.9 & 5.1 &  -- &  -- &  -- & -- &            --         & --  & 1,2,3\\
 43285& 2231 & --~~~~~ &  1 & abs &  230 & 6.0 & 4.0 &  -- & --  &  -- & -- &          --           & --  & --\\
 44458& 2284 & FR CMa &  1 & 2   &  180 &18.0 &10.5 & 150 & 570 & 460 & 23 & 3.00$\cdot$10$^{-12}$ & 5.0 & 3\\
 45542& 2343 & $\nu$ Gem&  1 & sh  &  160 & 5.1 & 3.5 & 175 & --  &  -- & -- &         --            & --  & 2,3,4\\
 47054& 2418 & --~~~~~&  1 & 1   &  180 & 4.0 & 3.1 & 140 & 500 & 400 & 27 & 3.80$\cdot$10$^{-13}$ & 5.8 & 3\\
 50658& 2568 & $\psi$09 Aur &  1 & sh  &  230 & 4.5 & 3.3 & 150 & 510 & 400 & 80 & 3.50$\cdot$10$^{-14}$ & 8.5 & 3,5\\
 50820& 2577 & --~~~~~&  1 & 2   &  150 & 7.4 & 4.6 &  -- & 550 & 440 & 20 & 7.20$\cdot$10$^{-14}$ & 13.8 & --\\
 52918& 2648 & 19 Mon &  1 & abs &  240 & 17.0&10.0 &  -- & --  & --  & -- &         --            & --   & --\\
 53416&--~~~ &--~~~~~ &  1 & 1   &  190 & 4.0 & 3.5 & 120 & 470 & 370 & 30 & 7.60$\cdot$10$^{-14}$ & 9.1 & -- \\
 58050& 2817 & --~~~~~&  3 & 1   &  110 & 9.0 & 5.3 &  -- & --  & --  & -- &         --            & --   & 3,5\\
 58343& 2825 & FW CMa &  1 & 2   &   50 & 7.5 & 4.7 &  -- & 550 & 440 &  6 & 1.10$\cdot$10$^{-13}$ & 13.7 & 2,3,4\\
 58715& 2845 & $\beta$ CMi &  3 & 1   &  210 & 3.8 & 3.0 & 130 & 490 & 390 & 32 & 1.25$\cdot$10$^{-13}$ & 7.0 & 2,3,4,5\\
 60855& 2921 & V378 Pup &  1 & 2   &  240 & 9.0 & 5.3 & 150 & --  &  -- &  --&      --               & --  & 2,3\\
 61224& 2932 & --~~~~~&  2 & 1   &  200 & 4.0 & 3.1 & 150 & 500 & 400 & 30 & 3.50$\cdot$10$^{-13}$ & 6.2 & 3\\
 65875& 3135 & V695 Mon &  3 & 2   &  180 & 12.0& 7.0 & --  & --  &  -- & -- &           --          & --  & 2,3\\
 71072&--~~~ & --~~~~~&  2 & abs &  100 & 6.5 & 4.5 & --  & --  & --  & -- &           --       & --  & --\\
 91120& 4123 & --~~~~~&  9 & 1   &  250 & 3.0 & 2.5 & 180 & 480 & 380 & 41 & 1.15$\cdot$10$^{-13}$ & 6.8 & 2,3\\  
109387& 4787 & $\kappa$ Dra &  13 & 1$\leftrightarrow$2   &  170 & 7.0 & 4.5 & 140 & 540 & 430 & 23 & 4.00$\cdot$10$^{-12}$ & 5.1& 3,5 \\
138749& 5778 & $\theta$ CrB &  11 & abs&  310 & 6.0 & 4.0 & --  &  -- & --  & -- &           --          & --  & 5\\
142926& 5938 & V839 Her &  10 & sh &  275 & 4.0 & 3.1 & 240 & 500 & 400 & 44 & 6.00$\cdot$10$^{-12}$ & 4.3 & 5\\
142983& 5941 & 48 Lib &  13 & sh &  390 & 4.1 & 3.2 &  -- & --  &  -- & -- &          --           &  -- & 2,3\\ 
143275& 5953 & $\delta$ Sco &  4 & 2   &  165 &18.0 &10.5 & --  & --  & --  & -- &           --          & --  & 6\\
162428&--~~~ & --~~~~~ &  5 & 1   &  240 & 4.0 & 3.0 & 180 & 500 & 400 & 36 & 9.00$\cdot$10$^{-12}$ & 6.1 &4 \\
162732& 6664 & $\zeta$ Her &  6 & sh  &   40 & 3.8 & 3.0 & 260 &  -- & --  & -- &           --          &  -- & 3,5\\
164284& 6712 & 66 Oph &  6 & 1   &  240 &10.5 & 6.1 & 150 & 570 & 460 & 32 & 8.80$\cdot$10$^{-14}$ & 9.3 &2,3\\
164447& 6720 & V974 Her &  4 & 1   &  180 & 4.0 & 3.5 & 140 & 470 & 370 & 28 & 1.75$\cdot$10$^{-13}$ & 5.8 &5\\
183362& 7403 & V558 Lyr &  4 & 2   &  220 & 9.0 & 5.3 & --  & --  &  -- & -- &         --            & --  &3\\
183656& 7415 & V923 Aql &  5 & sh  &  190 & 3.8 & 3.0 & 250 & --  & --  & -- &          --           & --  &2,3\\
183914& 7418 & $\beta$ Cyg B&  5 & 1   &  240 & 4.0 & 3.0 & 160 & 500 & 400 & 37 & 9.00$\cdot$10$^{-14}$ & 8.2 &3,4\\
187567& 7554 & V1339 Aql &  4 & 1$\rightarrow$2   &  200 & 13.0 & 7.7 & 140& 570 & 450 & 26 & 3.00$\cdot$10$^{-13}$ & 7.3 & --\\
189687& 7647 & 25 Cyg &  6 & sh  &  230 & 7.5 & 4.7 & 250  & 550 & 440 & 31 &  3.05$\cdot$10$^{-13}$& 3.6 & 3,4\\
191610& 7708 & 28 Cyg &  4 & 1   &  250 & 9.0 & 5.3 & 270 & 570 & 455 & 33 & 2.60$\cdot$10$^{-13}$ & 3.9 & 3,4\\
192044& 7719 & 20 Vul &  3 & 1   &  240 & 4.5 & 3.3 & 150 & 510 & 410 & 38 & 1.10$\cdot$10$^{-13}$ & 10.1& --\\
193911& 7789 & 25 Vul &  3 & 1   &  170 & 4.0 & 3.5 & 100 & 470 & 370 & 27 & 2.30$\cdot$10$^{-13}$ & 6.0 & 2,3,5\\
210129& 8438 & 25 Peg &  8 & 1   &  130 & 6.0 & 4.0 &  70 & 530 & 430 & 19 & 1.50$\cdot$10$^{-12}$ & 5.6 &1,2,3\\
212571& 8539 & $\pi$ Aqr&  3 & 1   &  220 & 13.0 & 7.7 & 280 & 570 & 450 & 41 & 6.00$\cdot$10$^{-13}$ & 3.9& 1,2,3,4 \\
214168& 8603 & 8 Lac B &  3 & 2   &  150 & 17.0 & 10.0&  -- &  -- &  -- & -- &         --           &  -- & --\\
216057& 8682 & --~~~~~&  4 & abs &  260 & 6.0 & 4.0 & --  &  -- & --  & -- &        --             &  -- & --\\
216200& 8690 & 14 Lac&  3 & sh  &  195 & 13.0& 7.7 & --  &  -- & --  & -- &       --              & --  & 5\\
217050& 8731 & --~~~~~&  1 & sh  &  240 & 6.0 & 4.0 & --  &  -- & --  & -- &       --              &  -- &3,4,5\\
217543& 8758 & V378 And &  2 & sh  &  305 & 7.5 & 4.7 & --  & --  & --  & -- &       --              &  -- & --\\
217675& 8762 & $o$ And &  2 & sh  &  200 & 6.0 & 4.0 & --  &  -- &  -- & -- &        --             &  -- & 4,5\\
217891& 8773 & $\beta$ Psc &  6 & 2   &   75 & 6.0 & 4.0 & --  & 535 & 430 & 10 & 5.20$\cdot$10$^{-13}$ & 6.1 &1,2,4,5  \\
\hline
\end{tabular}
\end{table*}

\section{Observation and data reduction}
The idea that underlies this catalog is to create a homogeneous data set of stars observed with the same instrument,
the telescope of the {\it M.G. Fracastoro} station of INAF-Catania Astrophysical Observatory. For this purpose,
we queried the ``Catalogue of Be stars'' compiled by Jaschek \& Egret (\cite{jaschek82}) selecting all the objects with 
V$\leq$7 and observable at the latitude of the observatory, which means all the stars with $\delta \ge -22^\circ$. The result 
of this query is the sample of 48 Be stars reported in Table~\ref{param}. The limiting magnitude was chosen to obtain a good 
compromise between the exposure time and the signal-to-noise ratio.

\begin{figure}
   \includegraphics[width=8cm,height=8cm]{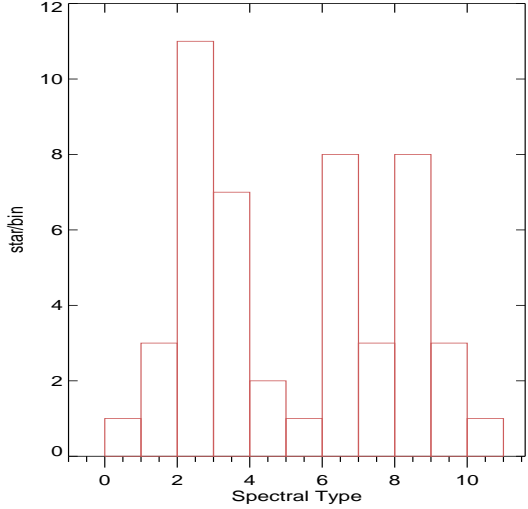}
      \caption{Histogram showing the distribution of program stars as a function of spectral type.
              }
         \label{spect_type}
   \end{figure}

The present catalog is based on new spectroscopic observations of all the stars in our sample, which spectral type are 
distributed between B1 and A0 (according to the histograms displayed in Fig.~\ref{spect_type}), and luminosities classes 
are III, IV, and V, as in the SIMBAD database\footnote{http://simbad.u-strasbg.fr/simbad/}. Some of these stars' spectra 
have never been published in other catalogs similar to ours, at least to our knowledge.late

All the spectra of our program stars have been acquired with the 91-cm telescope and FRESCO, the fiber-fed REOSC echelle 
spectrograph that allows spectra to be obtained in the range of 4300--6800 {\AA} with a resolution R=\,21\,000. The spectra were 
recorded on a thinned, back-illuminated (SITE) CCD with 1024$\times$1024 pixels of 24~$\mu$m size, whose typical readout noise 
is of about 8 e$^{-}$ and the gain is 2.5 e$^-$/ADU. All the spectra have been acquired during several observing runs spanning 
two years between 2008 and 2009.

The reduction of spectra, which included the subtraction of the bias frame, trimming, correcting for the flat-field and the 
scattered light, the extraction of the orders, and the wavelength calibration, was done by using the NOAO/IRAF package\footnote{IRAF 
is distributed by the National Optical Astronomy Observatory, which is operated by the Association of Universities for Research 
in Astronomy, Inc.}. The amount of scattered light correction was about 10 ADU. 
After dividing the extracted spectra by flat-field, the residual shape the spectrum was removed by dividing each spectral 
order by a Legendre function of a low order. Typical S/N of our spectra is $\sim$\,100. For some stars this limit has not been
reached because its apparent magnitude is close to V$\approx$7, in which case the S/N was about 50.

Finally, the IRAF package {\sf rvcorrect} was used to include the velocity correction due to the Earth's motion, which moved 
the spectra into the heliocentric rest frame. The task {\sf splot} and its facilities were used to measure the peaks separation 
in the H$\alpha$ profile. Errors in the pixels position were converted in errors on the separations, and were evaluated in 
$\approx$~20~km~s$^{-1}$.

For each spectrum we measured the equivalent widths (including underlying absorption) of both H$\alpha$ and H$\beta$, where a 
negative value means the corresponding line shows net emission. All the measured equivalent widths are reported in Table~\ref{eqw}.

\begin{table}
\caption{Program stars for which we have more than two spectra. We reported for both lines mean equivalent widths, their
         F values, computed according Eq.~\ref{Fval}, and the confidence levels of the detected variability.}
\label{eqw_test}
\begin{tabular}{r|rcc|rcc} 
\hline
\hline
        & \multicolumn{3}{c}{H$\alpha$}  &  \multicolumn{3}{c}{H$\beta$} \\
\hline
     HD &    EW(\AA)       &  F   & C    &     EW(\AA)     & F    & C    \\
\hline              
  6811  &    0.87$\pm$0.28 & 1.84 & 0.65 &   6.45$\pm$0.09 & 1.99 & 0.67 \\
 37202  &$-$15.26$\pm$1.17 & 1.41 & 0.49 &   2.34$\pm$0.01 & 1.41 & 0.49 \\
 37490  & $-$5.96$\pm$1.09 & 1.41 & 0.49 &   1.83$\pm$0.18 & 1.41 & 0.49 \\
 41335  &$-$31.92$\pm$0.78 & 1.41 & 0.49 &$-$1.44$\pm$0.52 & 1.41 & 0.49 \\
 58050  & $-$4.40$\pm$0.31 & 1.94 & 0.66 &   5.49$\pm$0.04 & 1.73 & 0.63 \\
 58715  & $-$1.19$\pm$0.36 & 1.94 & 0.66 &   7.96$\pm$0.06 & 1.98 & 0.66 \\
 61224  & $-$5.59$\pm$0.48 & 1.41 & 0.49 &   5.14$\pm$0.40 & 1.41 & 0.49 \\
 65875  &$-$43.00$\pm$0.89 & 1.90 & 0.66 &   0.04$\pm$0.60 & 1.77 & 0.64 \\
 71072  &    3.14$\pm$0.08 & 1.41 & 0.49 &   5.22$\pm$0.23 & 1.41 & 0.49 \\
 91120  & $-$0.40$\pm$0.50 & 3.49 & 0.92 &   8.07$\pm$0.15 & 3.21 & 0.91 \\
109387  &$-$19.17$\pm$0.44 & 3.58 & 0.94 &   3.04$\pm$0.06 & 2.85 & 0.90 \\
142926  & $-$1.63$\pm$0.45 & 3.68 & 0.93 &   7.05$\pm$0.13 & 3.06 & 0.90 \\
142983  &$-$24.66$\pm$0.86 & 3.31 & 0.93 &   3.09$\pm$0.14 & 3.85 & 0.95 \\
143275  &$-$10.13$\pm$2.29 & 2.34 & 0.76 &$-$0.81$\pm$0.20 & 2.30 & 0.75 \\
162428  &$-$15.57$\pm$0.14 & 2.44 & 0.80 &   4.78$\pm$0.38 & 2.40 & 0.80 \\
162732  & $-$6.61$\pm$0.88 & 2.52 & 0.83 &   5.92$\pm$0.50 & 2.89 & 0.85 \\
164284  &    1.37$\pm$1.57 & 2.71 & 0.84 &   5.18$\pm$0.08 & 2.64 & 0.84 \\
164447  & $-$0.95$\pm$1.13 & 2.64 & 0.81 &   6.68$\pm$0.15 & 2.68 & 0.82 \\
183362  &$-$26.39$\pm$0.15 & 1.88 & 0.65 &   1.25$\pm$0.15 & 1.79 & 0.64 \\
183656  & $-$7.68$\pm$0.72 & 2.50 & 0.80 &   5.15$\pm$0.18 & 2.28 & 0.80 \\
183914  & $-$0.94$\pm$0.22 & 2.78 & 0.82 &   8.10$\pm$0.07 & 2.53 & 0.80 \\
187567  &$-$21.82$\pm$0.66 & 2.14 & 0.74 &   0.56$\pm$0.19 & 2.21 & 0.74 \\
189687  &    0.93$\pm$0.25 & 2.93 & 0.86 &   4.76$\pm$0.05 & 3.11 & 0.87 \\
191610  &    0.12$\pm$0.24 & 2.03 & 0.72 &   4.69$\pm$0.15 & 2.04 & 0.72 \\
192044  & $-$9.96$\pm$0.13 & 1.97 & 0.66 &   5.25$\pm$0.05 & 1.96 & 0.66 \\
193911  & $-$5.45$\pm$0.13 & 1.95 & 0.66 &   5.10$\pm$0.05 & 1.89 & 0.65 \\
210129  &$-$14.17$\pm$0.21 & 3.26 & 0.90 &   5.11$\pm$0.15 & 3.13 & 0.89 \\
212571  & $-$5.40$\pm$0.04 & 1.99 & 0.67 &   1.98$\pm$0.14 & 1.98 & 0.66 \\
214168  &$-$14.79$\pm$0.35 & 2.00 & 0.67 &   1.87$\pm$0.52 & 1.82 & 0.65 \\
216057  &    4.62$\pm$0.15 & 2.36 & 0.76 &   7.19$\pm$0.10 & 2.21 & 0.74 \\
216200  &    1.15$\pm$0.43 & 1.98 & 0.66 &   5.08$\pm$0.21 & 1.97 & 0.66 \\
217543  &    2.35$\pm$0.76 & 1.41 & 0.49 &   5.31$\pm$0.11 & 1.41 & 0.49 \\
217675  &    3.05$\pm$0.28 & 1.41 & 0.49 &   5.64$\pm$0.18 & 1.41 & 0.49 \\
217891  &$-$12.94$\pm$0.64 & 2.70 & 0.84 &   4.70$\pm$0.19 & 2.70 & 0.84 \\
\hline                                                                           
\end{tabular}
\end{table}

\section{Classification and fit of the H$\alpha$ line profiles}
Almost all the stars in our sample were found with emission in the  H$\alpha$. Then, considering the shape of their profile and 
according to the classification scheme proposed by Hanuschik (\cite{hanu88}), we classified our stars as belonging to
\begin{itemize}
 \item class 1 when they exhibit a rather symmetrical double peak structure with V/R\,$\approx$\,1,
 \item class 2 when they have an asymmetric single peak or a dominant peak with a much weaker secondary peaked,
 \item shell stars when the central reverse is deeper than the continuum level,
 \item {\it abs} when there is not emission above the continuum level.
\end{itemize}

\noindent
Our classification is reported in the fourth column of Table~\ref{summary}.

Only for the stars belonging to class 1, plus stars from other classes but with V/R ratio close to unity, we attempt  
an estimation of the disk dimension. The approach we used was to minimize 
the difference between observed and synthetic profiles, computed in two separate steps. First of all, we calculated 
the photospheric H$\alpha$, then the contribution due to the net emission of the disk and then we added these two synthetic 
profiles obtained separately. These two steps are described in the following:

\begin{itemize}
\item {\it Computation of the photospheric profile}

We first computed the photospheric H$\alpha$ profiles for all our program stars. They were generated in three steps: 
{\it i)} first, we computed an LTE model atmosphere 
using the ATLAS9 code (Kurucz \cite{kur93}), {\it ii)} the stellar spectrum was then synthesized 
using SYNTHE (Kurucz \& Avrett \cite{kur81}), and {\it iii)} the spectrum was convolved with the instrumental and rotational 
profiles.

First of all we had to obtain an estimation of the effective temperature for each target. Considering that the continuum 
energy distribution of Be stars is typical of normal early-type stars both in the visual and UV, but not in the IR, where an excess could be
present because of the hot circumstellar dust (Zickgraf \cite{zick00}), effective temperatures were computed from Str\"omgren 
photometry (Hauck \& Mermilliod \cite{hauck98}) using the algorithm coded by Moon (\cite{moon85}), with the exception of eight stars 
for which photometry is not available. This method is allowed because it does not involve any IR filter. 
For seven of them we adopted the temperatures from the literature: HD\,10516, HD\,37202, and HD\,44458 from Soubiran et al. (\cite{soubiran10}),
HD\,58020 and HD\,71072 from Hohle et al. (\cite{hohle10}), HD\,162428 from Moujtahid et al. (\cite{mouj99}), 
and HD\,212571 from Wu et al. (\cite{wu11}), while for HD\,53416 we derived an estimation of temperature from spectral type and the 
calibration by Kenyon \& Hartman (\cite{kenyon95}). Since our targets have luminosity class IV/V (as reported in the SIMBAD
database), we fixed the surface gravity to $\log g $\,=\,4.0, except for HD\,11415, HD\,37490, HD\,45542, HD\,50658, 
HD\,109387, HD\,193911, and HD\,217050 (luminosity class III) 
for which $\log g$\,=\,3.0 has been preferred. Radii and masses were adopted following the calibration in Drilling \& Landolt (\cite{drill99}). 
Assuming these atmospheric parameters, we computed the v$\sin i$ of each star by spectral synthesis of the observed 
Mg{\sc i} $\lambda$4481 {\AA}. This line was chosen because in the spectral range of our targets, it reaches its maximum depth and therefore it 
is better suited to determining the rotational velocity. Errors on the projected rotational velocities are $\approx$~15~km~s$^{-1}$.

\item {\it Computation of net disk emission}

We have adopted the Be disk model approach of Hummel \& Vrancken (\cite{hummel00}) that is based on models developed by 
Horne \& Marsh (\cite{horne86}) and Horne (\cite{horne95}) for accretion disks in cataclysmic variables. The disk is assumed 
to be axisymmetric and centered over the equator of the underlying star, and the gas density varies as

 \begin{displaymath}
  \rho(R,Z)\,=\,\rho_0R^{-n} exp\left[-\frac{1}{2}\left(\frac{Z}{H(R)}\right)^2\right]
 \end{displaymath}

\noindent
where R and Z are the radial and vertical cylindrical coordinates (in units of stellar radii), $\rho_0$ is the base density 
at the stellar equator, {\it n} a radial density exponent, and H(R) the disk vertical scale height. The neutral hydrogen population 
within the disk is found by equating the photo-ionization and recombination rates (Gies et al. \cite{gies07}). The disk gas is 
assumed to be isothermal and related to the stellar effective temperature T$_{\rm eff}$ by T$_d$\,=\,0.6\,T$_{\rm eff}$ 
(Carciofi \& Bjorkman \cite{carciofi06}).

This approach take the contribution of the central star's finite size on the H$\alpha$ line formation process into account, 
i.e. the obscuration of the disk by the central star at any given inclination. The numerical model represents the disk by a large 
grid of azimuthal and radial surface elements, and the equation of transfer is solved along a ray through the center of each 
element according to 

\begin{displaymath}
 I_\lambda\,=\,S_\lambda^L\,(1-e^{-\tau_\lambda})\,+\,I_\lambda^S\,e^{-\tau_\lambda}
\end{displaymath}

\noindent
where I$_\lambda$ is the derived specific intensity, S$_\lambda^L$ the source function for the disk gas (taken as the Planck 
function for the disk temperature T$_d$), I$_\lambda^S$ the specific intensity for the H$\alpha$ of the star, and $\tau_\lambda$ 
the integrated optical depth along the ray. The first term applies to all the disk area elements that are unocculted by the star, 
while the second term applies to all elements that correspond to the projected photospheric disk of the star. The absorption line 
adopted in I$ _\lambda^S$ is Doppler-shifted according to solid-body rotation for the photospheric position in a star that is 
rotating at 80$\%$ of the critical value.

Electron scattering is not taken into account in the line profile computation. So we do not expect to reproduce the wings
of strong lines well, since for these lines the broadening of the wings due to the electron scattering can not be neglected. 
Disk kinematic is taken into account using a rotational velocity law written as 

\begin{displaymath}
 V(R)\,=\,V_{\rm rot}^*R^{-j}
\end{displaymath}

\noindent
(Hutchings \cite{hutch70}), where R represents the radial coordinate that has its origin at the center of the star, and $V_{\rm rot}^*$ 
denotes the actual rotational velocity at the stellar surface. The exponent ranges from $j\,=\,1/2$ for pure Keplerian rotation and
$j\,=\,1$ corresponding to conservation of angular momentum. Likewise the value of $j$ is still matter of debate, recent studies
seem to converge toward the Keplerian value (Hummel \& Vrancken \cite{hummel00}, Meilland et al. \cite{meilland07}).
Thus, in this study we assumed the disk to be in pure Keplerian rotation.

\end{itemize}

\noindent
Once we obtained and combined these two contributions, we started the minimization algorithm using as goodness-of-fit test 
the parameter

\begin{displaymath}
\chi^2 = \frac{1}{N} \sum (\frac{I_{\rm obs} - I_{\rm th}}{\delta I_{\rm obs}})^2
\end{displaymath}

\noindent
where $N$ is the total number of points, $I_{\rm obs}$ and $I_{\rm th}$ are the intensities
of the observed and computed profiles, respectively, and $\delta I_{\rm obs}$ is the photon noise.

As initial guesses for the inclination angle $i$ and for the disk radius, we used the equations

\begin{equation}
 v_{\rm eq} \sin i\,=\,0.8\,\sqrt{\frac{GM_*}{R_*}}\,\sin i 
\label{incl}
\end{equation}

\noindent
where $v_{\rm eq} \sin i$ is the value measured in our spectra, and 

\begin{equation}
 \frac{\Delta v_{\rm peak}}{2\,v \sin i}\,=\,r_d^{-j}
\label{rad}
\end{equation}

\noindent 
(equation [6] in Hanuschik et al. (\cite{hanu88})) where $\Delta v_{\rm peak}$ is the separation between violet and red peaks, 
as measured in each profile and reported in Table~\ref{summary}. 

Then, fine tuning was carried out using the {\it amoeba} minimization\footnote{The amoeba routine implements the simplex 
method of Nelder \& Mead (\cite{nelder65}).} algorithm between observed and computed profiles. In our
procedure two assumptions have been made: radial density exponent has been fixed to $n$\,=\,3 and, as  stated before, the Keplerian 
rotation of the disk has been considered ($j$\,=\,1/2). The first hypothesis, 
regarding the value of the density exponent can be justified by considering the work of Grundstrom \& Gies (\cite{grund06}). 
These authors computed several theoretical curves that described the dimension of the disk radius as a function of the H$\alpha$ 
equivalent width, for different values of the inclination angle i and different values of n. They concluded that the overall 
shape of those curves for different n and equal i are almost the same, since it is a small difference of $\approx 3 \%$ in 
correspondence of equivalent widths between -2 and -15 {\AA} when n change from 3 to 3.5. They then suggest that the 
particular choice of n is not as important as the choice of the right i. Moreover, Porter \& Rivinius (\cite{porter03}) 
from IR flux excess in Be stars suggested that n falls in the interval $n\,= \,2\div4$. Thus, on the basis of these results, 
we fixed the value of the density exponent to the middle value of n\,=\,3. Total H$\alpha$ profiles, star+disk, are presented 
in Figs.~\ref{halpha1},~\ref{halpha2}, and~\ref{halpha3}.

To derive an estimation of the disk radius, we used the method developed by Grundstrom \& Gies (\cite{grund06}) to form a 
synthetic image of the system star+disk in the plane of sky by summing the intensity over a 2.8 nm band centered on H$\alpha$. 
We collapsed this image along the projected major axis to get the summed spatial intensity, and we adopted the value for which the 
summed intensity drops to half its maximum value as effective disk radius. 

We estimated the errors on the disk dimensions and on the inclination angles to be $\pm$2\,R$_*$ and $\pm$3$^\circ$, respectively.
These determinations have been estimated by varying in Eqs.~\ref{incl} and in~\ref{rad} the observed quantities 
$v \sin i$ and peak separation by their experimental errors and considering as uncertainties the semi-amplitude of this variation.

All the adopted and derived parameters are reported in Table~\ref{summary}.

  \begin{figure}
   \includegraphics[width=8cm]{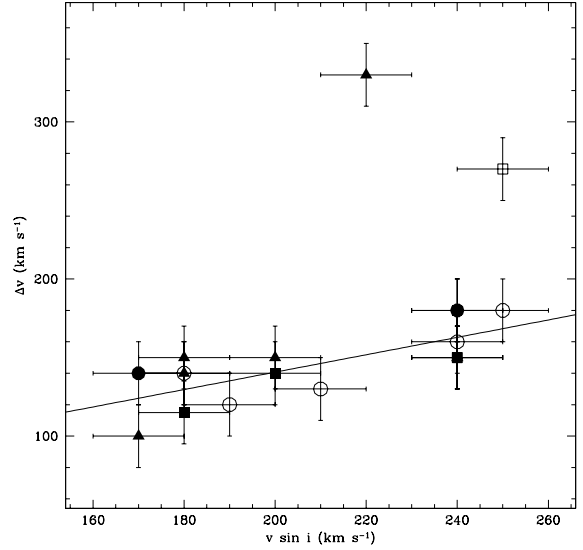}
      \caption{Correlation between measured v $\sin i$ and peaks separation. The points have been grouped on the basis of the 
       mean equivalent width: EW\,$\le$\,-20 {\AA} (filled squares), -20\,{\AA}\,$<$\,EW\,$\le$\,-15\,{\AA} (filled circles),
       -10\,{\AA}\,$<$\,EW\,$\le$\,-5\,{\AA} (filled triangles), -5\,{\AA}\,$<$\,EW\,$\le$\,0\,{\AA} (open circles), and
       EW\,$\ge$\,0 {\AA} (open squares).
              }
         \label{corr}
   \end{figure}

\section{H$\alpha$ and H$\beta$ Variability}
\label{variab}
Usually Be stars display variability in their equivalent width (EW) and/or in their spectral profile.

To find whether a star presents equivalent width variation, we applied to both H$\alpha$ and H$\beta$ equivalent widths 
the statistical method called F-test. When more than one observation was present for a given target, we calculated the amplitude 
of the variation, $\Delta\,EW$, and the standard deviation of the sample using

\begin{displaymath}
 \sigma\,=\,\sqrt{ \frac{1}{N-1} \sum (EW_i - \overline{EW})^2 }\\
\end{displaymath}

\noindent
where N is the number of points, and the (N-1) corresponds to the degree of freedom used for the F-test. Having obtained $\Delta EW$ 
and $\sigma$, we calculated a simple observable to assess the variability for a given target using a ratio of the form:

\begin{equation}
  F\,=\,\frac{\Delta\,EW}{\sigma} .
\label{Fval}
\end{equation}

\noindent
This simple ratio represents the number of times that the amplitude of the variation is greater than the standard deviation.

To determine whether or not this number is meaningful and whether the star shows variability, we evaluated the corresponding 
confidence level; see last column of Table~\ref{eqw_test}. We considered all the stars for which C\,$\ge$\,80$\%$ as definitively 
variable (13 stars, that is 41$\%$ of the sample), while we cannot say anything for the target with C\,$\le$\,50$\%$ 
(22$\%$ of the sample) due to the small number of points. The 37$\%$ of the stars, are probably variable but within 
a confidence level ranging from 50$\%$ to 80$\%$. In any case, all the targets show the same behavior in both spectral lines.

In this atlas we show the full set of our profiles (H$\alpha$ and H$\beta$) for the 48 observed stars.
In Figs.~\ref{variab1} to~\ref{variab5}, we show the Balmer profiles observed for program stars with multiple measurements,
In each panel we reported for each profile also the last four digits of the heliocentric julian date of the observation.
In some cases to improve the visualization, profiles have been blown up according to

   \begin{figure*}
   \includegraphics[width=8cm,height=7cm]{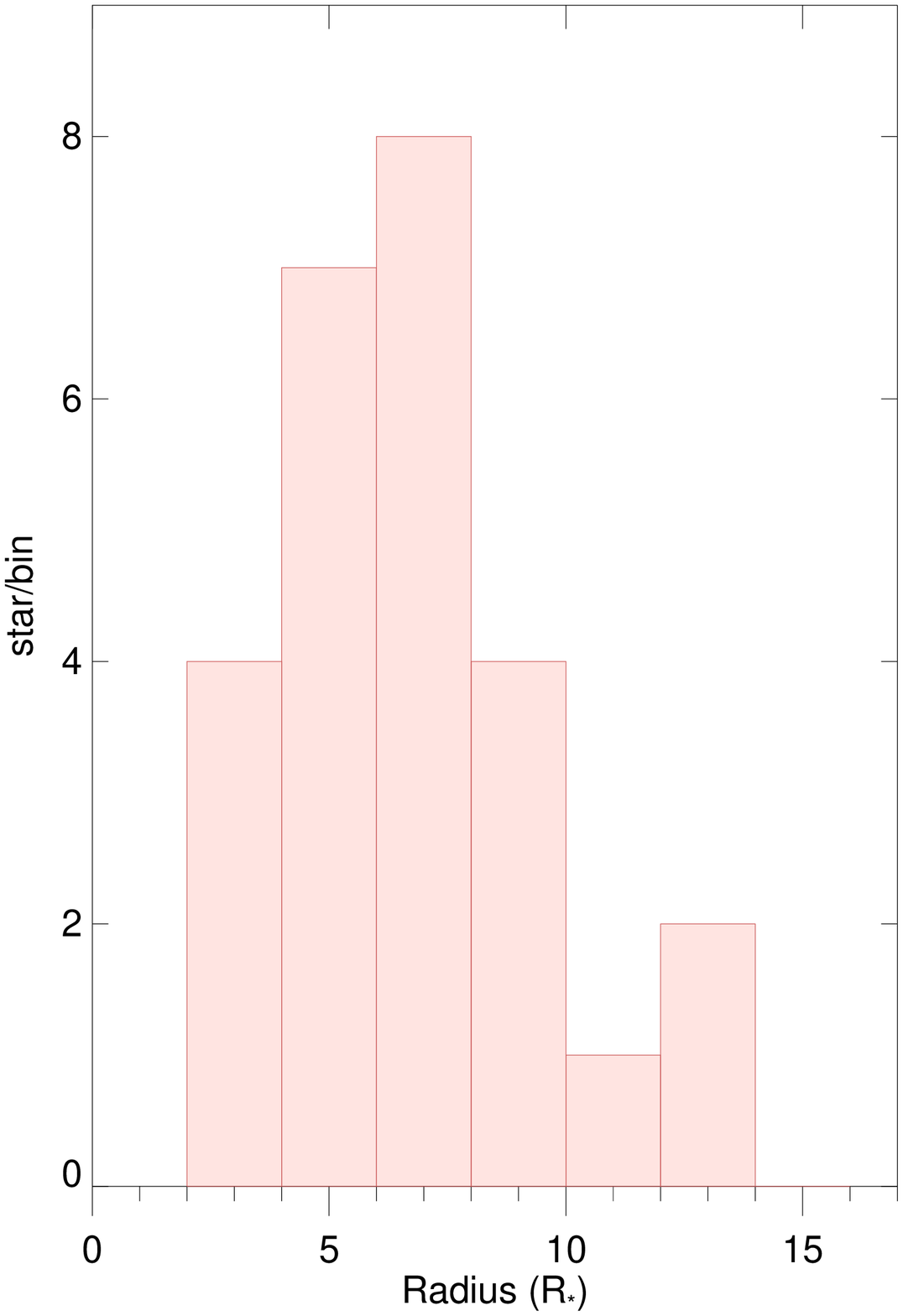}
   \includegraphics[width=8cm,height=7cm]{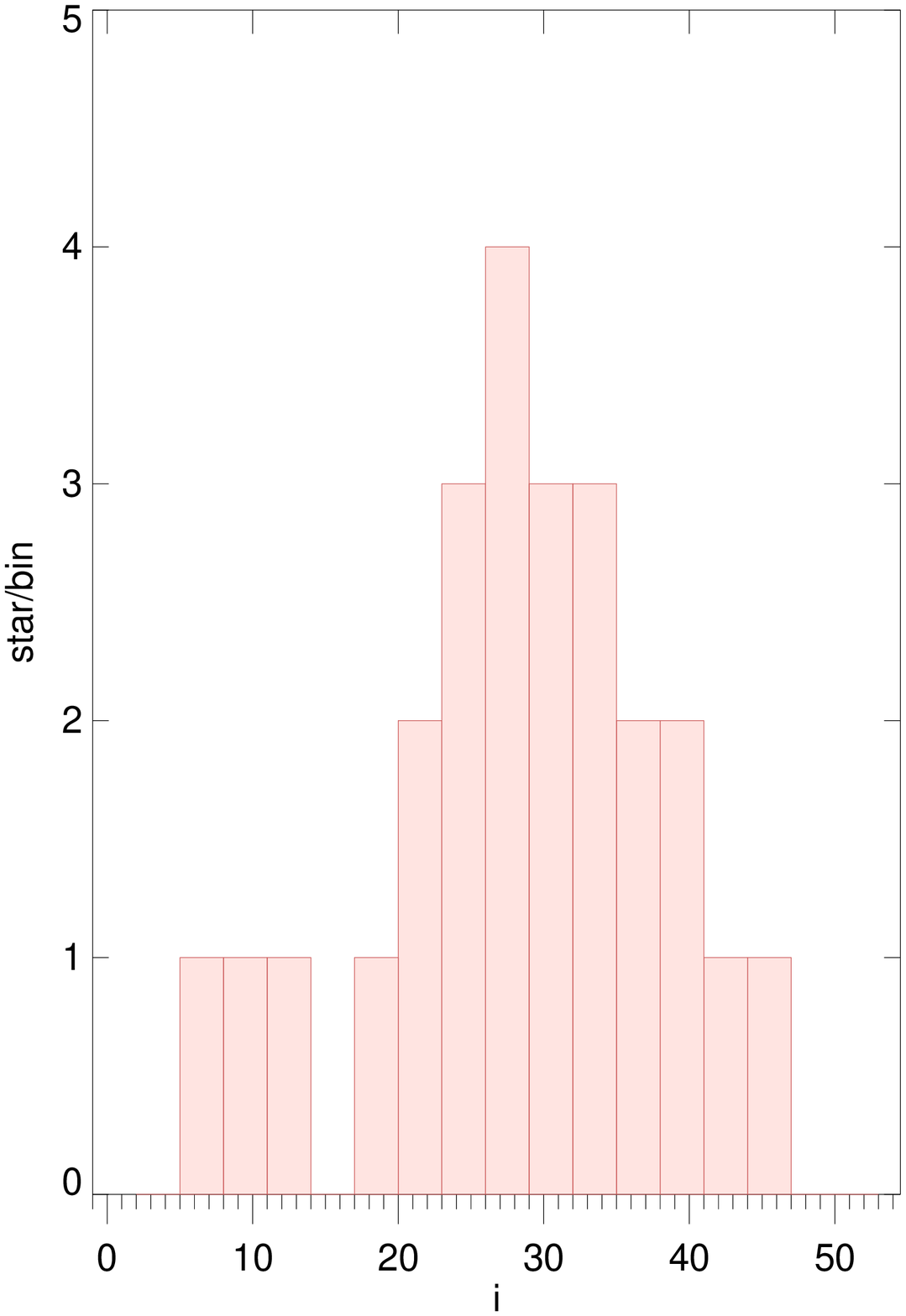}
      \caption{Histograms of distribution of stars as a function of the dimension of their disks (left panel) and of the inclination 
               angles (right panel). To build these histograms we chose a bin size of 2 R$_*$ and 3$^\circ$, respectively. }
         \label{hist}
   \end{figure*}

\begin{equation}
 F'_\lambda = [(F_\lambda - 1) * k] + 1
\end{equation}
 \noindent
where F$_\lambda$ denotes the observed flux, F$'_\lambda$ the displayed flux, and k is the magnification factor.

In Fig.~\ref{one_shell} we show all the stars for which we collected one spectrum only. Most of the stars have been 
measured several times during the 2008/2009 period, but only few stars showed significant variations
in their spectral profiles. These objects are discussed separately in the following:

{\it HD\,37202 -} This star has been observed in two nights separated by 205 days. It is clearly visible 
an increase in flux in the violet peak in both spectral lines, even if it is more evident in H$\beta$. 

{\it HD\,41335 -} As for the previous object, only two observations have been acquired for this star in a
range of 207 days. The star exhibits V/R variations in both lines.

{\it HD\,58050 -} The triple-peaked structure visible in the H$\alpha$ observed in the first night is missing
in the other two spectra. No features are seen in the H$\beta$.

{\it HD\,109387 -} This star was observed for 13 nights in a period spanning 110 days. The top part
of the H$\alpha$ emission profiles shows irregular variability, with back and forth changing from class 2 toward class 1, 
but no evident sign of day-by-day variability
has been observed. No variations have been detected in H$\beta$ double-peaked profile.  

{\it HD\,142926 -} During 110 days this shell star was observed 10 times showing a slight V/R variability
in the H$\alpha$.

{\it HD\,143275 -} In the four spectra acquired by us, $\delta$\,Sco shows important changes in the emission
of H$\alpha$. The V/R changes over the observational period and it shows a flat core in the first spectrum. 

{\it HD\,164284 -} Observed for six times in 536 days, this star exhibits equivalent width variability. The 
double-peaked emission of the H$\alpha$ profile decrease with time, although the V/R remains constant and 
$\approx$\,1. Also the H$\beta$ shows a change in the shape, since it is the last profile without any peak.

{\it HD\,183362 -} This star shows an increased emission level on the red side of its profile.

{\it HD\,183656 -} V/R variability has been observed, both in H$\alpha$ and in H$\beta$ profiles, in the six spectra 
acquired in a range of 358 days.

{\it HD\,187567 -} This star show variability in the H$\alpha$ profile, and evolves from class 1 toward class 2.

{\it HD\,189687 -} In the H$\alpha$, this star does not show any sign of variability for the first month of observations. 
In the last spectrum taken after 32 days from the second to last, it starts to show an increase of the flux in the red peak. 
No H$\beta$ variability has been detected. 

{\it HD\,191610 -} This star shows a gradual increment of the flux in the red peak.

\section{Discussion and conclusions}
In this paper we presented a homogeneous sample of H$\alpha$ and H$\beta$ line profiles observed in 32 Be stars, which show 
emission at least in the H$\alpha$ line. According to Hanuschik (\cite{hanu88}), we classified our targets on the basis of
the following scheme:

\begin{itemize}
 \item 16 stars, 33$\%$ of the sample, in class 1;

 \item 13 stars, 27$\%$ of the sample, in class 2;

 \item 11 stars, 23$\%$ of the total, has been classified as shell stars;

 \item 6 stars, 12$\%$ of the total, do not show net emission.
\end{itemize}

\noindent
In this list we do not include the two stars that show any phase transition between classes 1 and 2.

This frequency distribution shows that the majority of our sample of 48 Be stars, randomly distributed in spectral type, 
belongs to class 1 profiles. Regarding the 13 stars classified as class 2, seven of them are single peak, while six show 
structured profiles. 

Two stars showed variability from one class to another. HD\,187567 has undergone an evolution from class 1 to class 2, while
the behavior of HD\,109387 is more complicated. In 110 days, this star has been observed 13 times, and it showed a transition
from class 2 to class 1 and back again to class 2.

In Fig.~\ref{corr} we compared the behavior of the H$\alpha$ peaks separation to $v \sin i$ for the stars of the class 1
(included the class 2 HD\,60855). A linear correlation seems to exist, as expected from the work of Hanuschik et al. (\cite{hanu88}), 
although two stars discarded from this trend, namely HD\,191610 and HD\,212571. To verify this correlation we computed 
the Pearson {\it r} coefficient, obtaining {\it r}\,=\,0.72 and the linear fit given by the equation:

\begin{equation}
 y = (0.56 \pm 0.13) \cdot x + (29.94 \pm 26.80) .
\end{equation} 

\noindent
Thus, we confirm that $v \sin i$ and H$\alpha$ peaks separation are in linear correlation, confirming the disk-like
geometry of Be star envelopes and, probably this assumption is not valid for HD\,191610 and HD\,212571.

For all the stars belonging to the class 1, we attempted to model the emission with the purpose of deriving some
parameters such inclination angle, base density at the stellar equator and disk radius. In Fig.~\ref{hist} we show the 
distribution of stars as a function of the disk radius (right panel) and of the inclination angle (left panel). The histograms
were built considering a binning equal to the estimated errors, that is, 2R$_*$ and 3$^\circ$ in the disk dimension and
inclination angle, respectively. They show as there is a major concentration of stars for disk around $6 \div 8$ R$_*$ 
(about 17$\%$ of our sample) and for angles around $23^\circ \div 35^\circ$ (about 28$\%$ of the sample).

Moreover, with the aim of inferring line profile variability, for most of the stars of our sample we obtained more than one 
spectrum in a period spanning two years between 2008 and 2009. All but seven stars, those discussed in Sect.~\ref{variab}, 
do not show any evident sign of variability in both Balmer lines.

\begin{acknowledgements}
This research made use of the SIMBAD database, operated at the CDS, Strasbourg, France.
\end{acknowledgements}

\newpage
\appendix

\section{Fit emission}
   \begin{figure*}
   \begin{center}$
   \begin{array}{cc}

   \includegraphics[width=8cm]{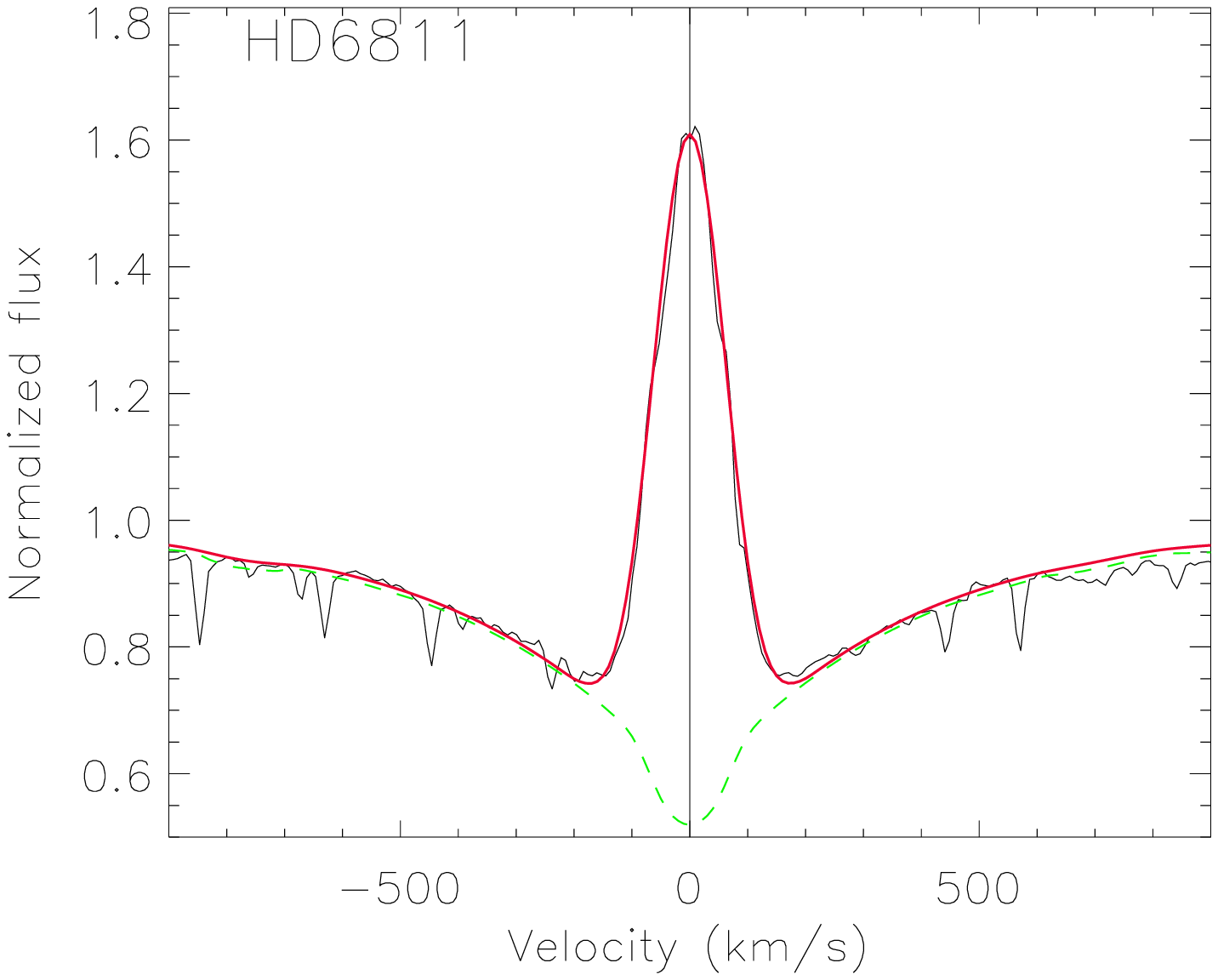} &
   \includegraphics[width=8cm]{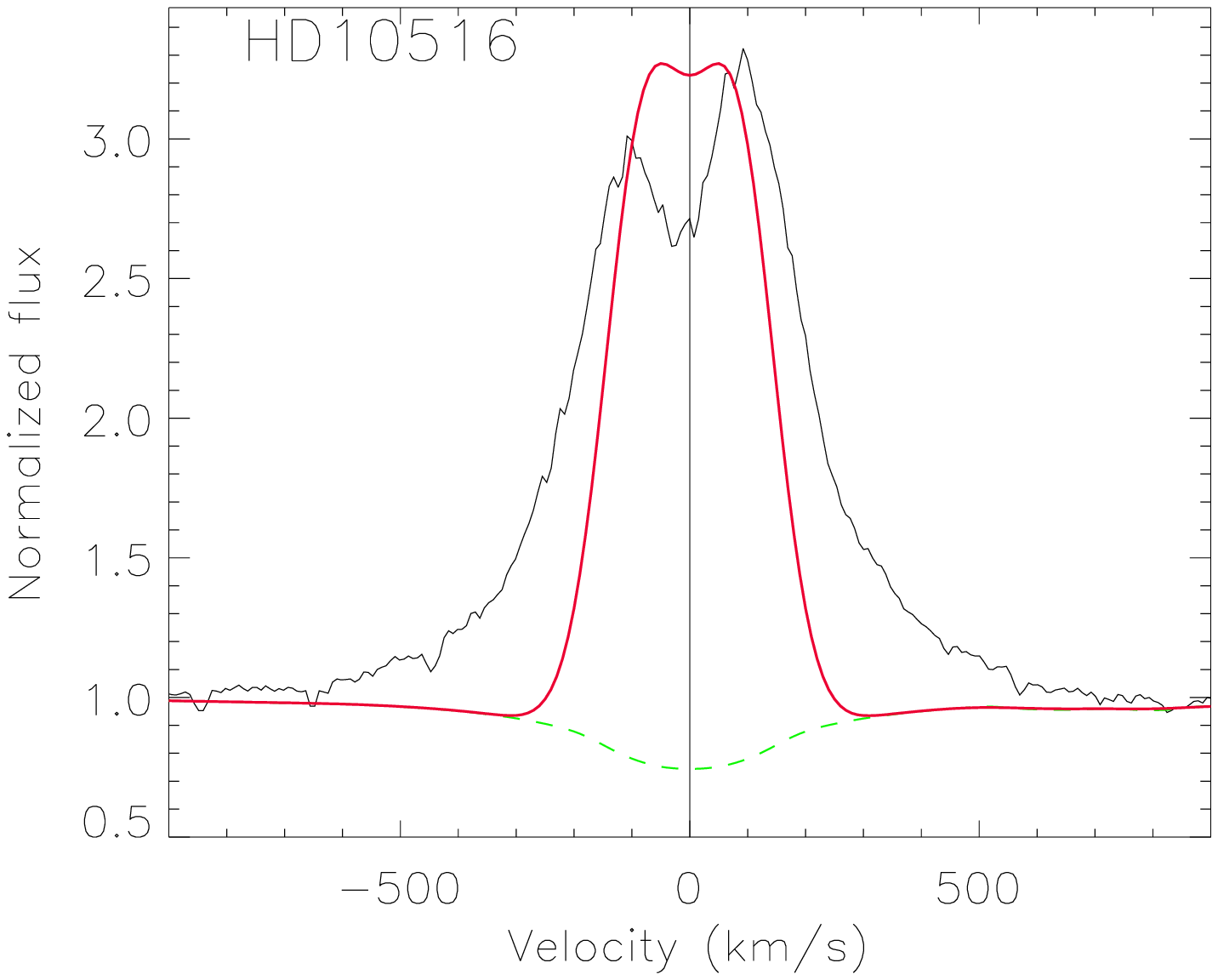} \\
   \includegraphics[width=8cm]{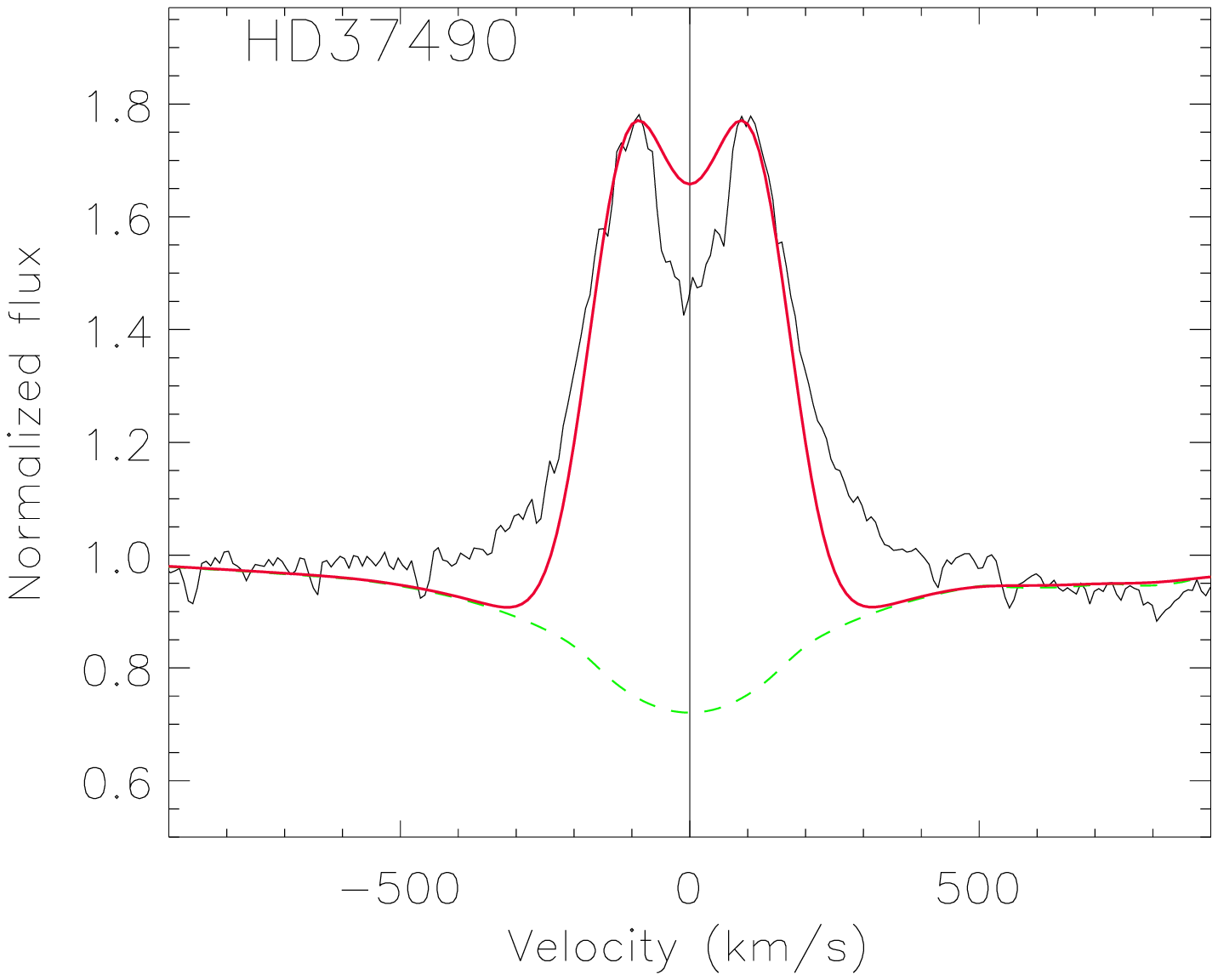} & 
   \includegraphics[width=8cm]{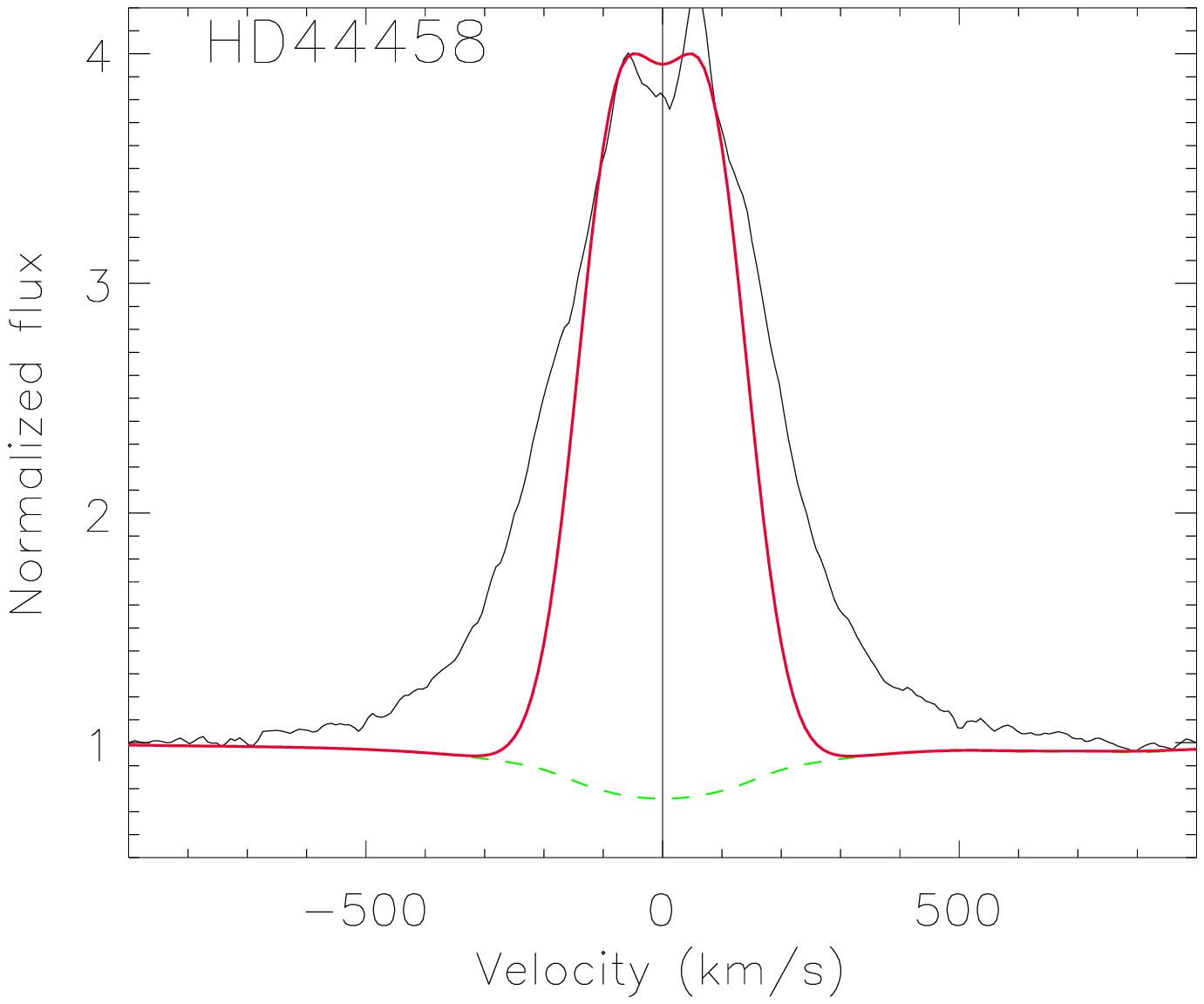} \\
   \includegraphics[width=8cm]{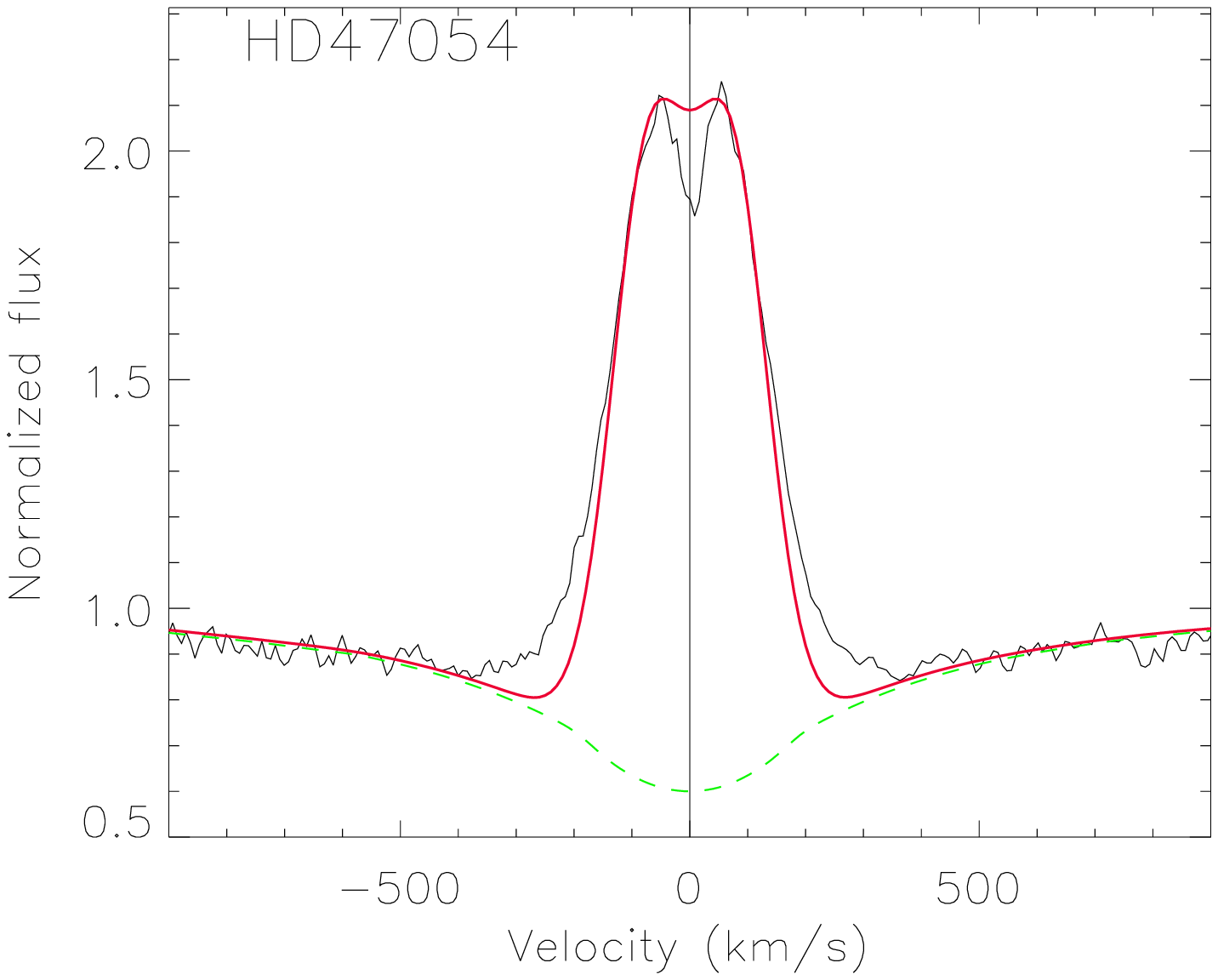} &
   \includegraphics[width=8cm]{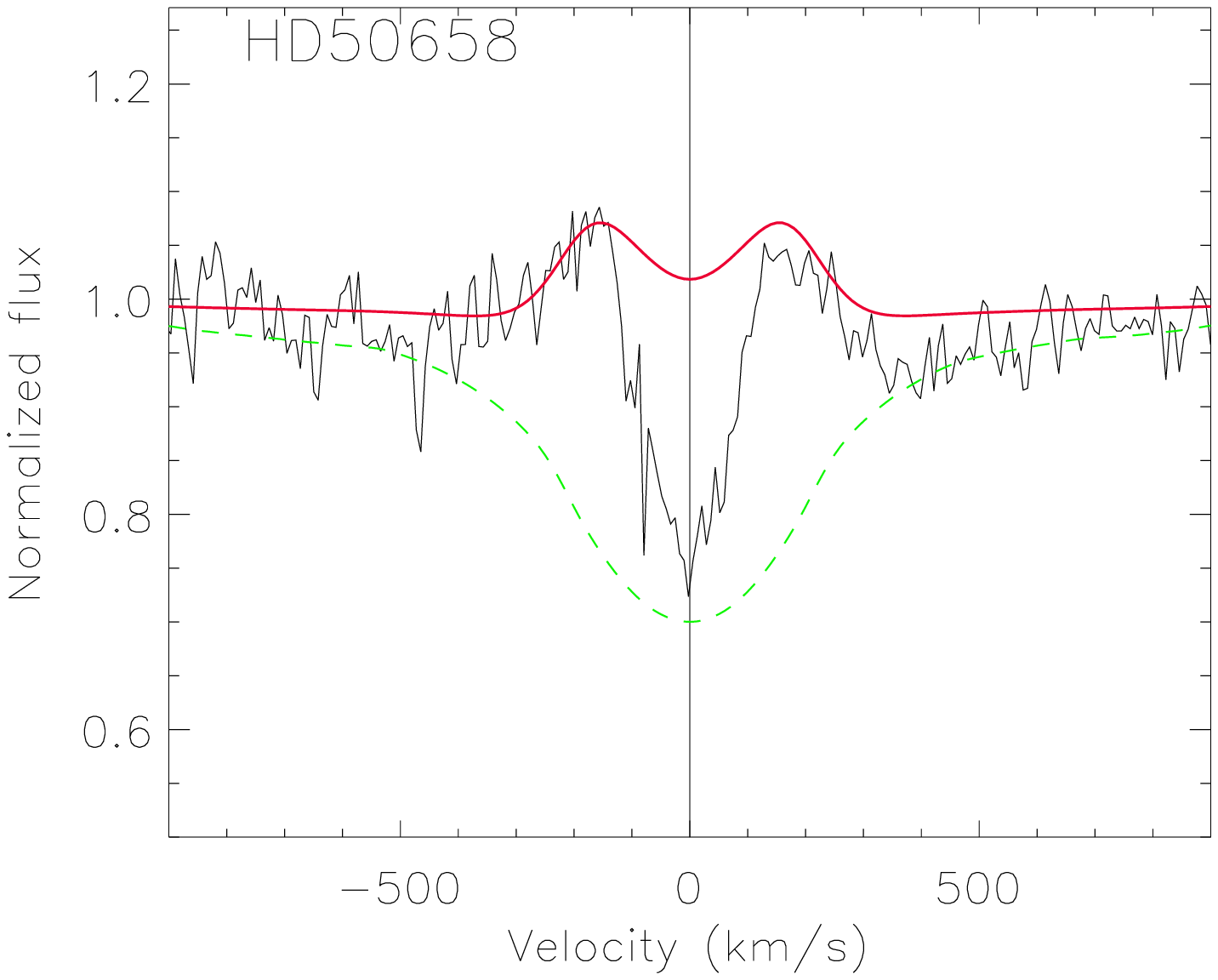}\\
   \includegraphics[width=8cm]{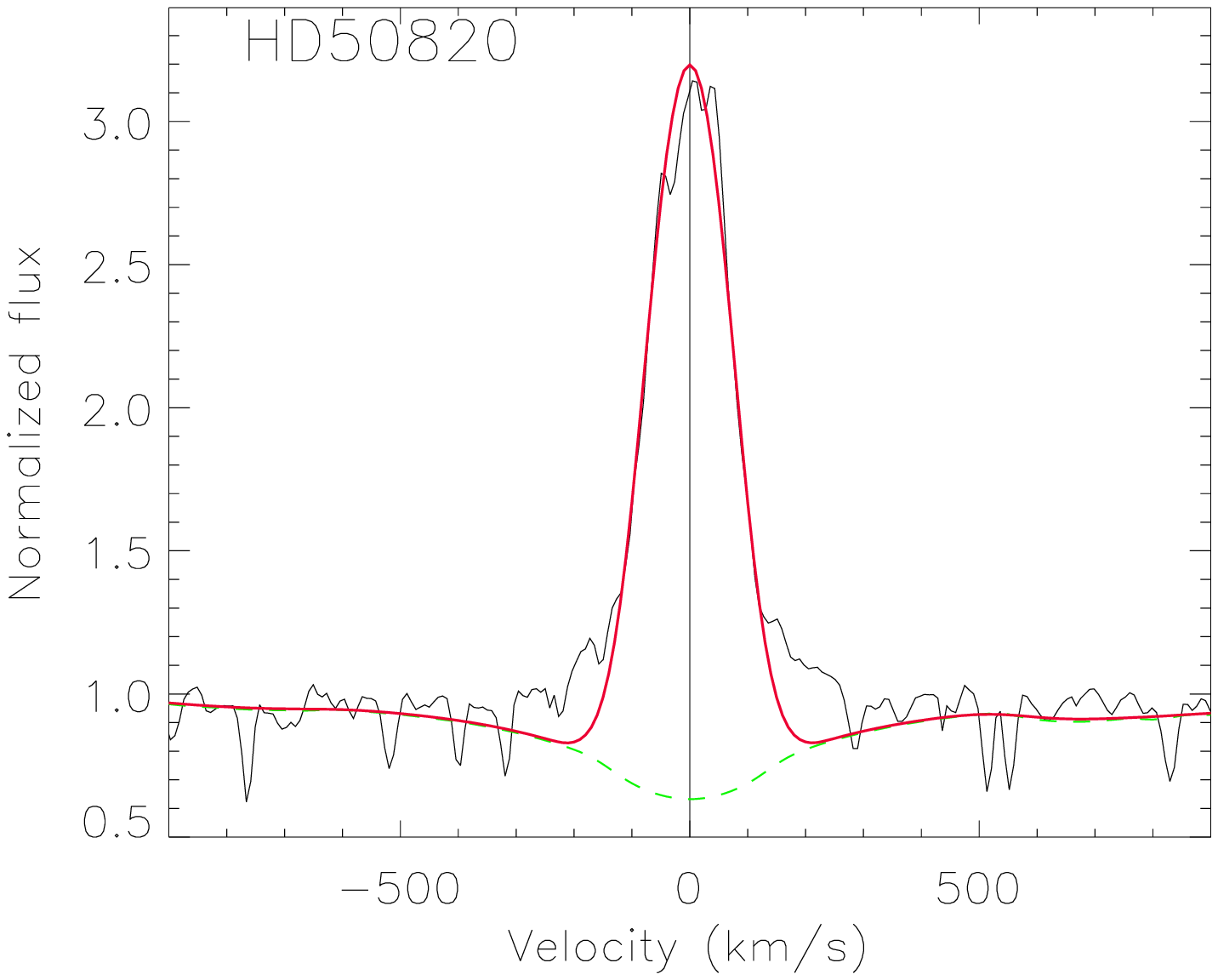} &
   \includegraphics[width=8cm]{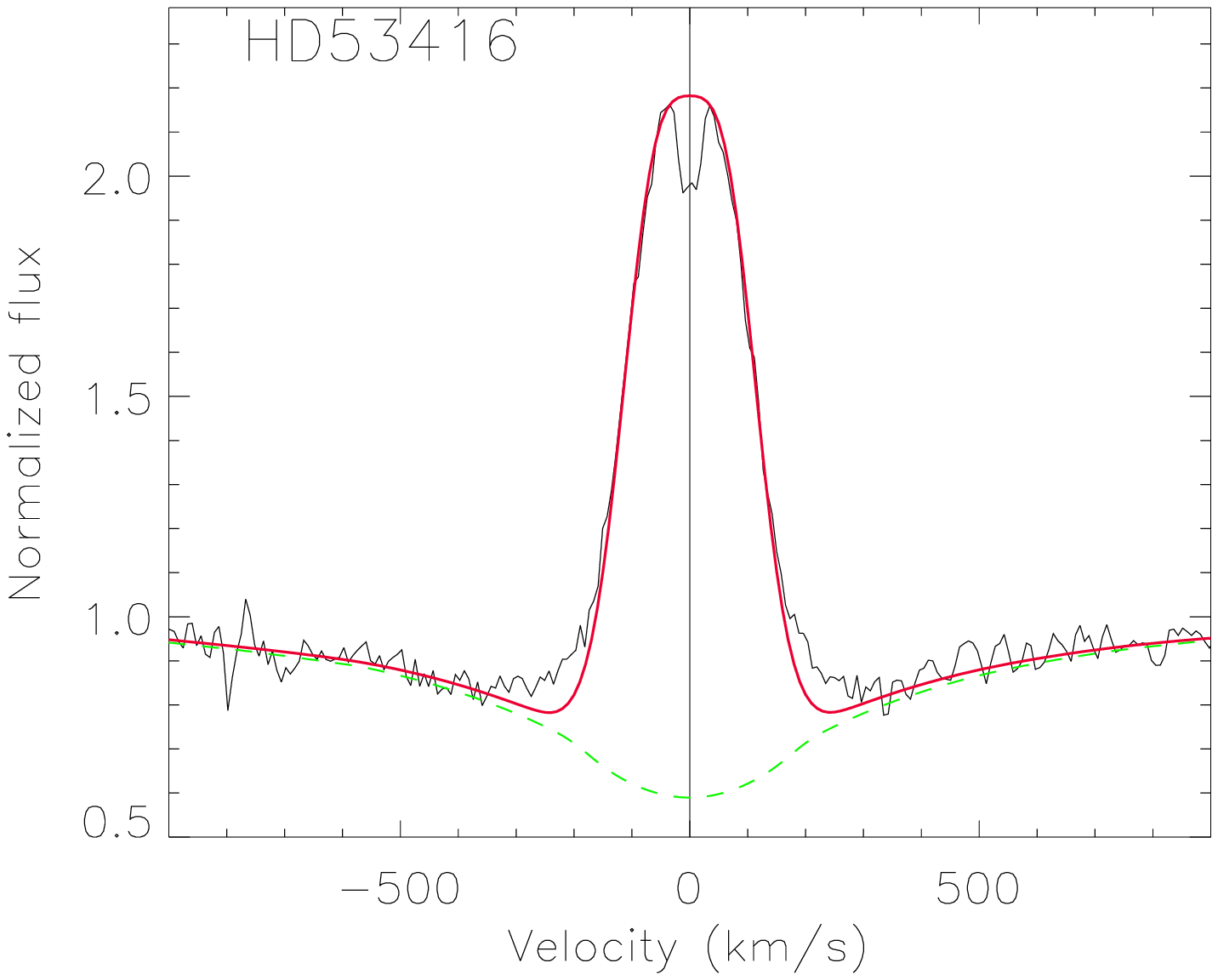}
   \end{array}$
   \end{center}

      \caption{Observed H$\alpha$ of Be stars with both stellar absorption (dashed green lines) and
               total emission profiles (solid red lines) overimposed. Best fit parameters for stellar and disk synthetic profiles
               are reported in Table~\ref{summary}. The circumstellar emission have been calculated by fixing the radial density
               exponent (n\,=\,3) and considering Keplerian rotation of the disk (i.e. j\,=\,1/2)
              }
         \label{halpha1}
   \end{figure*}
   \begin{figure*}
   \begin{center}$
   \begin{array}{cc}

   \includegraphics[width=8cm]{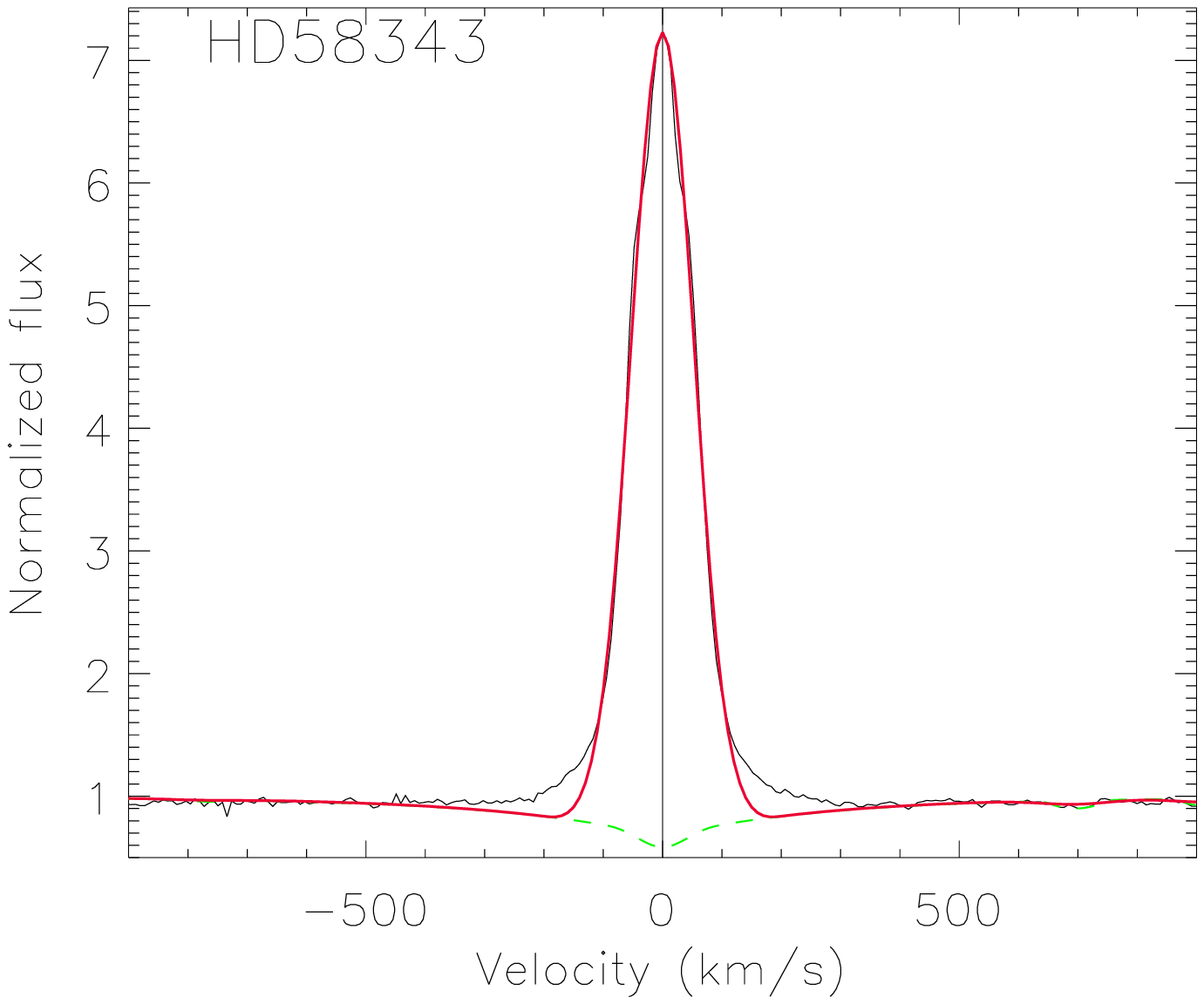} &
   \includegraphics[width=8cm]{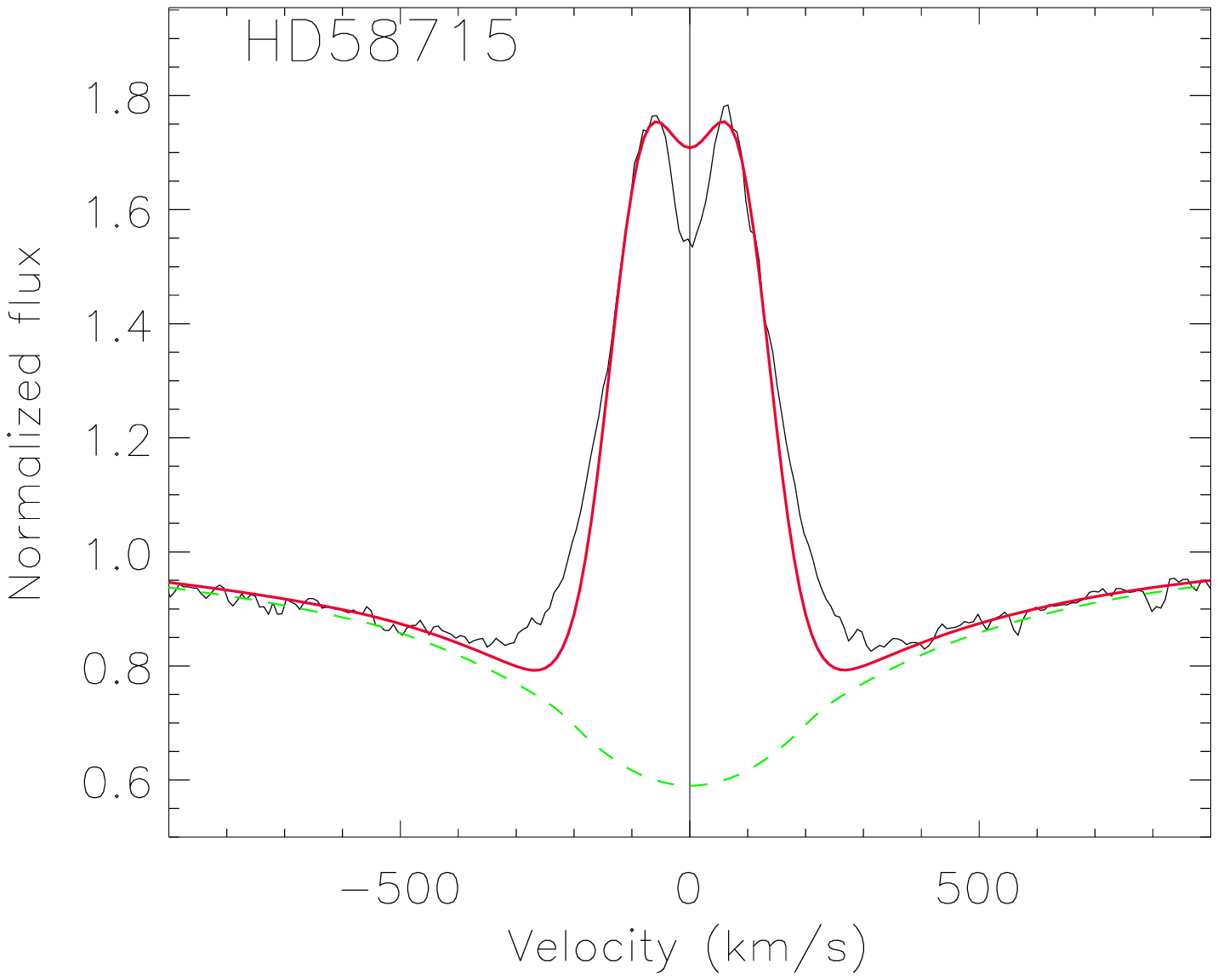} \\
   \includegraphics[width=8cm]{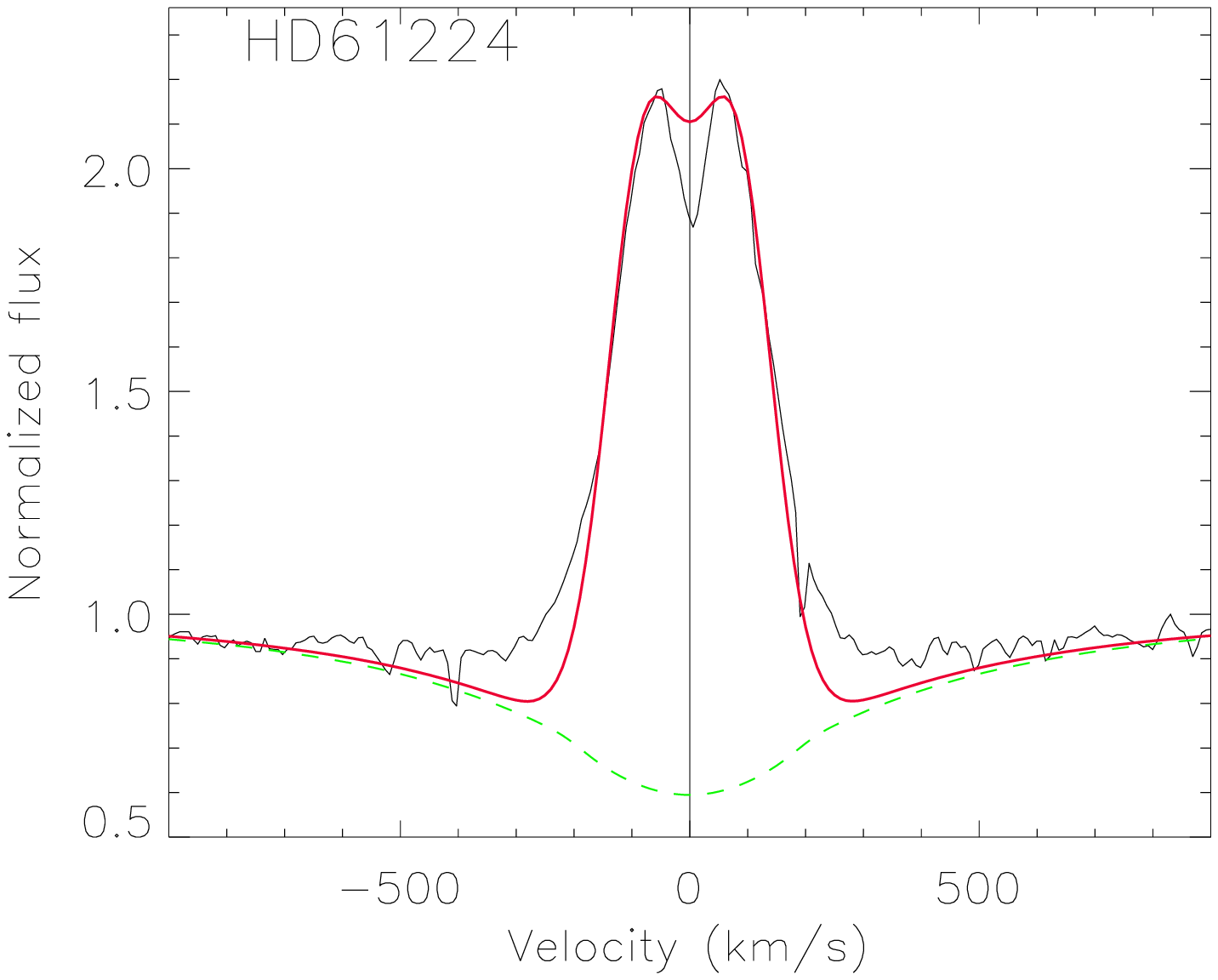} & 
   \includegraphics[width=8cm]{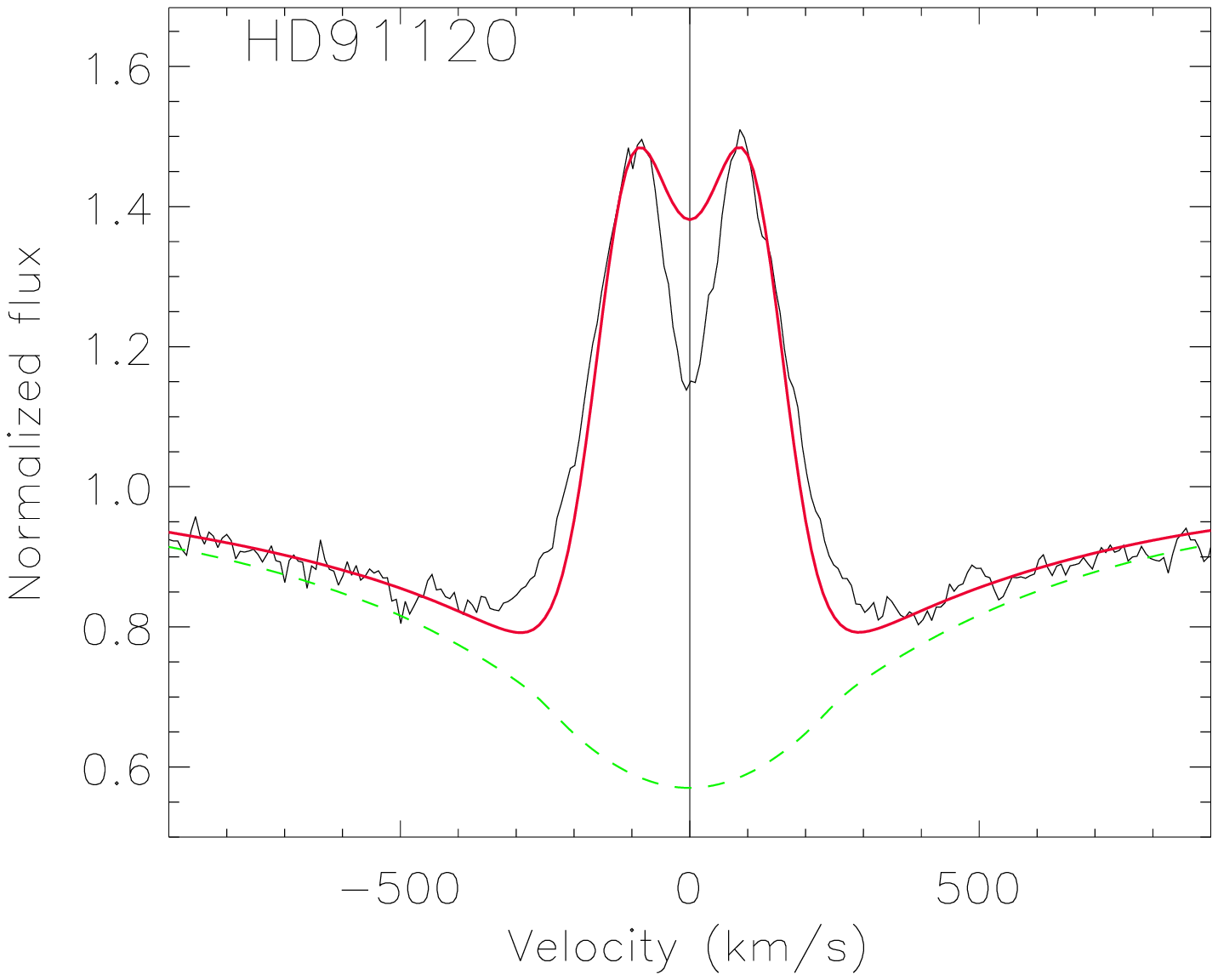} \\
   \includegraphics[width=8cm]{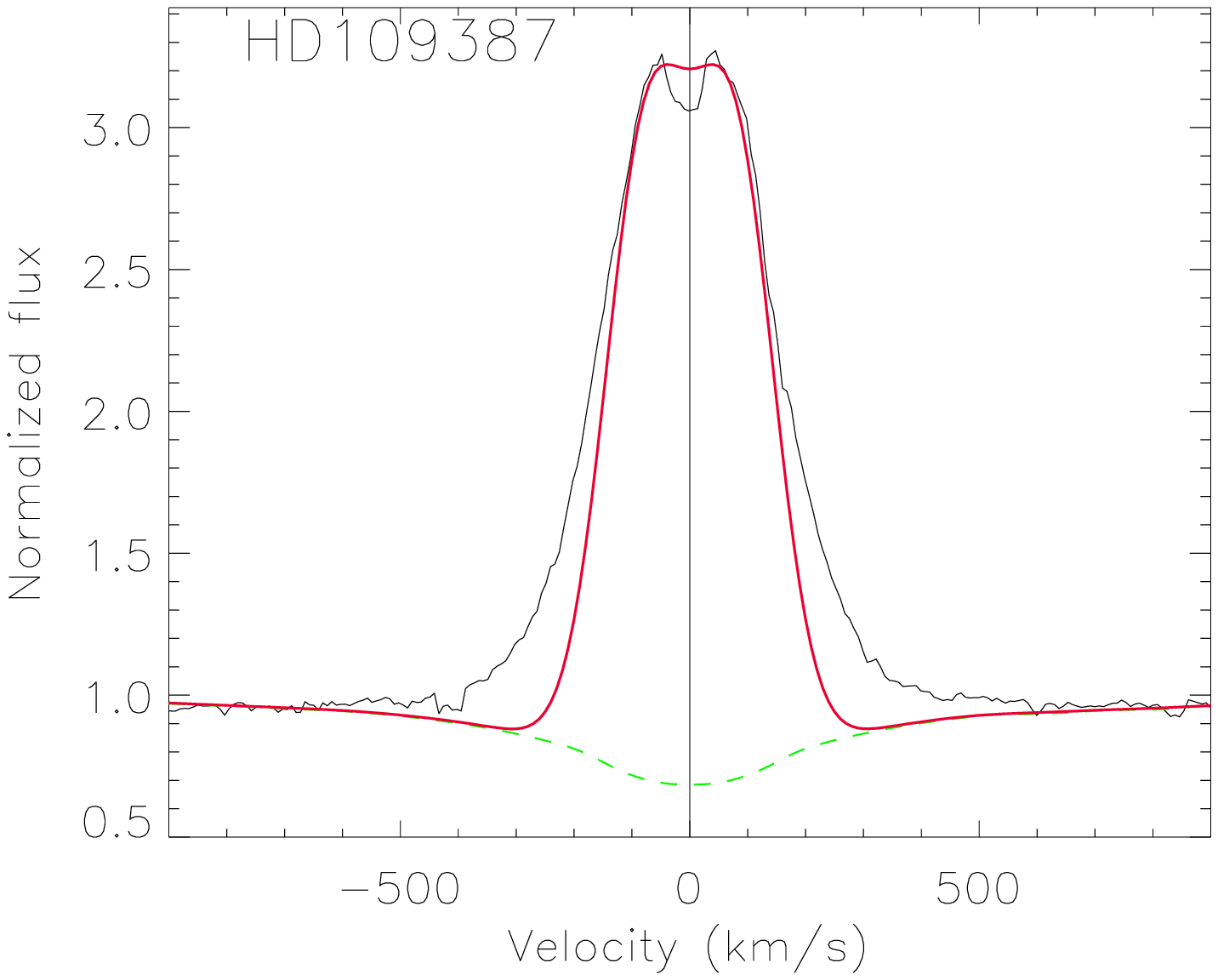} &
   \includegraphics[width=8cm]{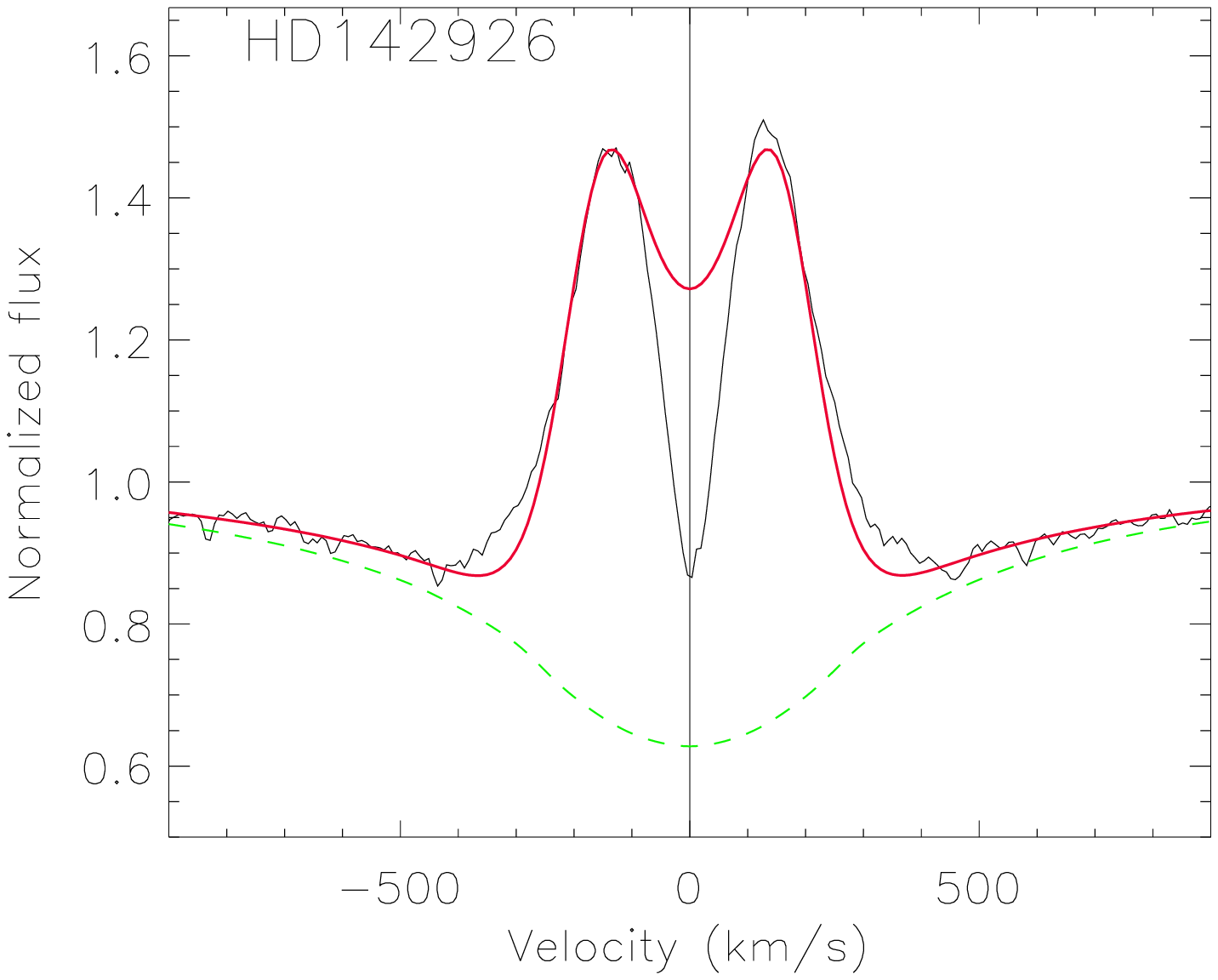}\\
   \includegraphics[width=8cm]{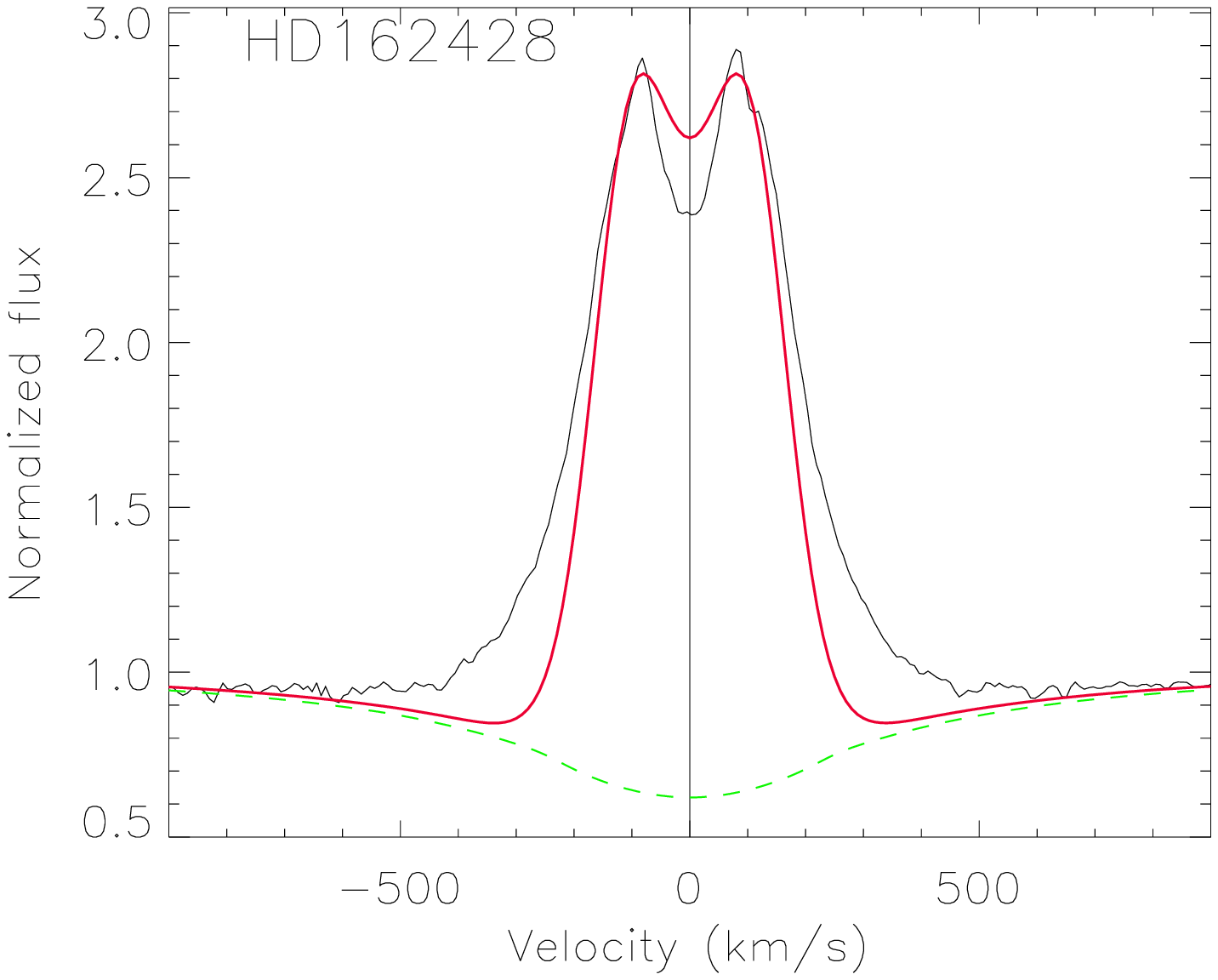} &
   \includegraphics[width=8cm]{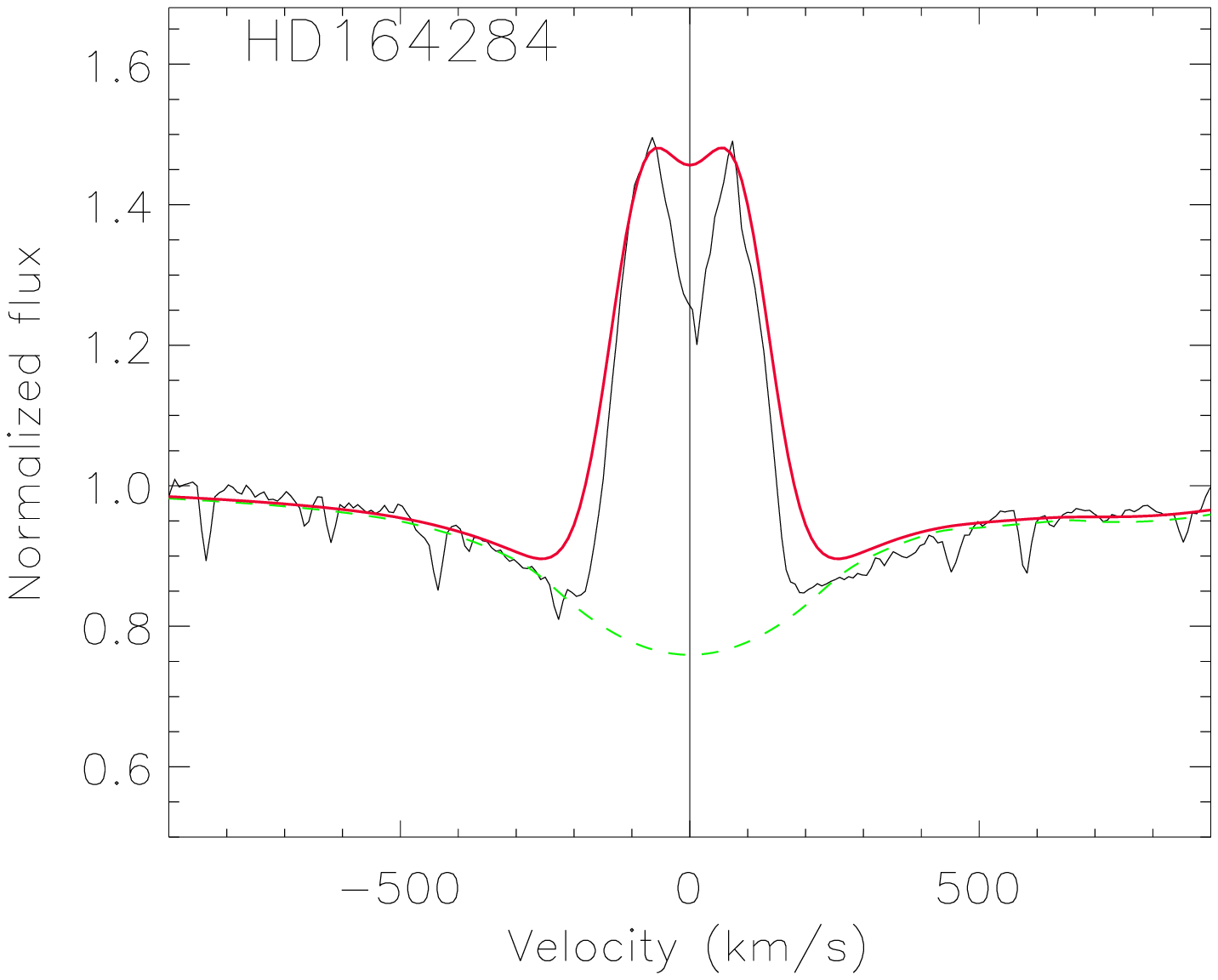}
   \end{array}$
   \end{center}

      \caption{As in Fig.~\ref{halpha1}
              }
         \label{halpha2}
   \end{figure*}
   \begin{figure*}
   \begin{center}$
   \begin{array}{cc}

   \includegraphics[width=8cm]{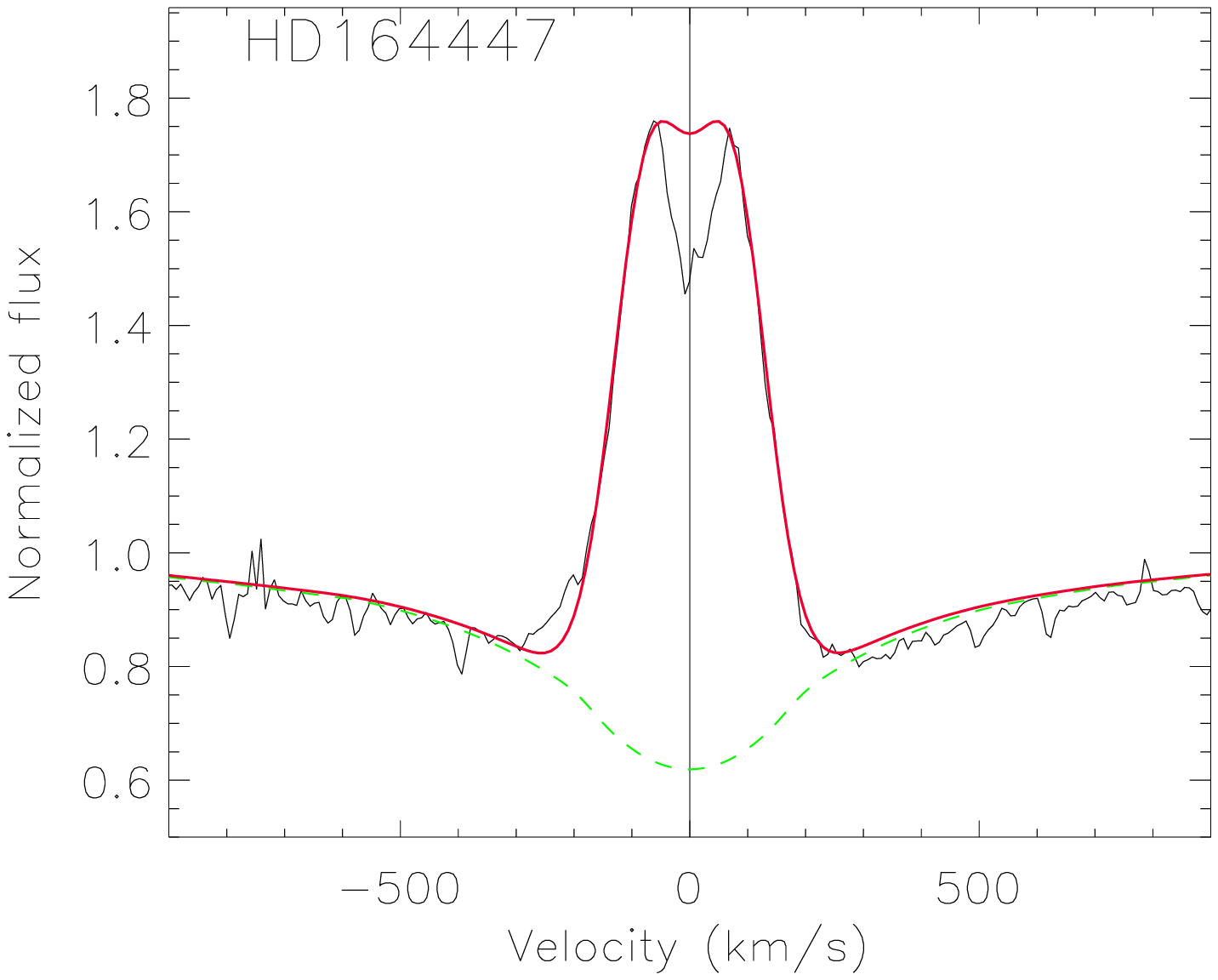} &
   \includegraphics[width=8cm]{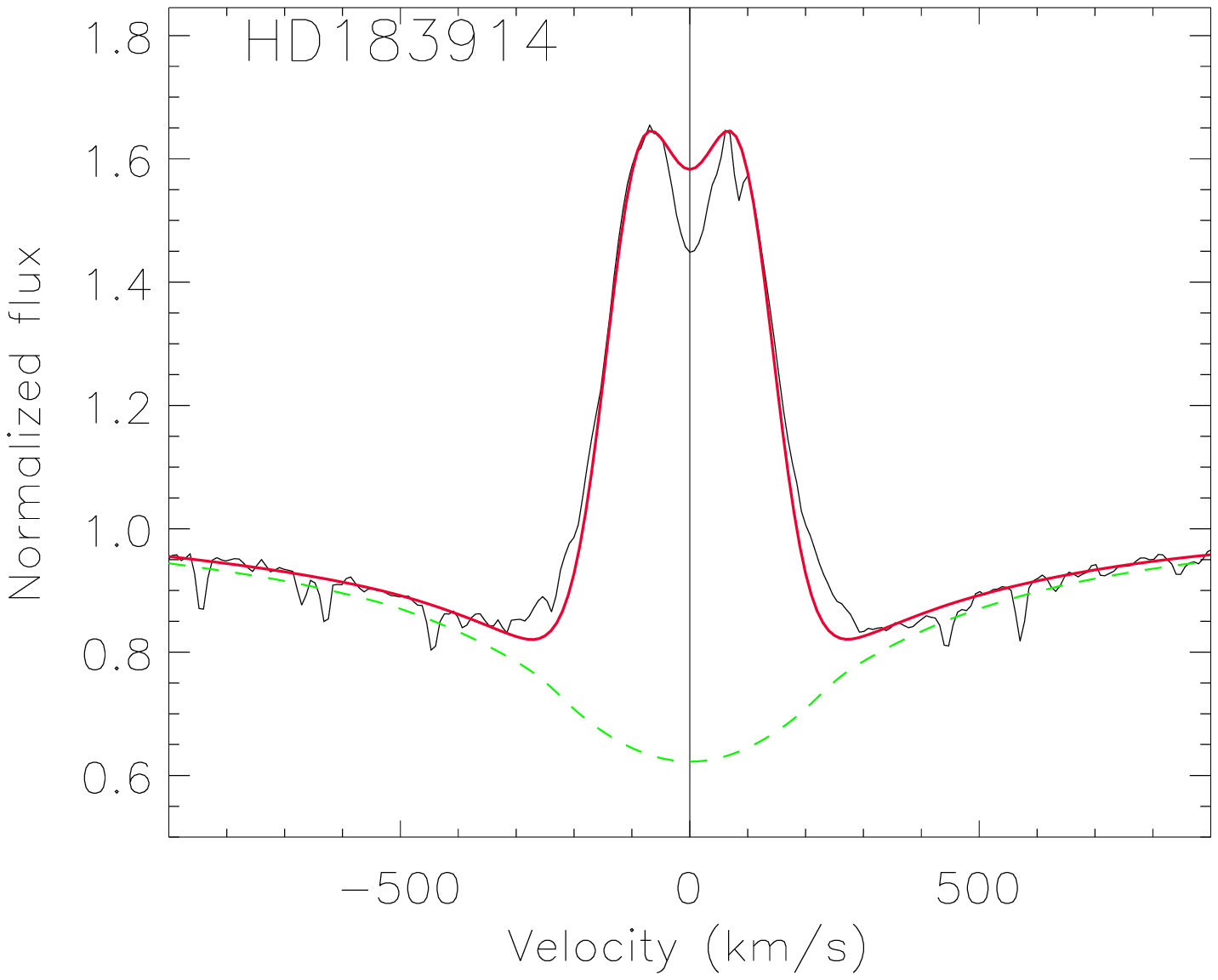} \\
   \includegraphics[width=8cm]{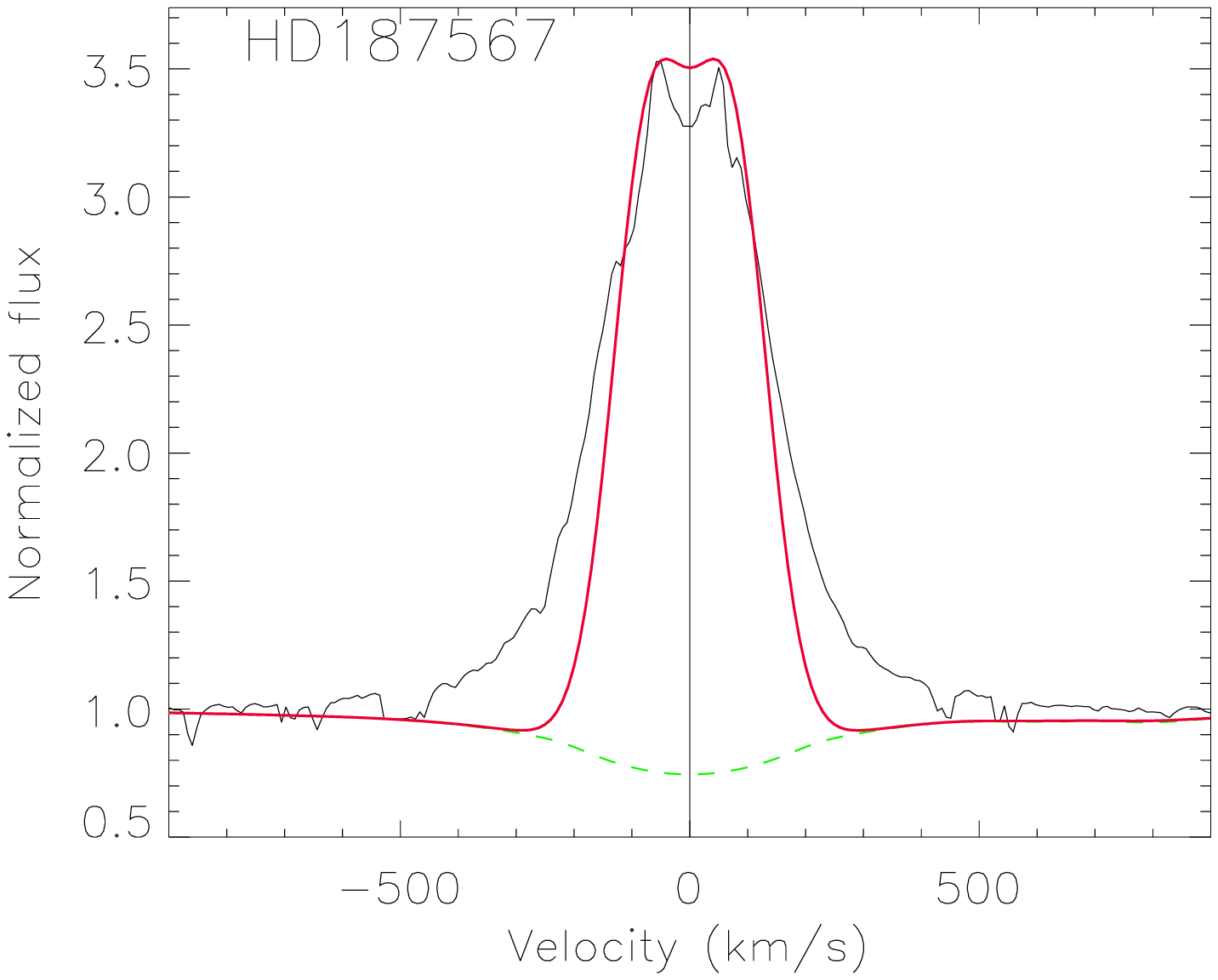} & 
   \includegraphics[width=8cm]{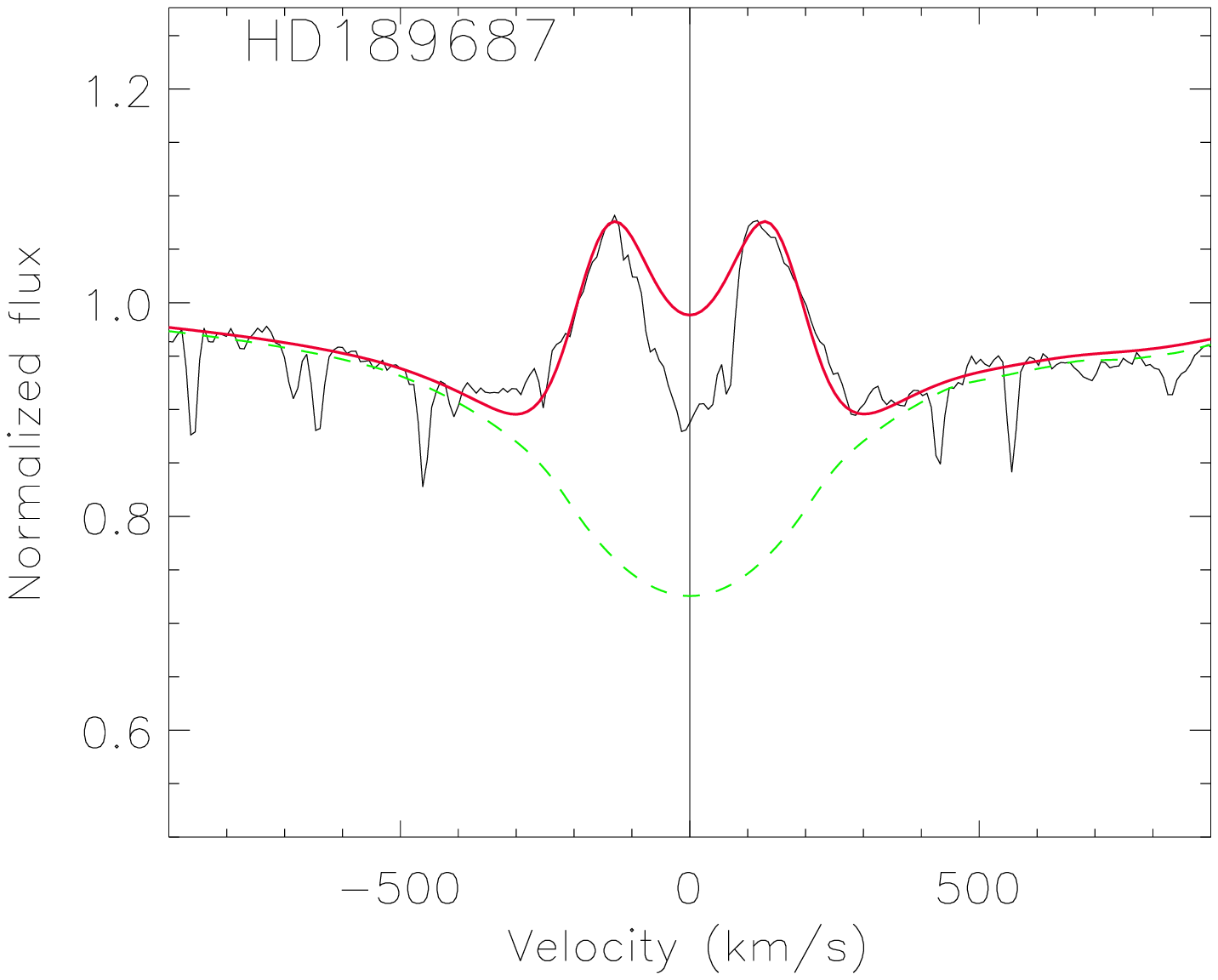} \\
   \includegraphics[width=8cm]{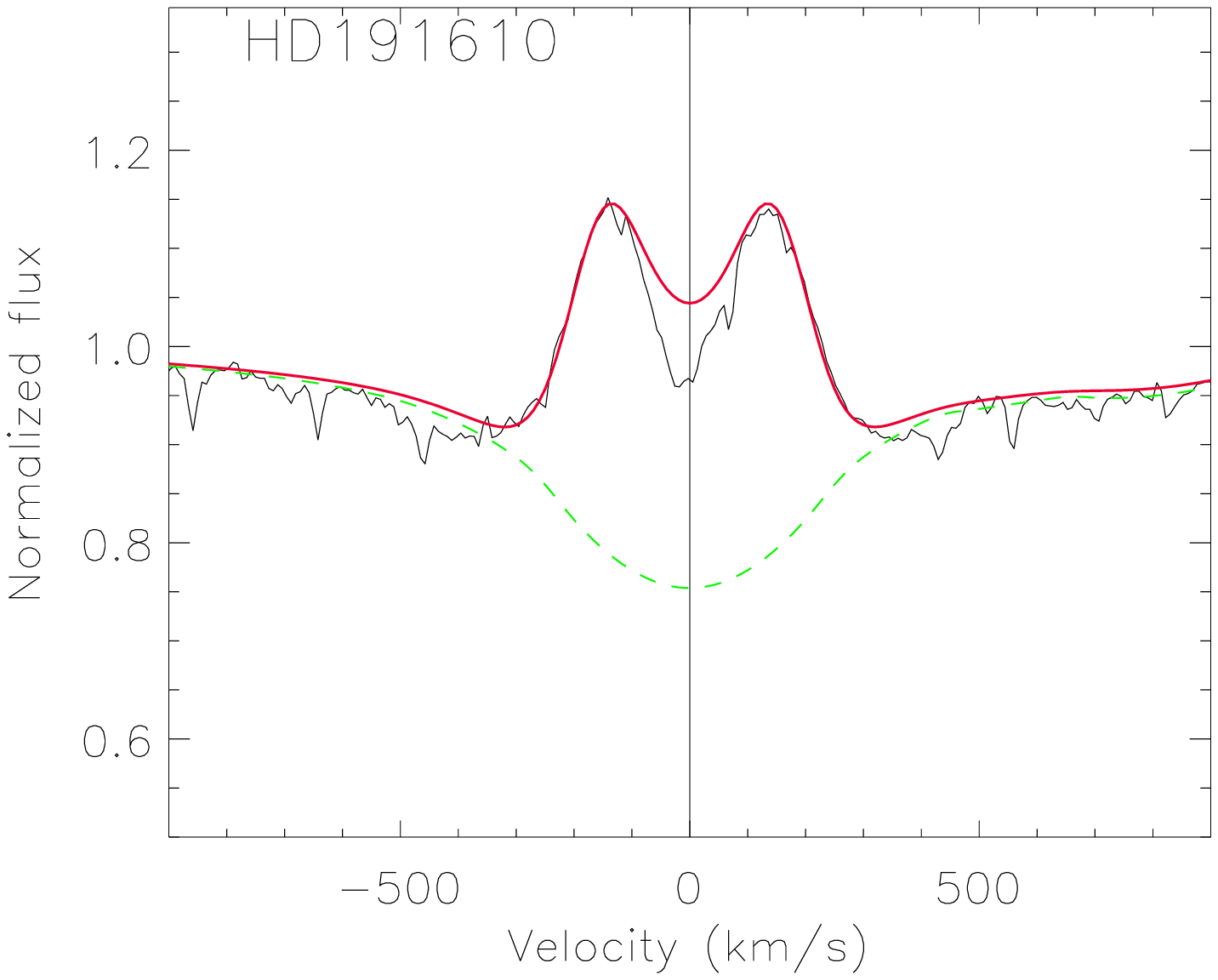} &
   \includegraphics[width=8cm]{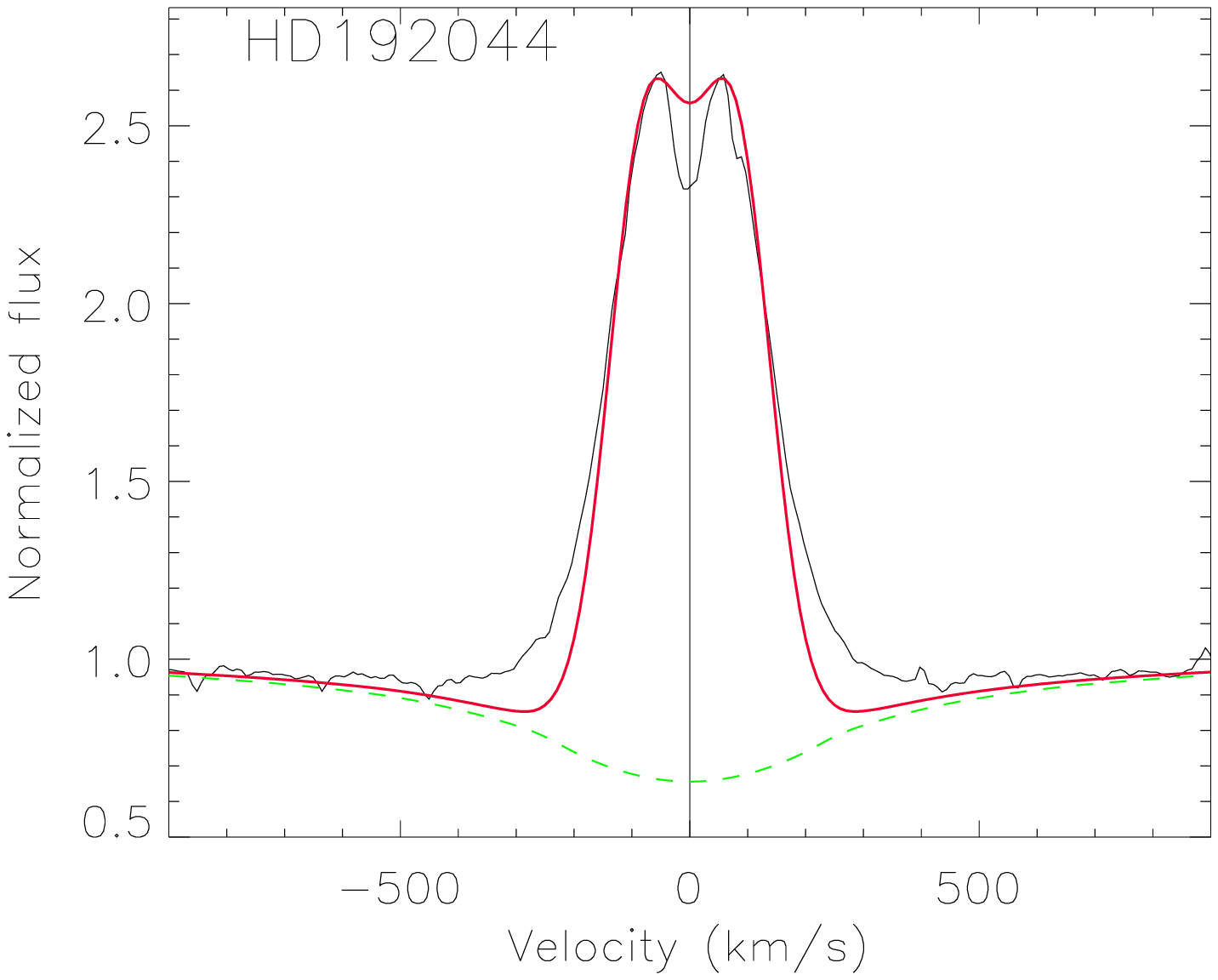}\\
   \includegraphics[width=8cm]{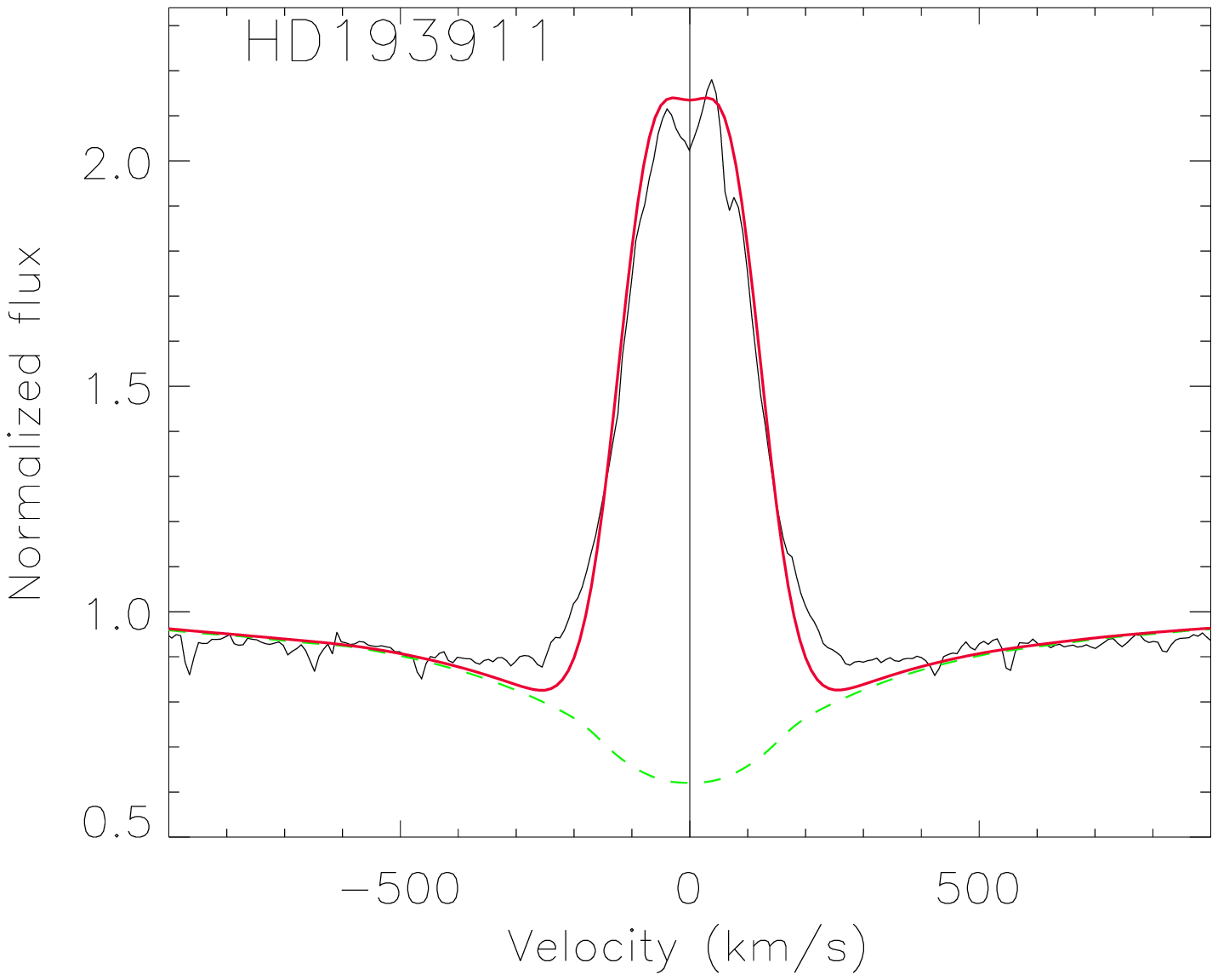} &
   \includegraphics[width=8cm]{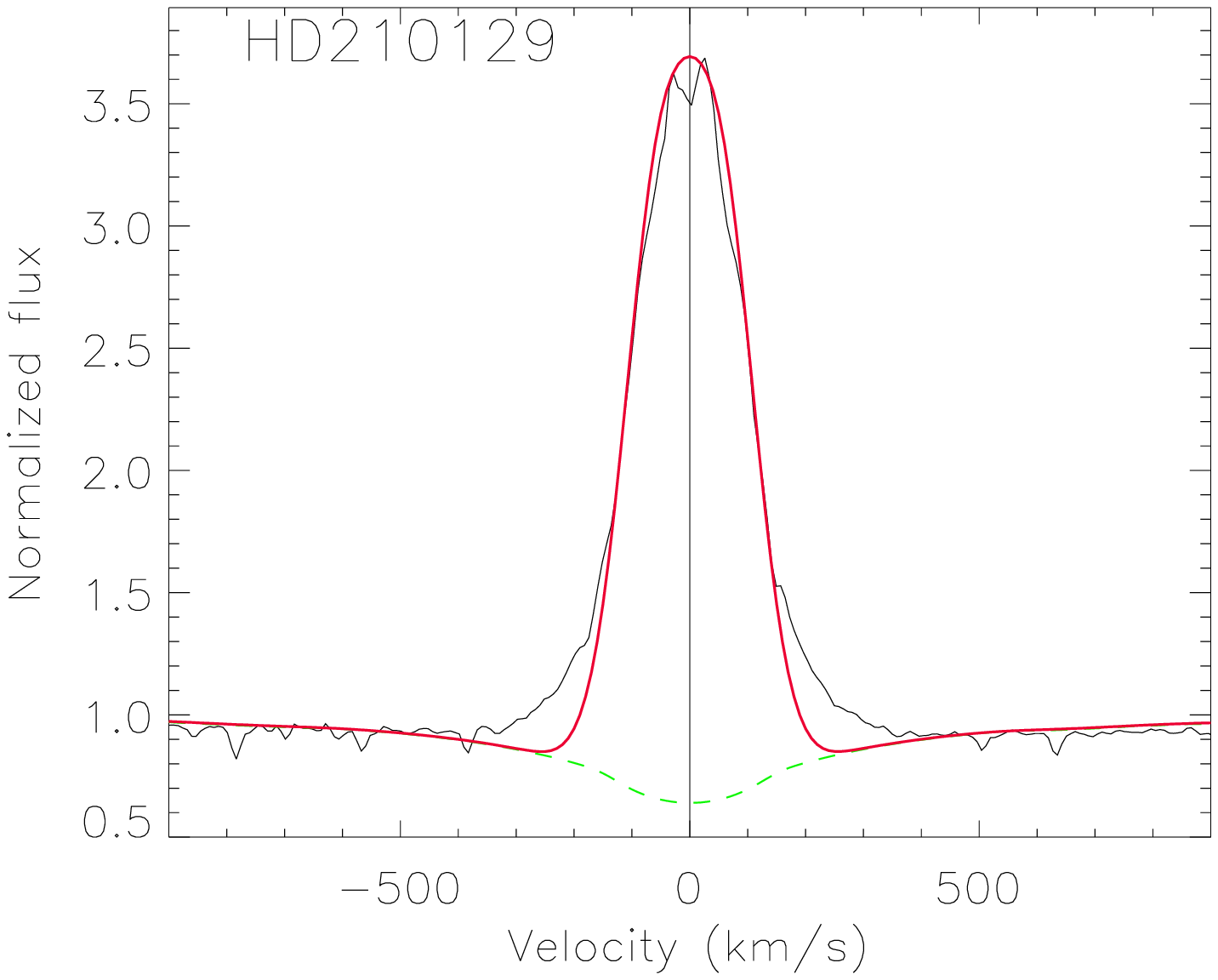}
   \end{array}$
   \end{center}

      \caption{As in Fig.~\ref{halpha1}
              }
         \label{halpha3}
   \end{figure*}

   \begin{figure*}
   \begin{center}$
   \begin{array}{cc}

   \includegraphics[width=8cm]{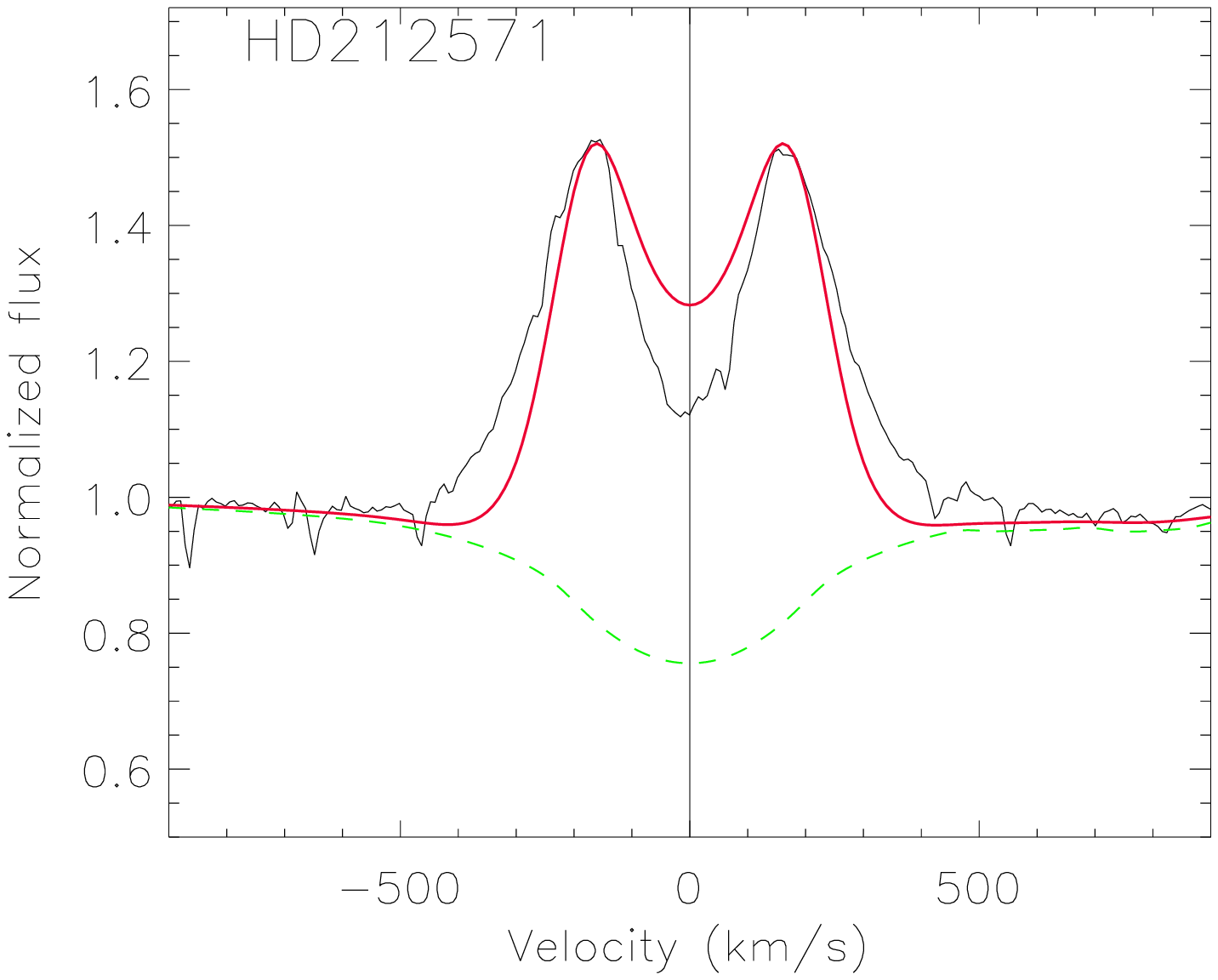} &
   \includegraphics[width=8cm]{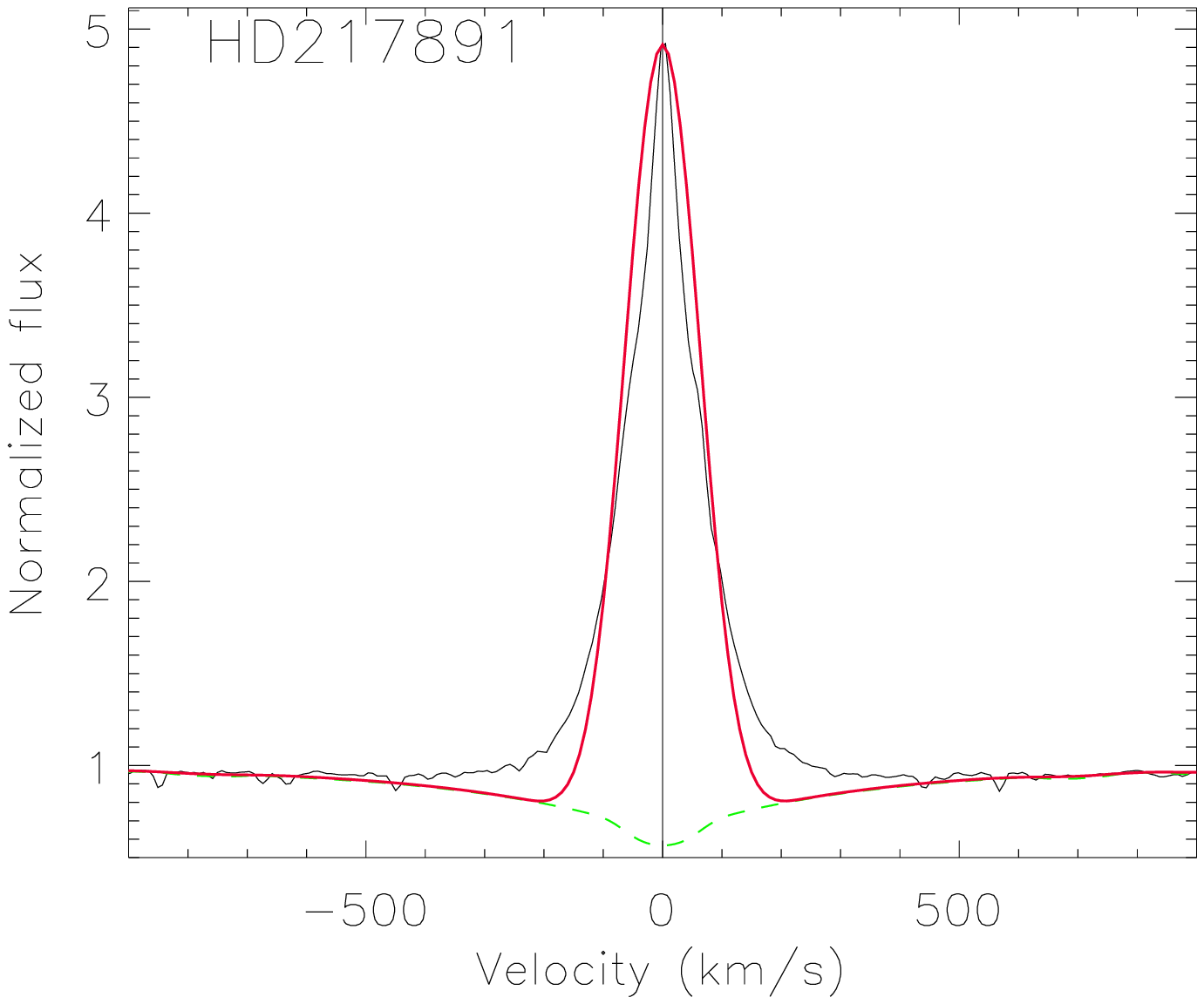} 
  \end{array}$
   \end{center}

      \caption{As in Fig.~\ref{halpha1}
              }
         \label{halpha4}
   \end{figure*}

\section{Variability}
\begin{table*}
\caption{H$\alpha$ and H$\beta$ equivalent width as measured in our spectra. Negative values mean that the lines show net emission.}
\label{eqw}
\begin{tabular}{rcrr|rcrr|rcrr} 
\hline
\hline             
HD~~~  &    JD       &  EW({\AA})  &  EW({\AA})   & HD~~~ & JD         &   EW({\AA}) &  EW({\AA}) & HD~~~ & JD     &   EW({\AA}) &  EW({\AA}) \\
       &  (2450000+) &  H$\alpha$~~&  H$\beta$~~  &       & (2450000+) &  H$\alpha$~~     & H$\beta$~~   &       & (2450000+) &  H$\alpha$~~     & H$\beta$~~ \\
\hline
  6811 & 4714.6282 &     1.06 &   6.53 & 142926 & 4543.6113 &   $-$2.82 &   7.14 & 187567 & 4654.5327 & $-$ 21.09 &   0.81 \\
       & 4747.5640 &     0.99 &   6.46 &        & 4554.6243 &   $-$1.61 &   7.04 &        & 4655.5890 & $-$ 21.46 &   0.59 \\
       & 4749.6417 &     0.55 &   6.36 &        & 4555.6541 &   $-$1.68 &   6.84 &        & 4678.5234 & $-$ 22.24 &   0.40 \\
 10516 & 4749.6602 & $-$25.21 &   0.37 &        & 4556.6487 &   $-$1.40 &   6.80 &        & 4714.4153 & $-$ 22.50 &   0.44\\
 11415 & 4749.6753 &     4.25 &   5.83 &        & 4575.5709 &   $-$1.61 &   7.11 & 189687 & 4654.5758 &      0.76 &   4.70 \\
 37202 & 4544.2610 & $-$16.09 &   2.33 &        & 4576.5771 &   $-$1.68 &   7.14 &        & 4655.5411 &      0.93 &   4.75  \\
       & 4749.7298 & $-$14.44 &   2.35 &        & 4586.5501 &   $-$1.49 &   7.12 &        & 4677.5324 &      0.62 &   4.77 \\
 37490 & 4544.2799 &  $-$5.19 &   1.95 &        & 4588.5402 &   $-$1.38 &   7.05 &        & 4681.5430 &      1.01 &   4.76 \\
       & 4749.7175 &  $-$6.73 &   1.70 &        & 4616.4682 &   $-$1.46 &   7.03 &        & 4682.5777 &      0.89 &   4.76 \\
 41335 & 4542.3012 & $-$31.37 &$-$1.81 &        & 4653.3880 &   $-$1.18 &   7.20 &        & 4714.4602 &      1.35 &   4.84 \\
       & 4749.7436 & $-$32.48 &$-$1.07 & 142983 & 4543.5309 &  $-$26.32 &   3.24 & 191610 & 4712.3812 &      0.48 &   4.81 \\ 
 43285 & 4543.3009 &     5.06 &   7.57 &        & 4555.5462 &  $-$23.71 &   3.15 &        & 4713.3568 &   $-$0.00 &   4.63 \\
 44458 & 4544.3009 & $-$31.82 &$-$0.80 &        & 4556.5508 &  $-$24.48 &   3.08 &        & 4747.3386 &   $-$0.01 &   4.81 \\
 45542 & 4542.3231 &  $-$0.91 &   5.93 &        & 4575.5004 &  $-$24.74 &   3.03 &        & 4749.4744 &   $-$0.00 &   4.51 \\
 45725 & 4542.3417 &     9.38 &  11.36 &        & 4576.4808 &  $-$24.61 &   3.15 & 192044 & 4712.4081 &   $-$9.93 &   5.21 \\
 47054 & 4543.3275 &  $-$6.86 &   6.17 &        & 4585.4945 &  $-$25.81 &   3.07 &        & 4713.3805 &  $-$10.10 &   5.24 \\ 
 50658 & 4542.3850 &     1.34 &   5.67 &        & 4586.4283 &  $-$25.08 &   3.16 &        & 4747.3594 &   $-$9.84 &   5.30 \\ 
 50820 & 4544.3397 &  $-$8.32 &$-$2.73 &        & 4587.4507 &  $-$24.20 &   2.75 & 193911 & 4712.4526 &   $-$5.42 &   5.04 \\ 
 52918 & 4542.4174 &     2.77 &   3.78 &        & 4588.4201 &  $-$23.85 &   3.24 &        & 4713.4085 &   $-$5.34 &   5.12 \\  
 53416 & 4556.2808 &  $-$4.66 &   6.42 &        & 4599.4307 &  $-$24.62 &   3.03 &        & 4747.3837 &   $-$5.60 &   5.14 \\ 
 58050 & 4544.3849 &     4.31 &   5.44 &        & 4616.3757 &  $-$25.63 &   3.05 & 210129 & 4654.5979 &  $-$13.86 &   4.85 \\ 
       & 4554.2984 &     4.74 &   5.51 &        & 4653.4399 &  $-$23.49 &   3.29 &        & 4677.5640 &  $-$14.28 &   5.02 \\ 
       & 4556.3476 &     4.14 &   5.51 &        & 4654.4038 &  $-$24.03 &   2.97 &        & 4678.5672 &  $-$14.30 &   5.24 \\  
 58343 & 4543.3636 & $-$19.29 &   2.75 & 143275 & 4677.3348 &  $-$11.01 &$-$0.78 &        & 4681.5694 &  $-$14.18 &   5.04 \\  
 58715 & 4542.4392 &  $-$0.79 &   7.91 &        & 4681.3050 &   $-$6.94 &$-$0.64 &        & 4682.6018 &  $-$14.14 &   5.11 \\  
       & 4554.3634 &  $-$1.29 &   7.95 &        & 4682.3327 &  $-$10.28 &$-$0.73 &        & 4714.4878 &  $-$14.01 &   5.23 \\  
       & 4556.3986 &  $-$1.49 &   8.02 &        & 4714.2604 &  $-$12.30 &$-$1.09 &        & 4747.4096 &  $-$14.54 &   5.33 \\  
 60855 & 4543.3894 & $-$38.59 &$-$0.95 & 162428 & 4586.5791 &  $-$15.50 &   5.03 &        & 4749.5015 &  $-$14.04 &   5.04 \\  
 61224 & 4555.2948 &  $-$5.93 &   5.42 &        & 4588.5667 &  $-$15.72 &   4.12 & 212571 & 4712.4805 &   $-$5.37 &   2.11 \\  
       & 4556.4231 &  $-$5.25 &   4.85 &        & 4677.3624 &  $-$15.69 &   4.99 &        &  4714.5117 &  $-$5.44 &   2.00 \\  
 65875 & 4543.4116 & $-$42.00 &$-$0.28 &        & 4681.3258 &  $-$15.58 &   4.80 &        &  4749.5255 &  $-$5.40 &   1.83 \\  
       & 4544.4331 & $-$43.32 &   0.74 &        & 5011.4665 &  $-$15.38 &   4.96 & 214168 &  4713.4584 & $-$15.13 &   2.12 \\   
       & 4554.4361 & $-$43.69 &$-$0.33 & 162732 & 4586.6250 &   $-$7.24 &   6.13 &        &  4747.4338 & $-$14.44 &   2.22 \\  
 71072 & 4555.3404 &     3.20 &   5.05 &        & 4616.5402 &   $-$6.95 &   5.91 &        &  4749.5512 & $-$14.80 &   1.27 \\  
       & 4576.3047 &     3.09 &   5.38 &        & 4654.4306 &   $-$6.00 &   6.40 & 216057 &  4713.4828 &   4.48 &   7.25 \\    
 91120 & 4543.4858 &  $-$1.39 &   8.32 &        & 4677.4028 &   $-$7.04 &   4.94 &        &  4714.5419 &   4.60 &   7.15 \\    
       & 4554.3985 &  $-$0.31 &   8.19 &        & 4682.3762 &   $-$7.31 &   6.07 &        &  4747.4580 &   4.84 &   7.29 \\    
       & 4555.4708 &  $-$0.40 &   8.15 &        & 5011.5115 &   $-$5.09 &   6.04 &        &  4749.5750 &   4.58 &   7.07 \\   
       & 4556.4743 &  $-$0.52 &   8.02 & 164284 & 4575.5993 &   $-$0.01 &   5.19 & 216200 &  4713.5068 &   1.07 &   5.04 \\     
       & 4576.3563 &  $-$0.66 &   7.90 &        & 4576.6058 &   $-$0.00 &   5.04 &        &  4714.5693 &   1.61 &   5.31 \\     
       & 4585.3822 &  $-$0.66 &   8.02 &        & 4616.5681 &      0.91 &   5.23 &        &  4749.6007 &   0.76 &   4.90 \\     
       & 4586.3393 &  $-$0.01 &   7.84 &        & 4677.4774 &      1.43 &   5.12 & 217050 &  4747.5172 & $-$27.42 &   1.82 \\   
       & 4587.3192 &  $-$0.01 &   8.05 &        & 4682.4266 &      1.64 &   5.26 & 217543 &  4713.5577 &   1.81 &   5.38 \\    
       & 4588.3168 &     0.37 &   8.16 &        & 5011.5620 &      4.25 &   5.23 &        &  4747.5433 &   2.89 &   5.23 \\   
109387 & 4543.5060 & $-$20.01 &   3.05 & 164447 & 4575.6219 &   $-$2.03 &   6.62 & 217675 &  4713.6001 &   3.24 &   5.51 \\   
       & 4554.4814 & $-$19.83 &   3.01 &        & 4677.4415 &   $-$1.24 &   6.64 &        &  4747.4769 &   2.85 &   5.77 \\   
       & 4555.5156 & $-$19.34 &   3.07 &        & 4681.3697 &   $-$1.44 &   6.68 & 217891 &  4678.6041 & $-$12.16 &   4.93 \\ 
       & 4556.5191 & $-$19.09 &   2.97 &        & 4749.3442 &   $-$1.02 &   6.93 &        &  4681.6008 & $-$12.46 &   4.86 \\ 
       & 4575.4348 & $-$19.43 &   3.02 &        & 5011.5811 &      0.96 &   6.52 &        &  4713.6187 & $-$12.85 &   4.76 \\ 
       & 4576.4464 & $-$19.42 &   2.98 & 183362 & 4653.5050 & $-$ 26.22 &   1.35 &        &  4714.6005 & $-$12.80 &   4.66 \\ 
       & 4585.4577 & $-$19.09 &   3.02 &        & 4682.5161 & $-$ 26.45 &   1.08 &        &  4747.4941 & $-$13.47 &   4.57 \\ 
       & 4586.3921 & $-$19.00 &   2.98 &        & 4714.3334 & $-$ 26.50 &   1.33 &        &  4749.6231 & $-$13.90 &   4.42 \\ 
       & 4587.3898 & $-$18.78 &   2.99 & 183656 & 4653.5999 &   $-$8.27 &   5.24 &        &            &          &        \\ 
       & 4588.3867 & $-$18.77 &   3.07 &        & 4681.4527 &   $-$7.59 &   5.25 &        &            &          &        \\ 
       & 4599.3878 & $-$18.45 &   3.14 &        & 4714.3630 &   $-$8.01 &   5.35 &        &            &          &        \\
       & 4616.3342 & $-$19.11 &   3.13 &        & 4749.3742 &   $-$8.07 &   4.98 &        &            &          &         \\
       & 4653.3303 & $-$18.84 &   3.12 &        & 5011.6055 &   $-$6.48 &   4.94 &        &            &          &        \\
       &           &          &        & 183914 & 4653.5454 &   $-$1.01 &   8.03 &        &            &          &       \\
       &           &          &        &        & 4681.4958 &   $-$0.91 &   8.09 &        &            &          &       \\
       &           &          &        &        & 4682.5518 &   $-$0.96 &   8.20 &        &            &          &        \\
       &           &          &        &        & 4714.3886 &   $-$0.61 &   8.14 &        &            &          &        \\
       &           &          &        &        & 4749.4339 &   $-$1.23 &   8.06 &        &            &          &        \\
\hline                                                                           
\end{tabular}
\end{table*}

   \begin{figure*}
   \begin{center}$
   \begin{array}{cc}

   \includegraphics[width=8cm]{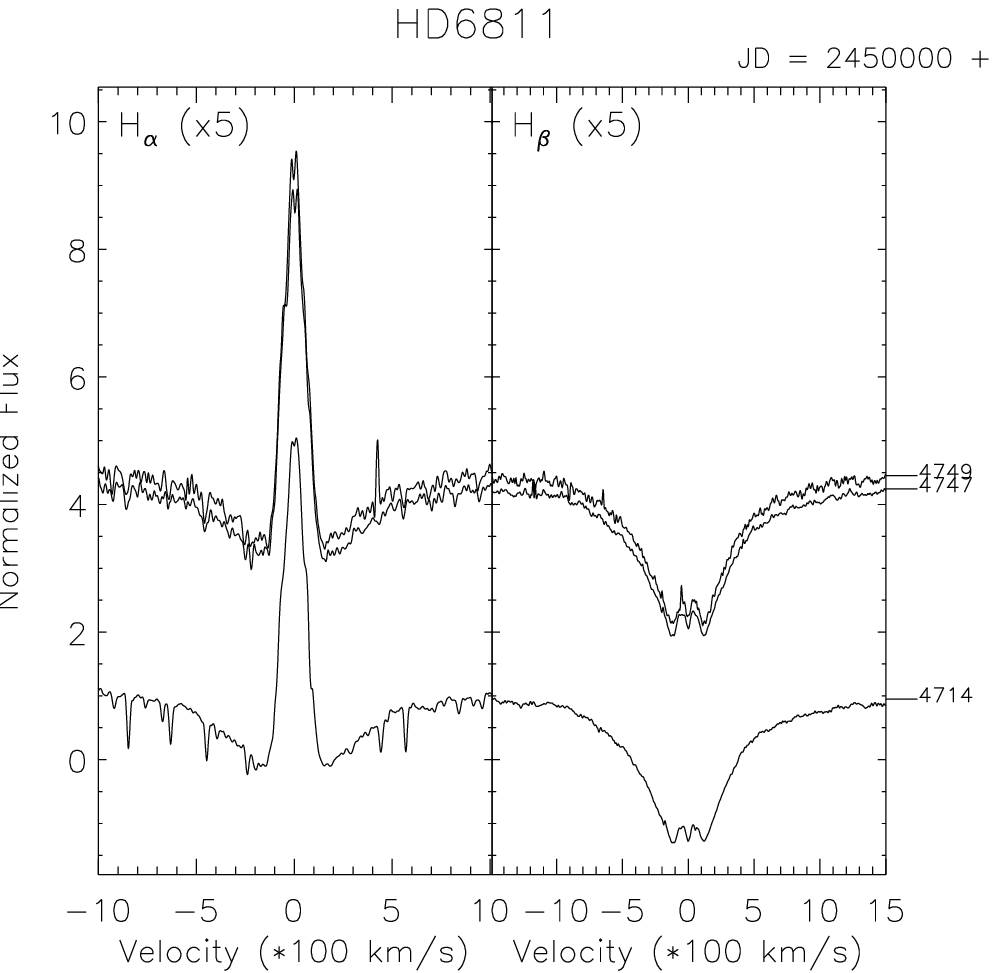} &
   \includegraphics[width=8cm]{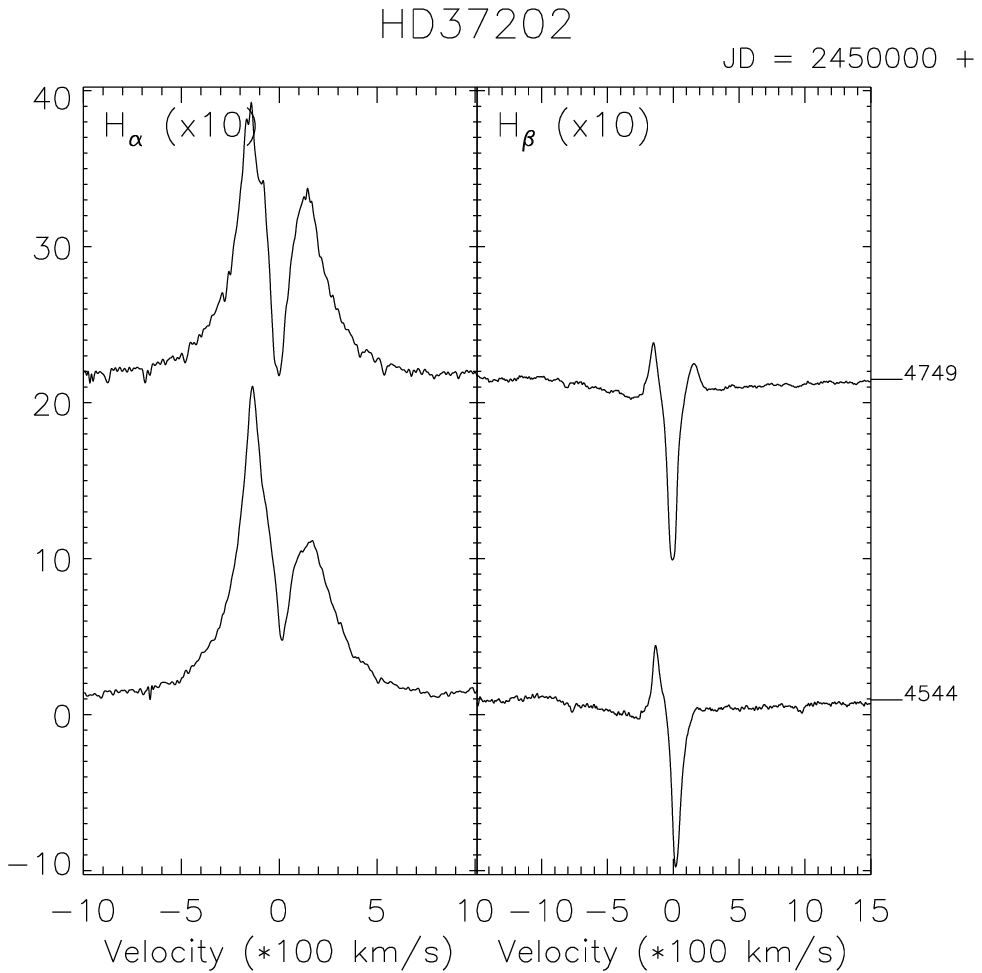} \\
   \includegraphics[width=8cm]{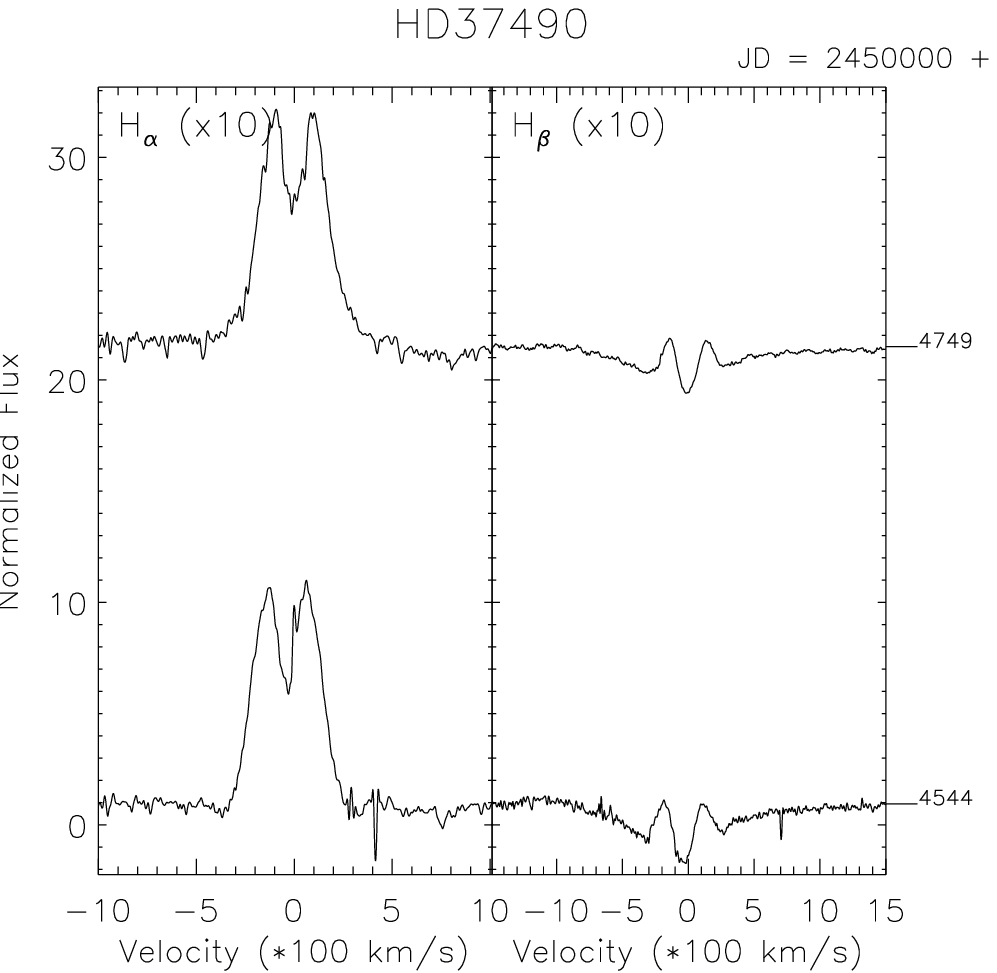} & 
   \includegraphics[width=8cm]{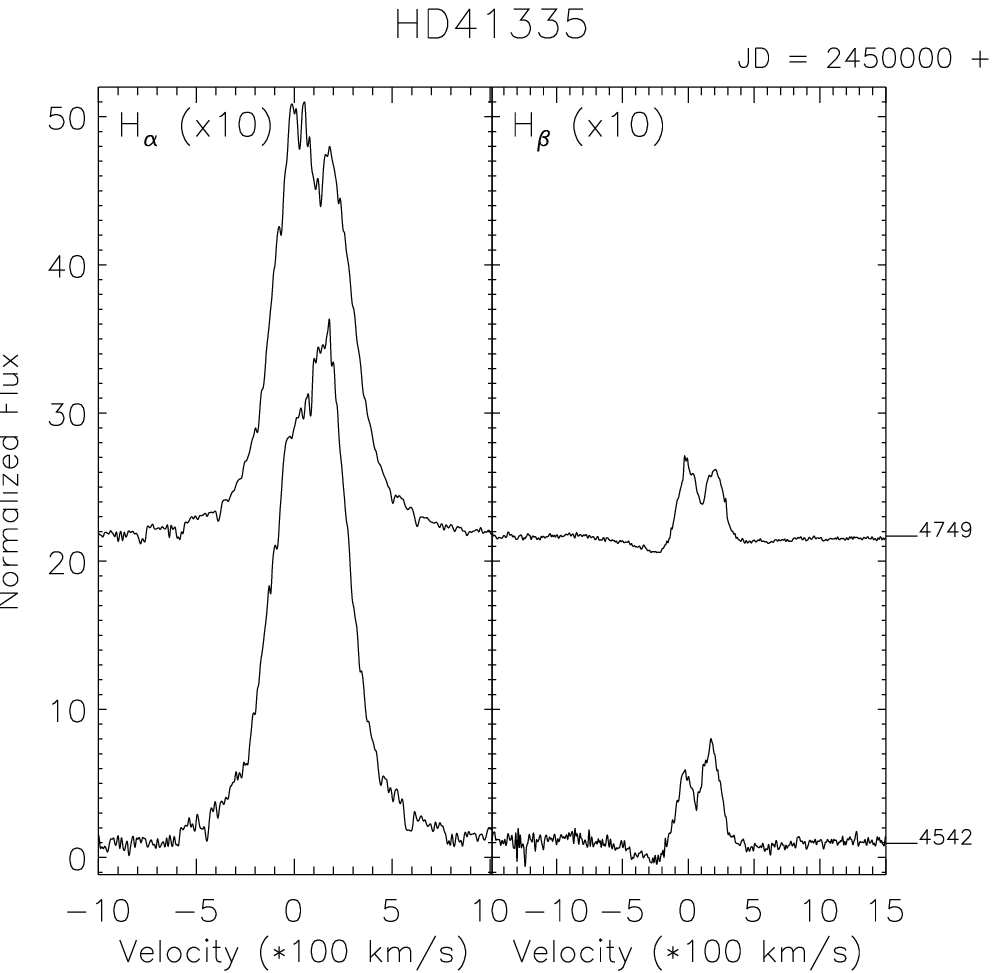} \\
   \includegraphics[width=8cm]{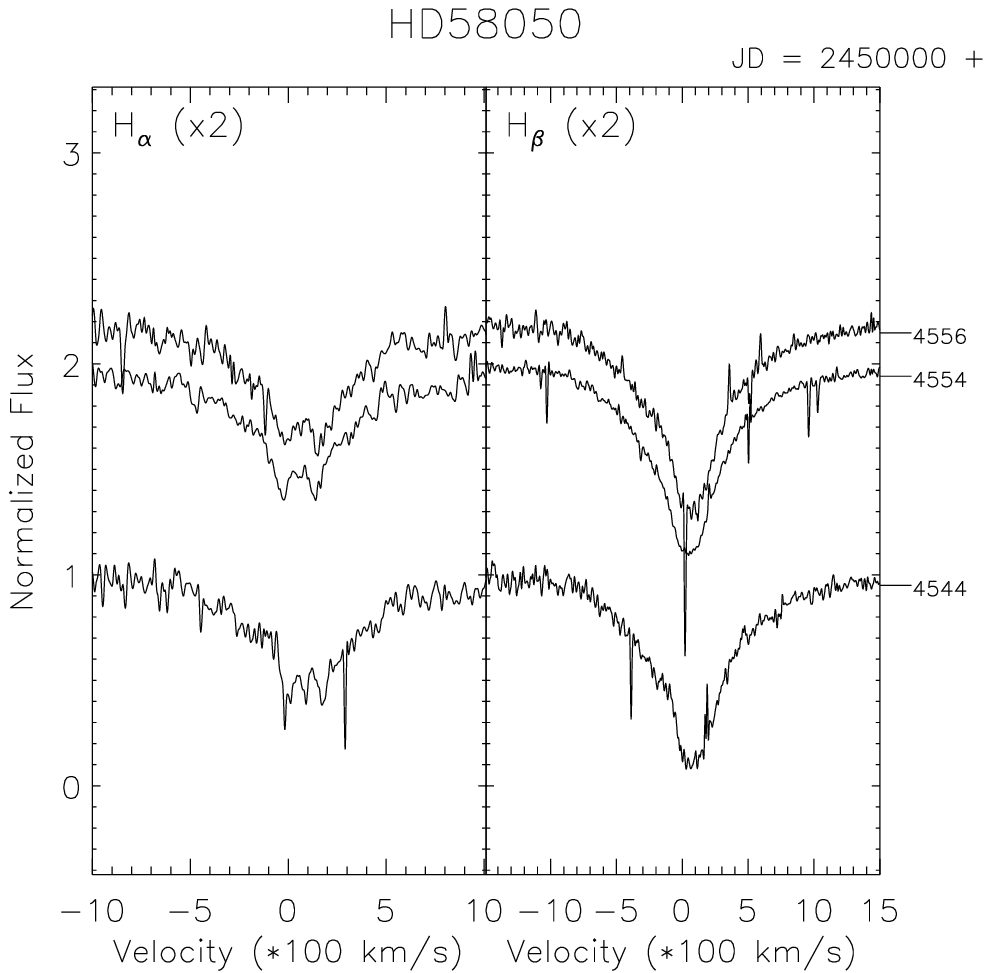} &
   \includegraphics[width=8cm]{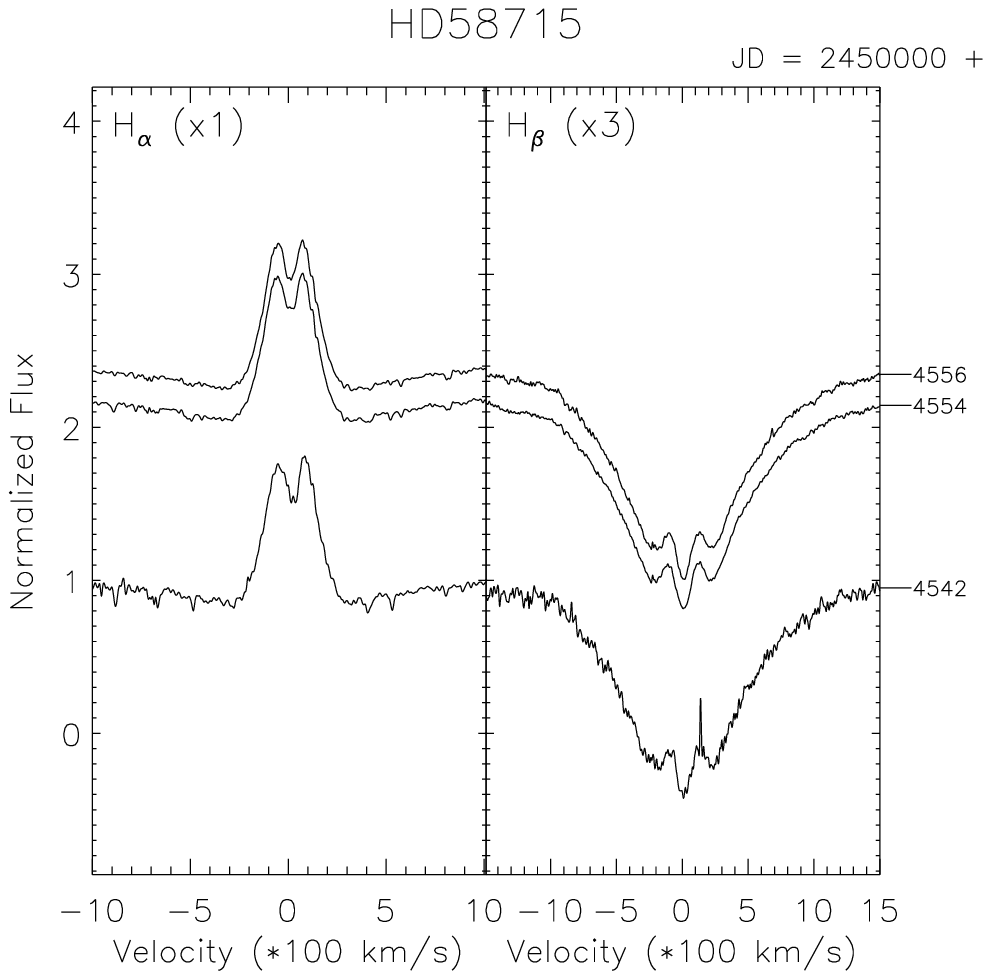}
   \end{array}$
   \end{center}

      \caption{Line profiles of H$\alpha$ and H$\beta$ observed for our program stars. Spectra have been shifted along the vertical
               axis according to the Julian Day of their observation, whose last four digits are reported. In parenthesis we report 
               the magnification factors applied to each profiles in order to improve the visualization.
              }
         \label{variab1}
   \end{figure*}
   \begin{figure*}
   \begin{center}$
   \begin{array}{cc}

   \includegraphics[width=8cm]{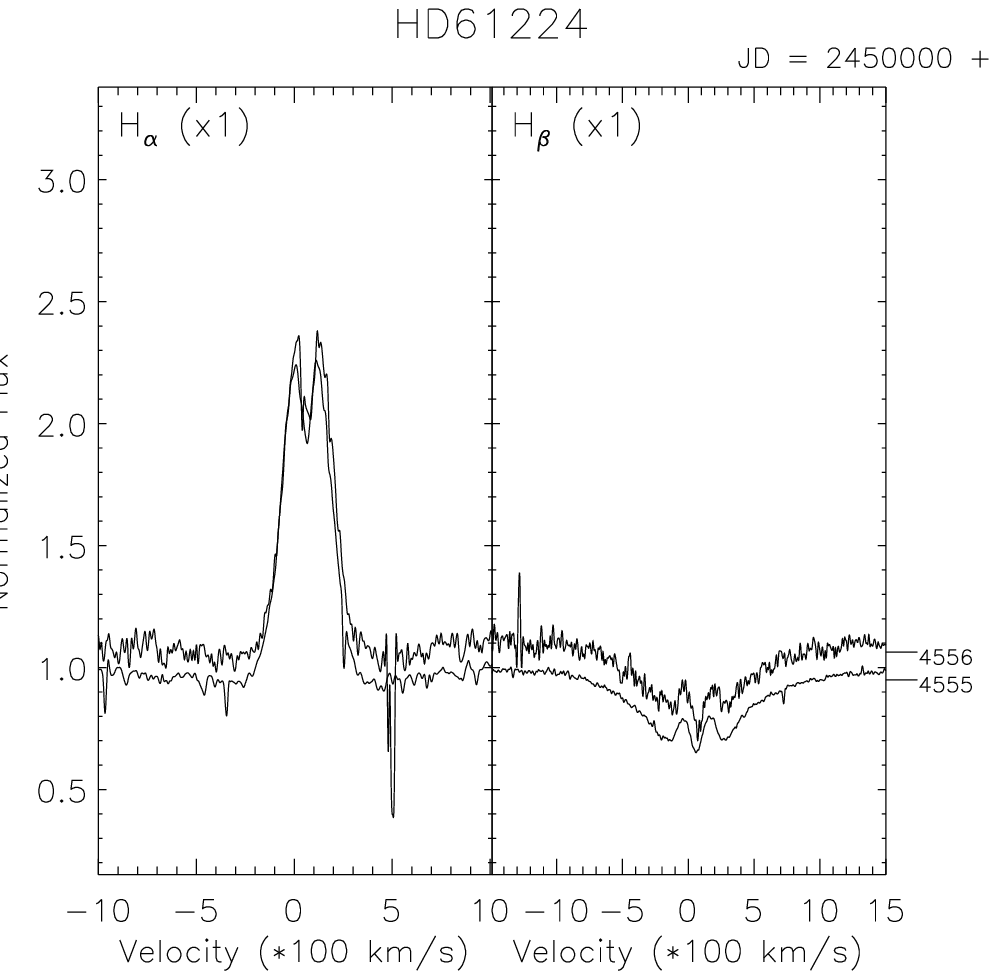} &
   \includegraphics[width=8cm]{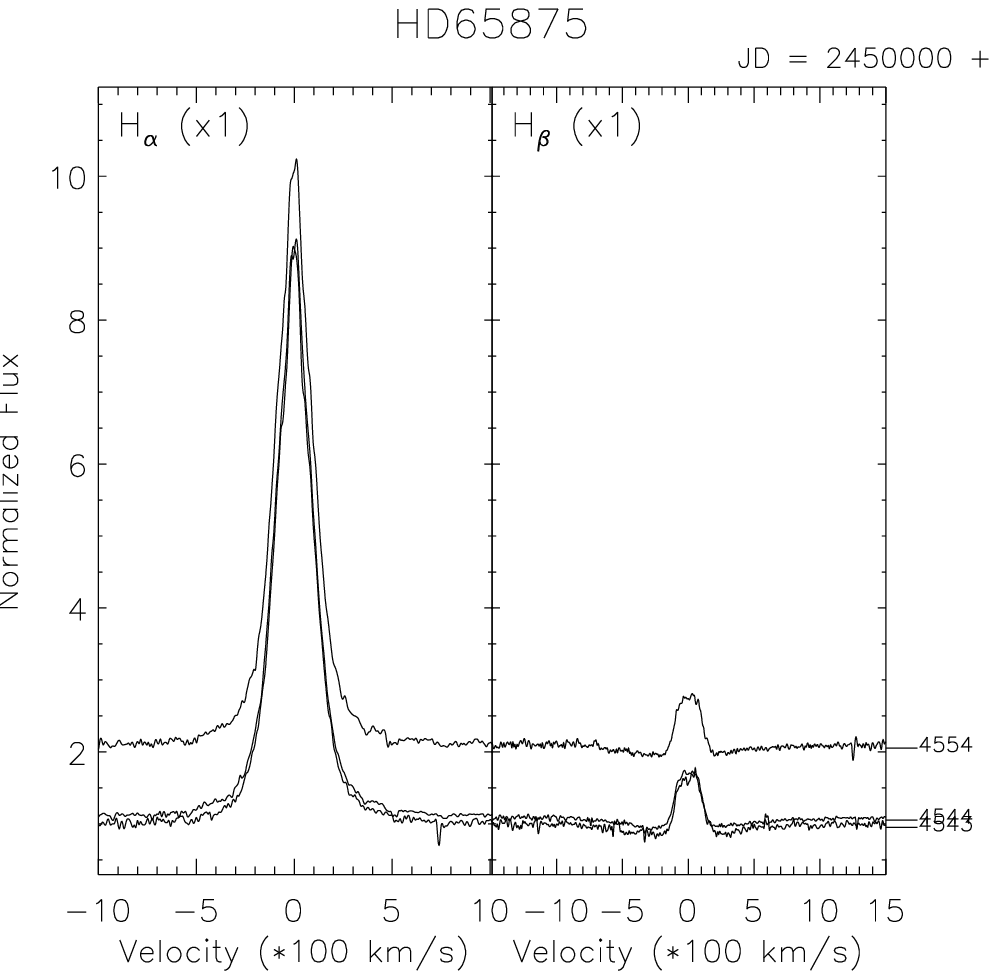} \\
   \includegraphics[width=8cm]{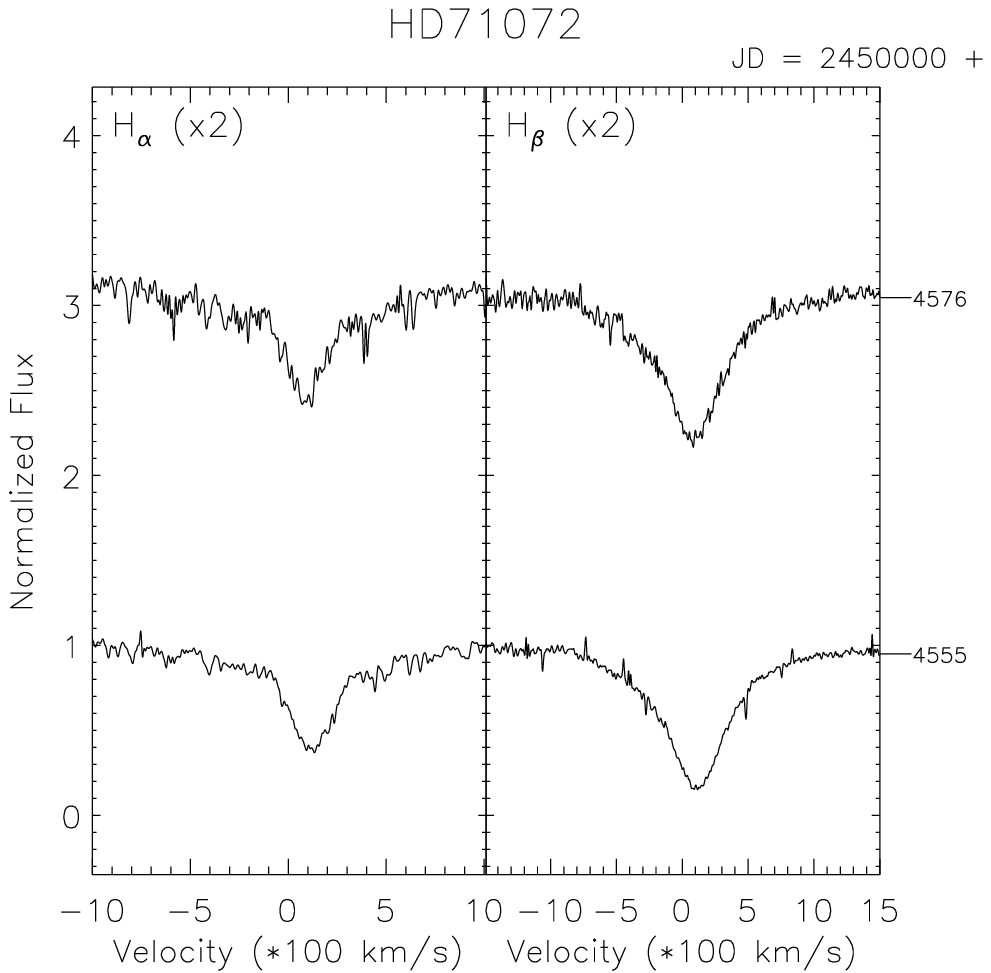} & 
   \includegraphics[width=8cm]{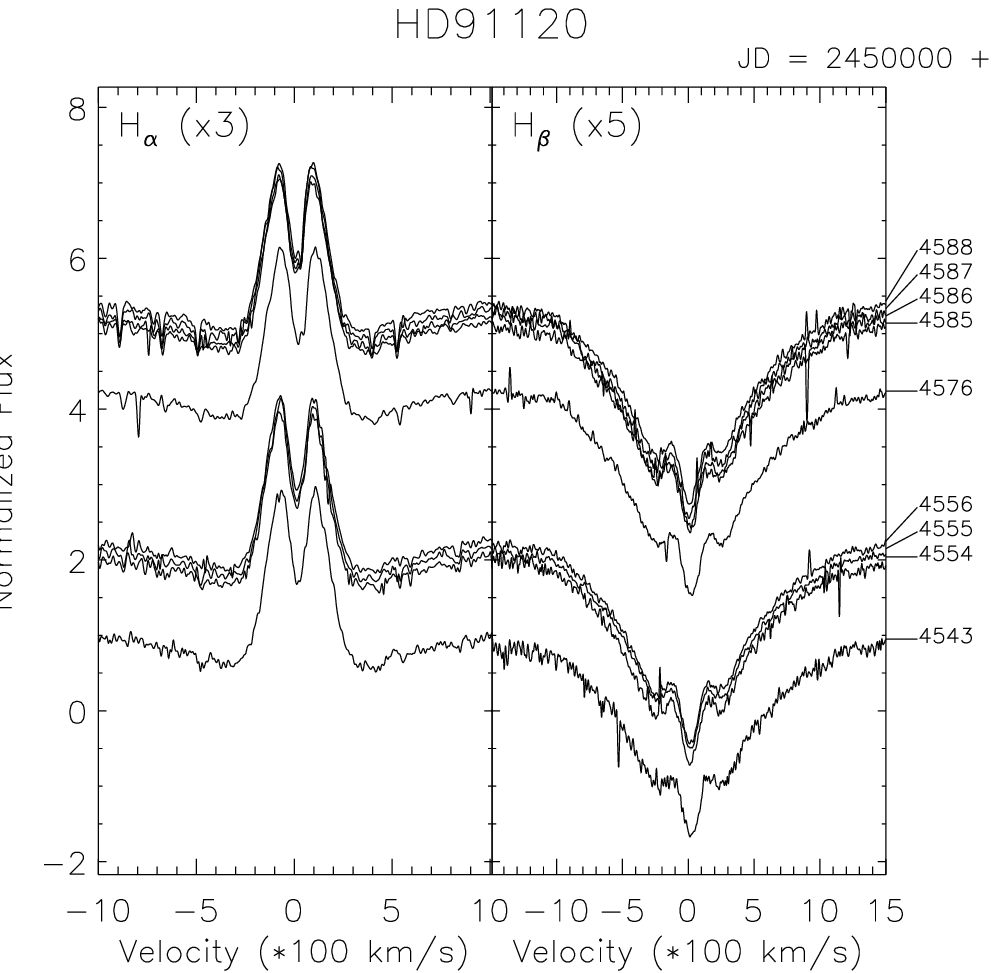} \\
   \includegraphics[width=8cm]{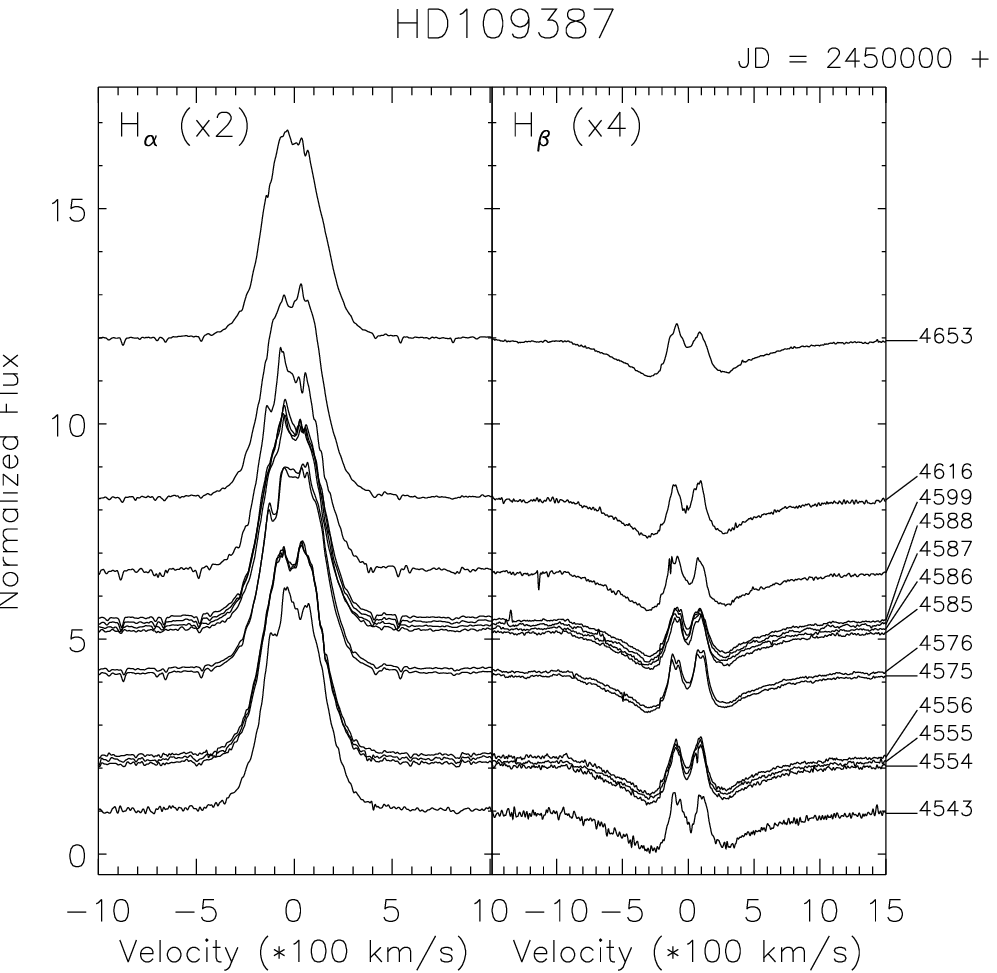} &
   \includegraphics[width=8cm]{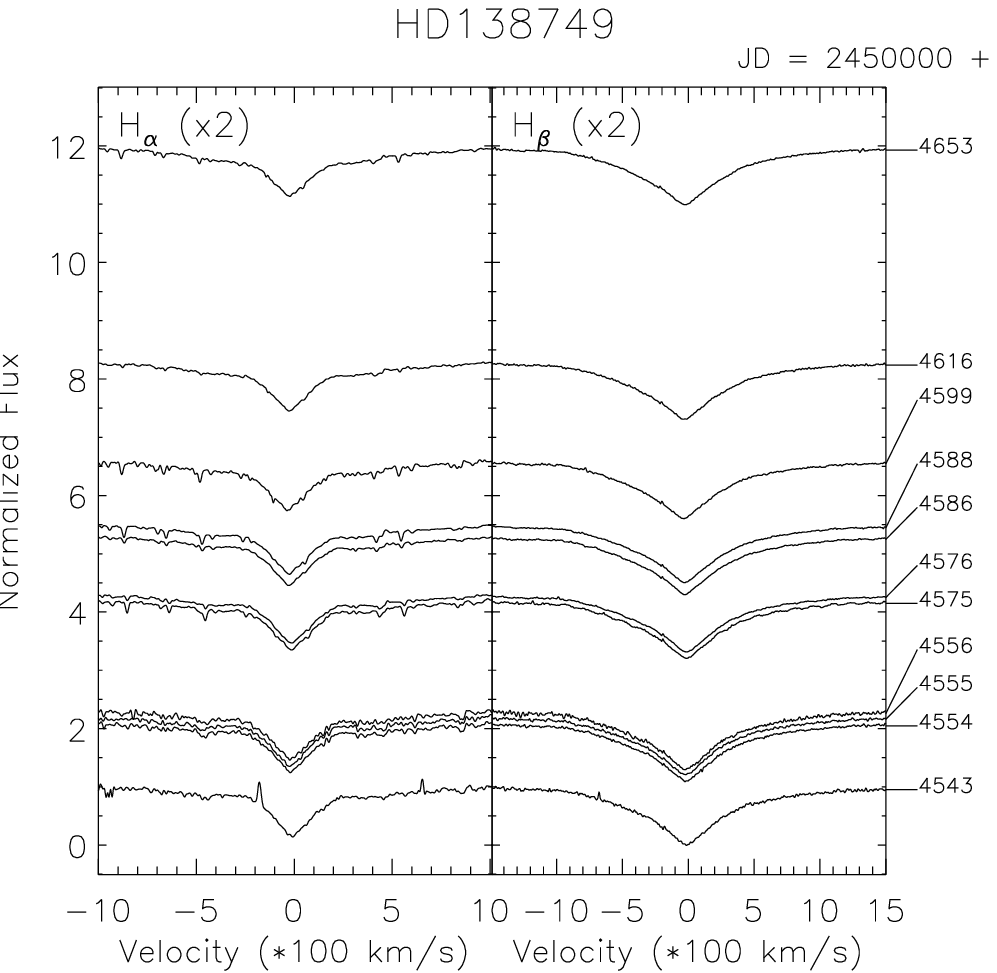}
   \end{array}$
   \end{center}

      \caption{As in Fig.~\ref{variab1}
              }
         \label{variab2}
   \end{figure*}
   \begin{figure*}
   \begin{center}$
   \begin{array}{cc}   
   \includegraphics[width=8cm]{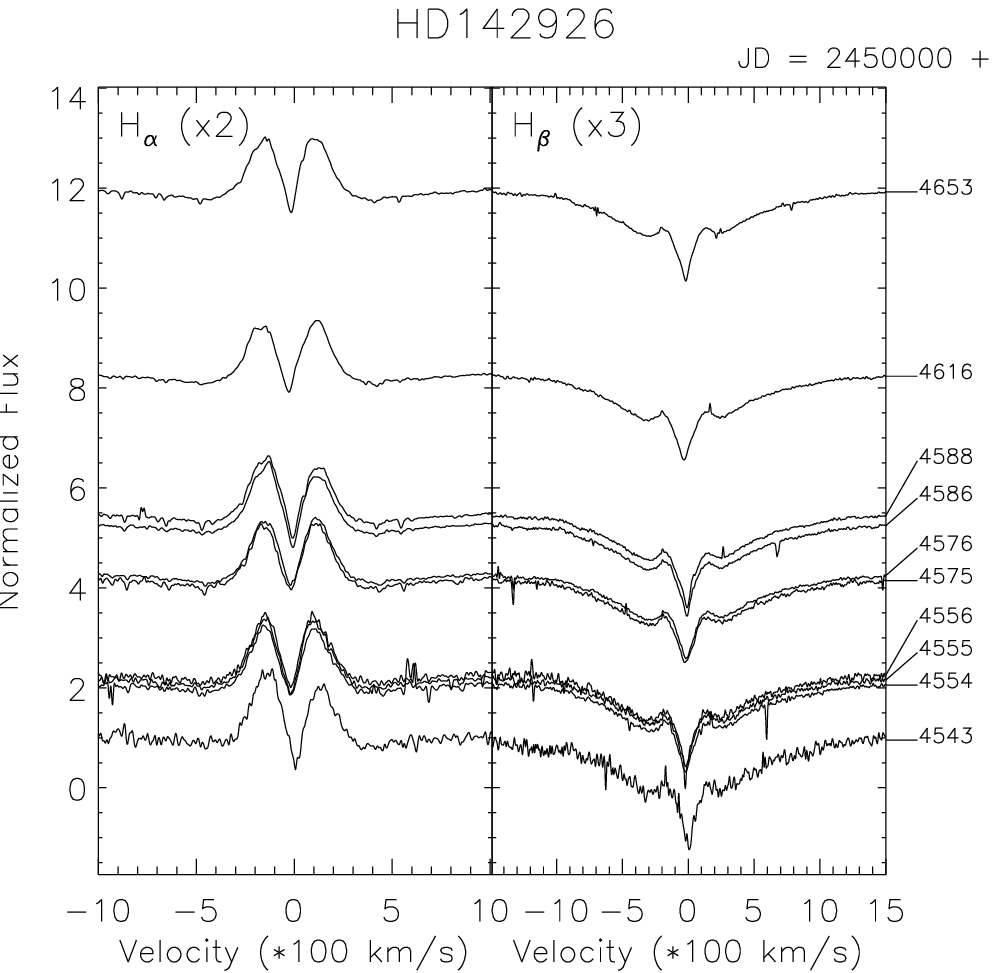} &
   \includegraphics[width=8cm]{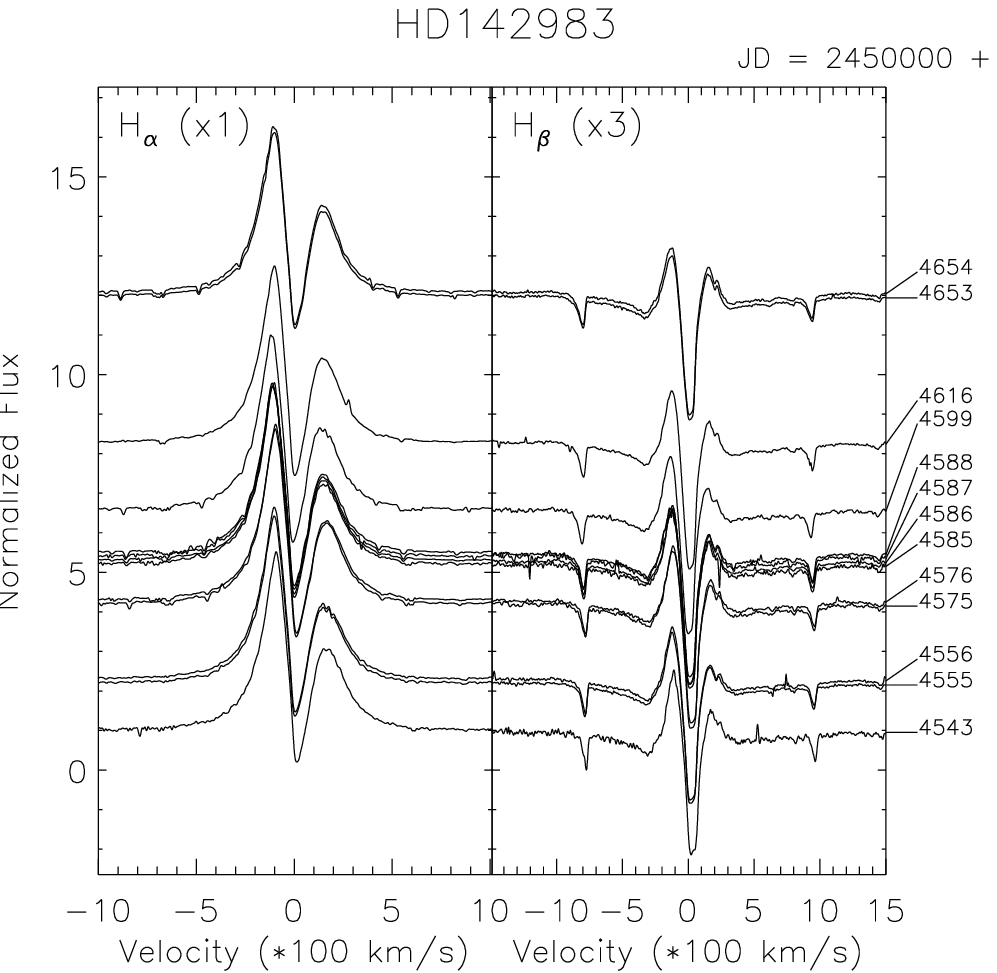} \\
   \includegraphics[width=8cm]{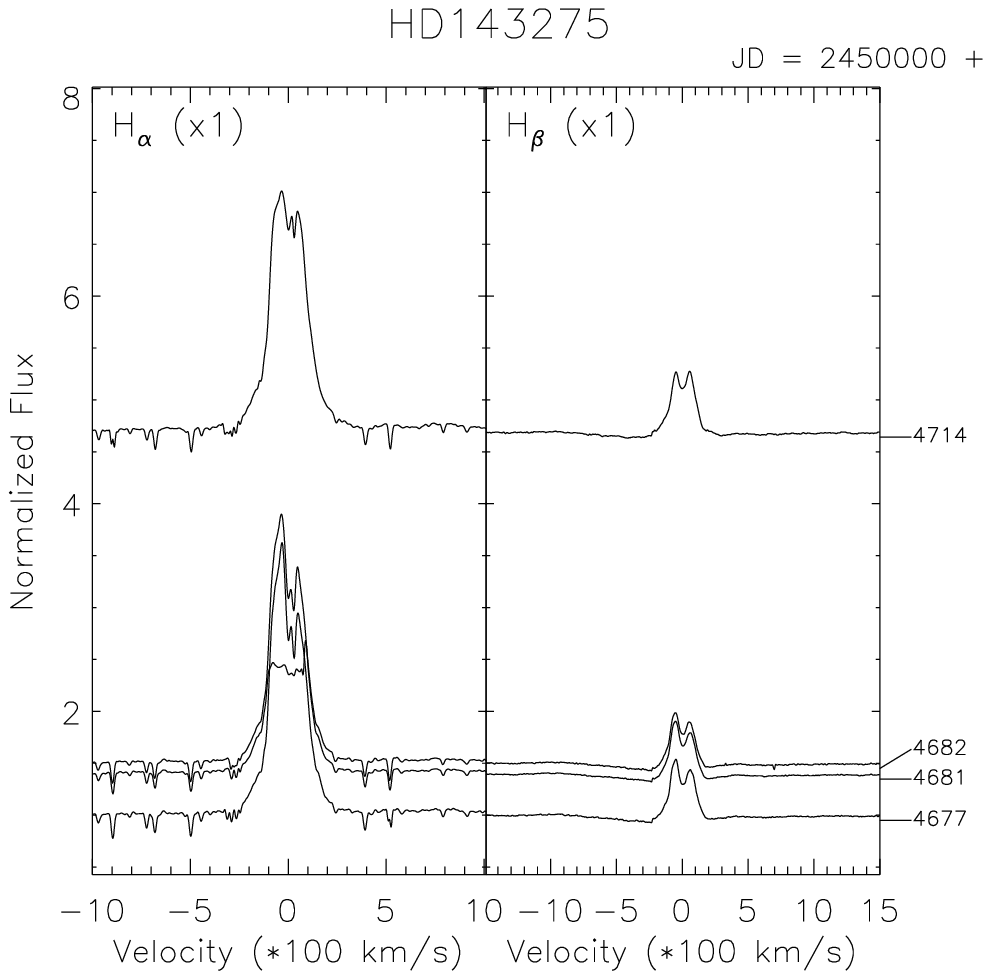} & 
   \includegraphics[width=8cm]{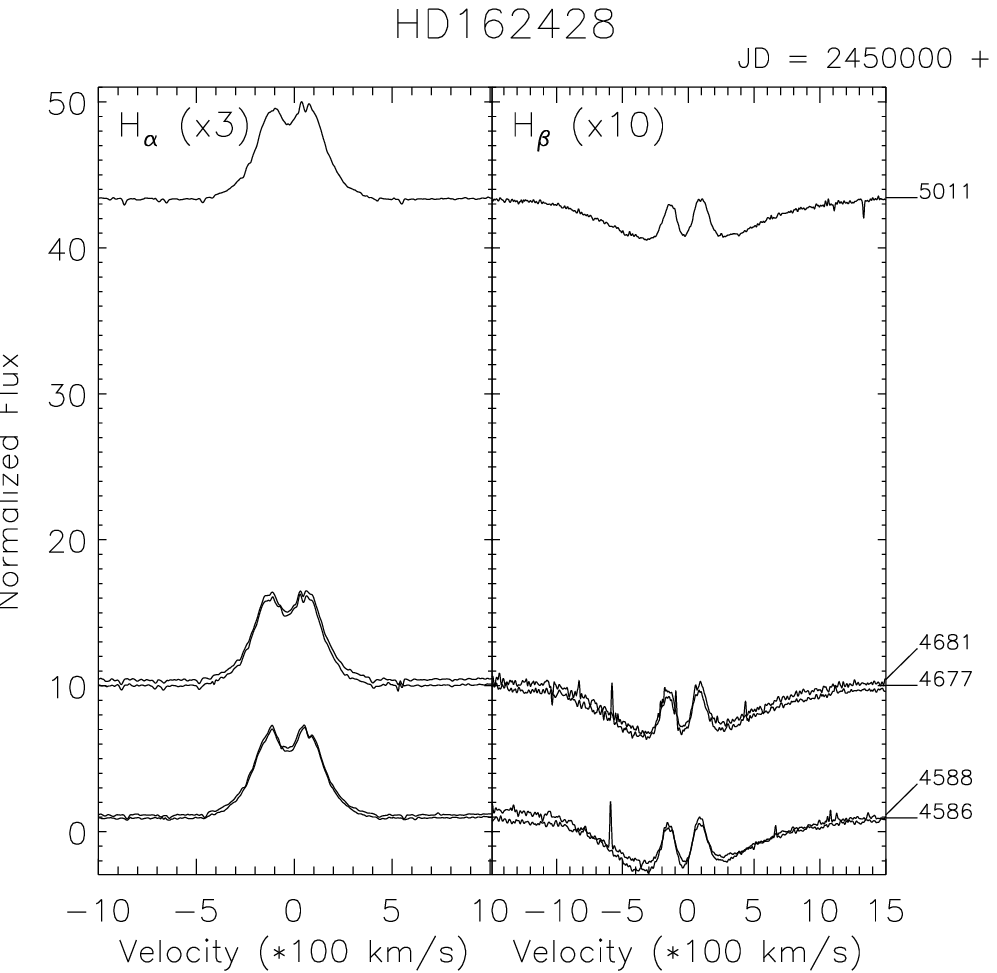} \\
   \includegraphics[width=8cm]{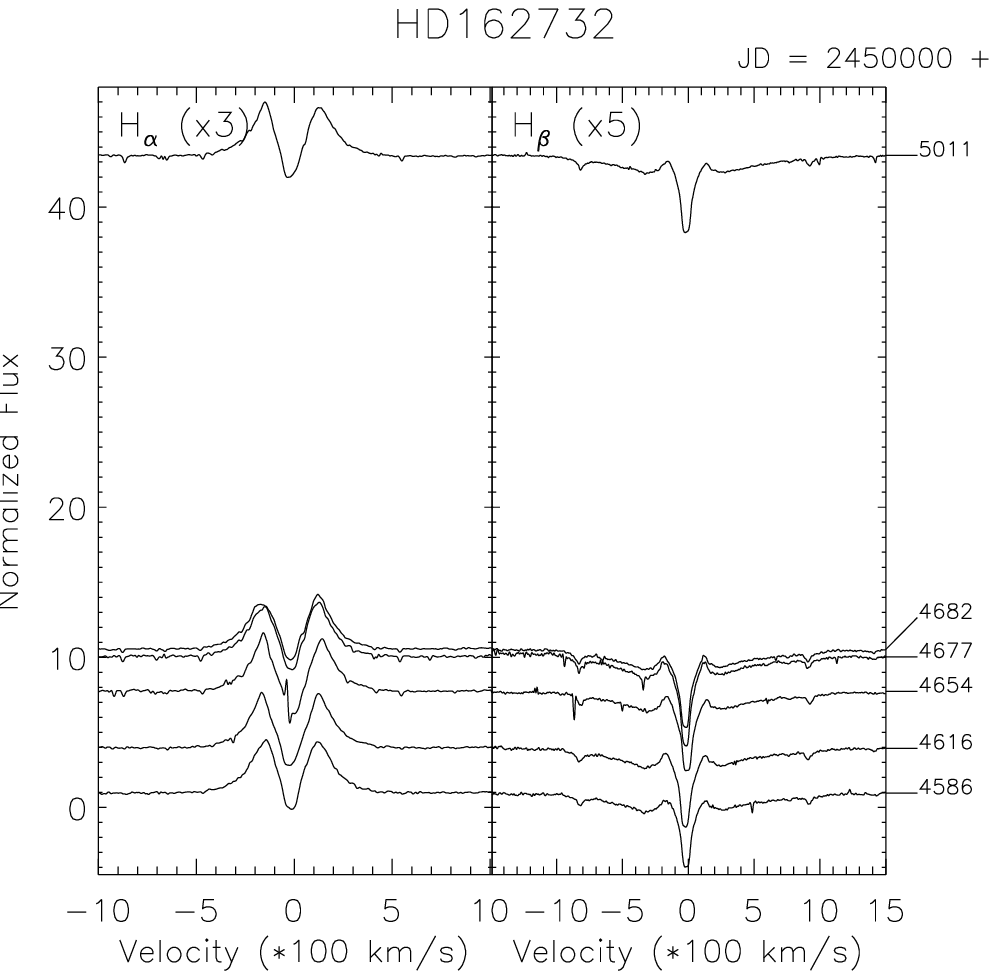} &
   \includegraphics[width=8cm]{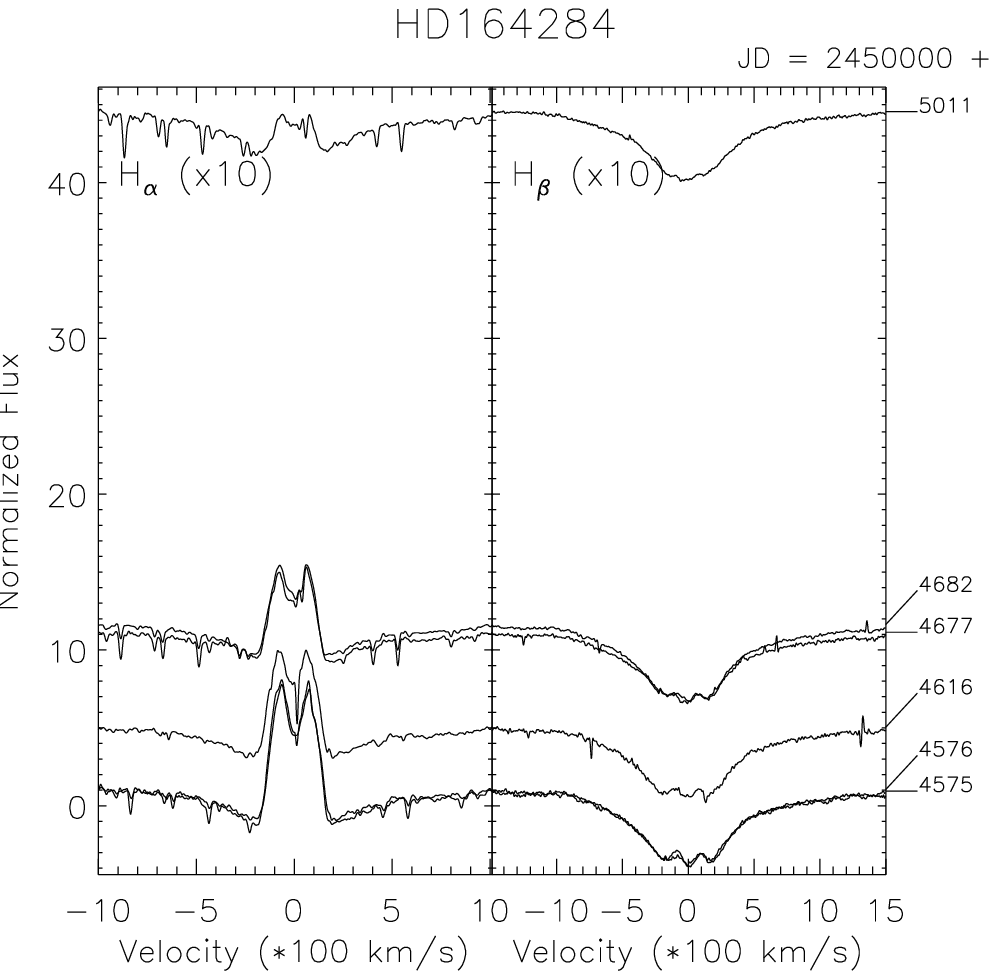}  
   \end{array}$
   \end{center}

      \caption{As in Fig.~\ref{variab1}
              }
         \label{variab3}
   \end{figure*}
   \begin{figure*}
   \begin{center}$
   \begin{array}{cc}
   \includegraphics[width=8cm]{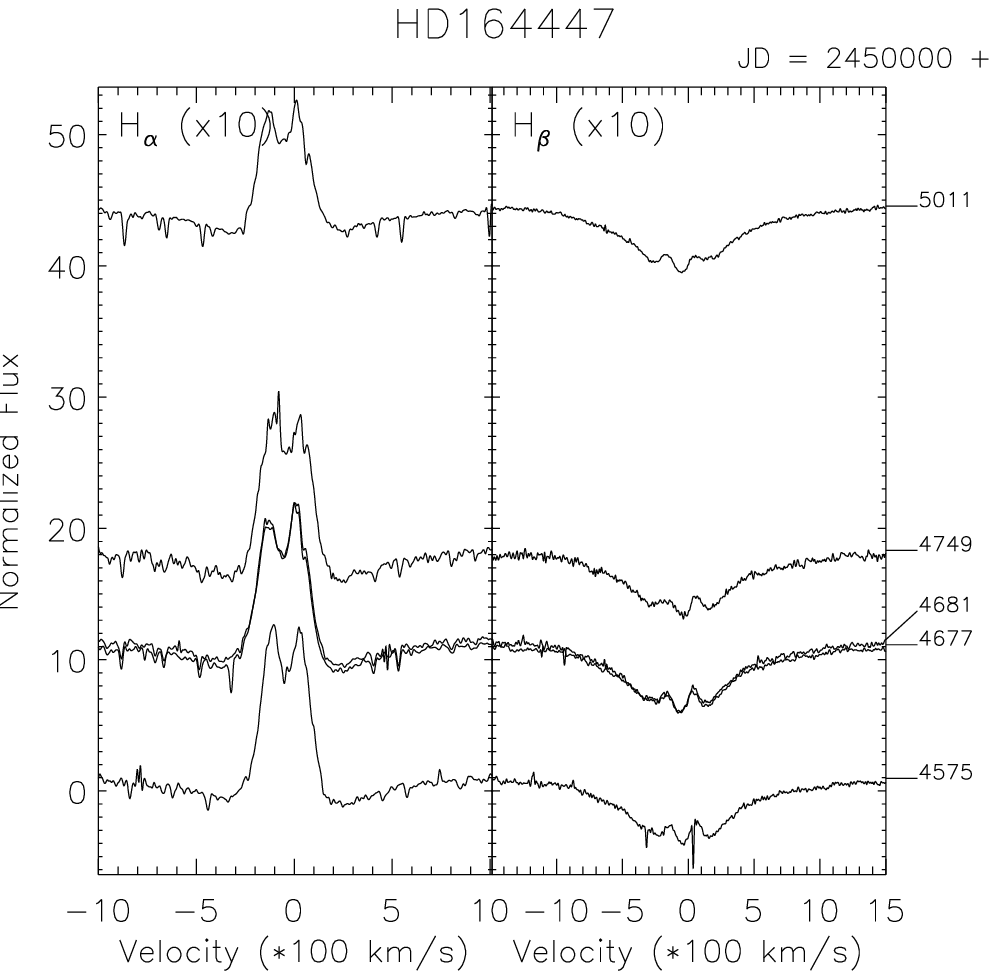} &
   \includegraphics[width=8cm]{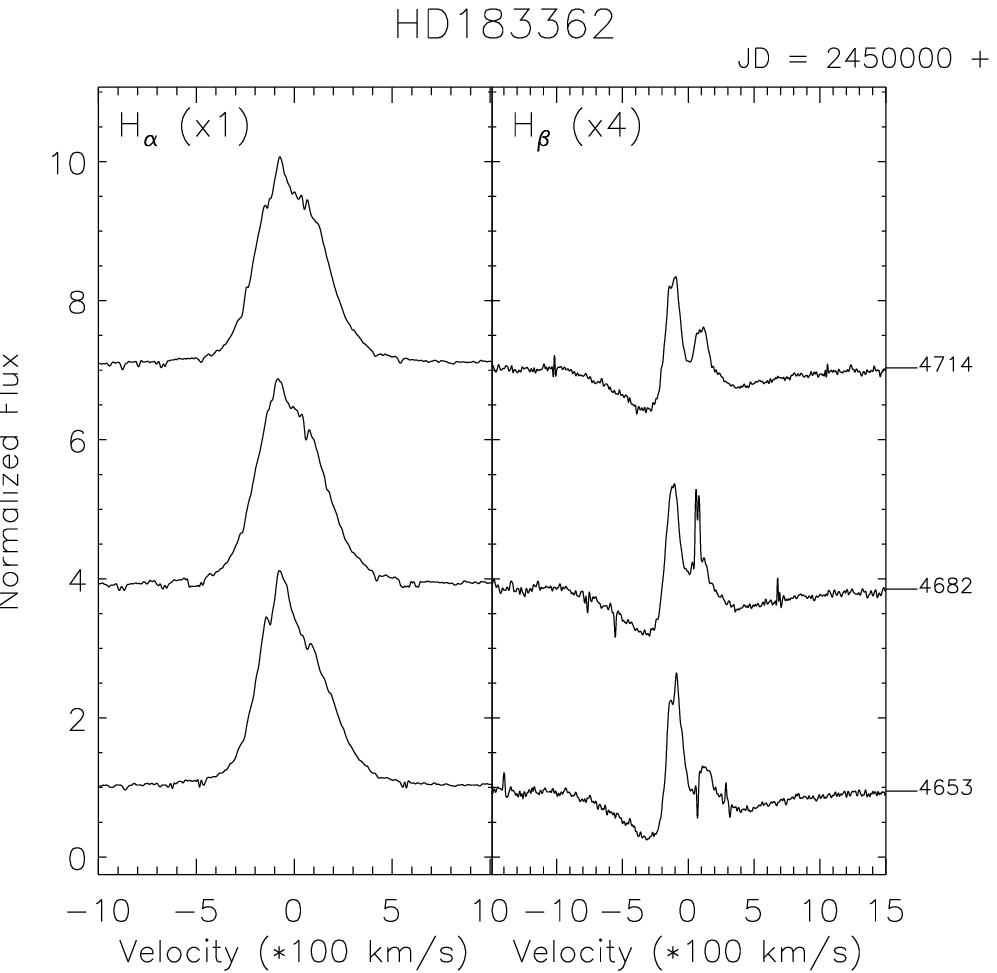} \\
   \includegraphics[width=8cm]{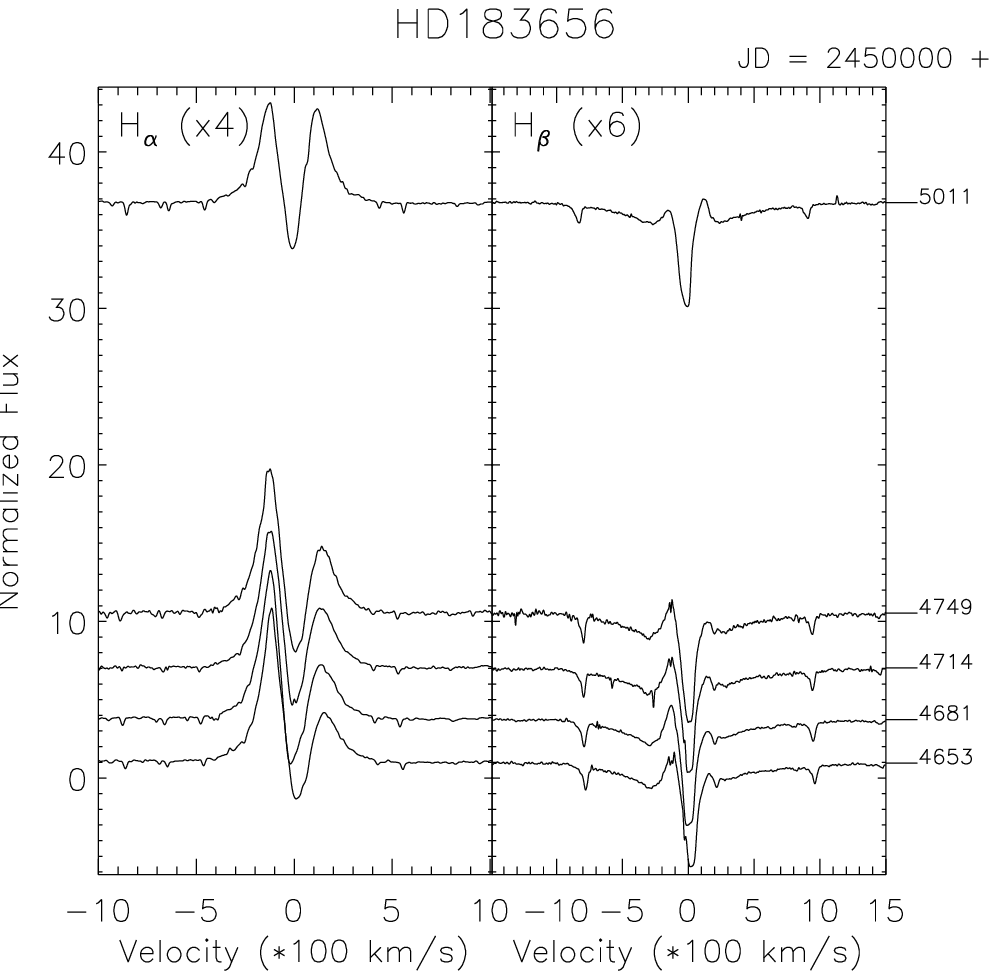} & 
   \includegraphics[width=8cm]{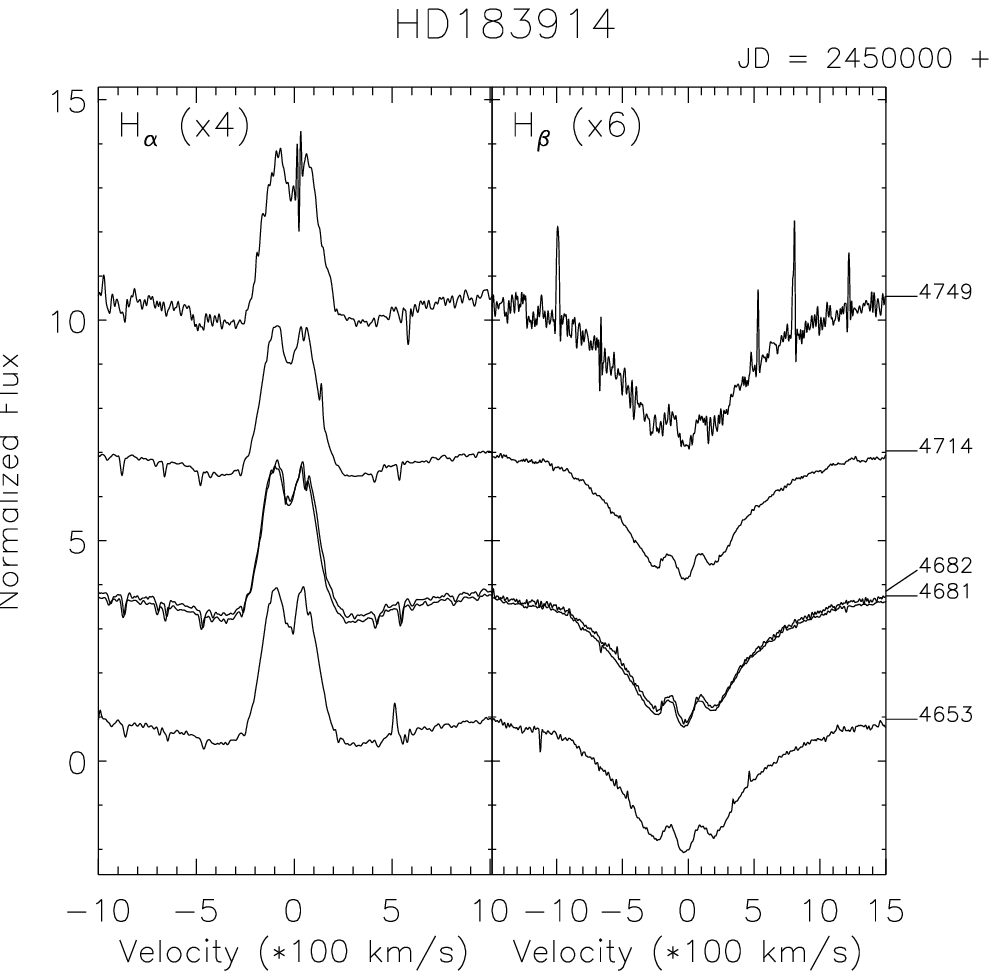} \\
   \includegraphics[width=8cm]{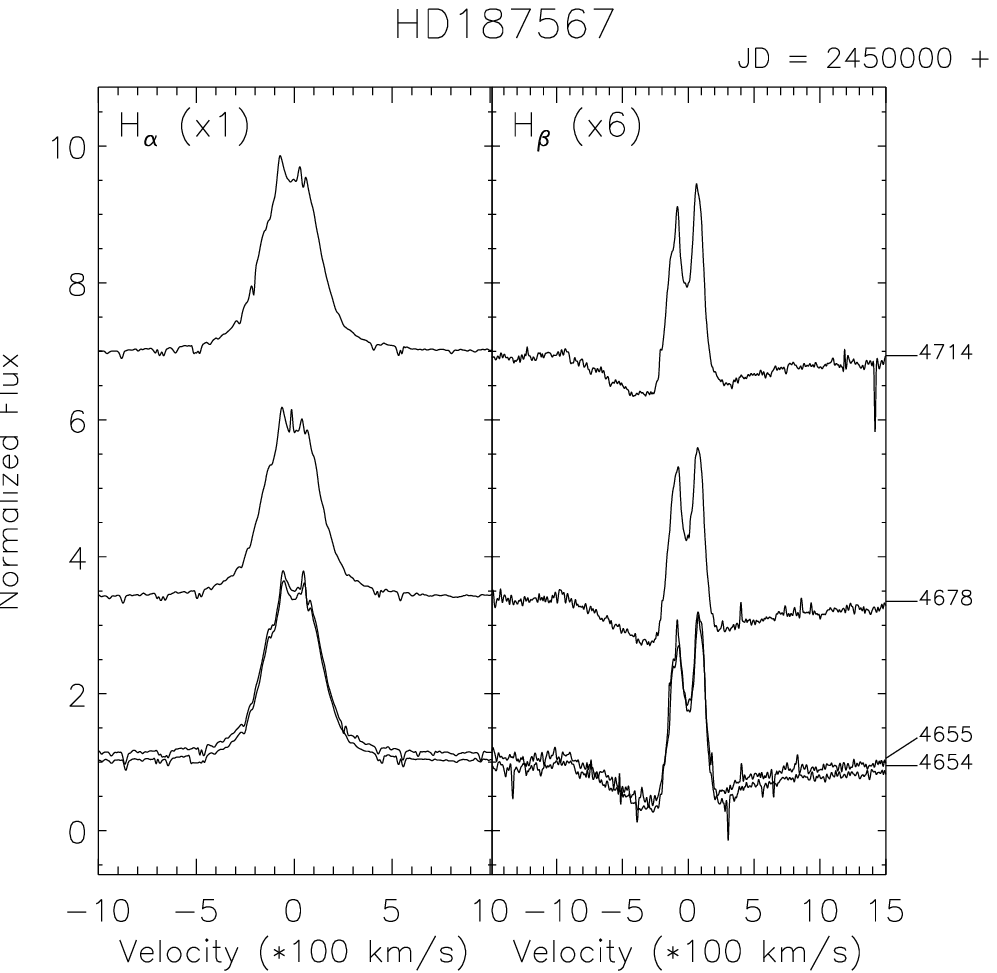} &
   \includegraphics[width=8cm]{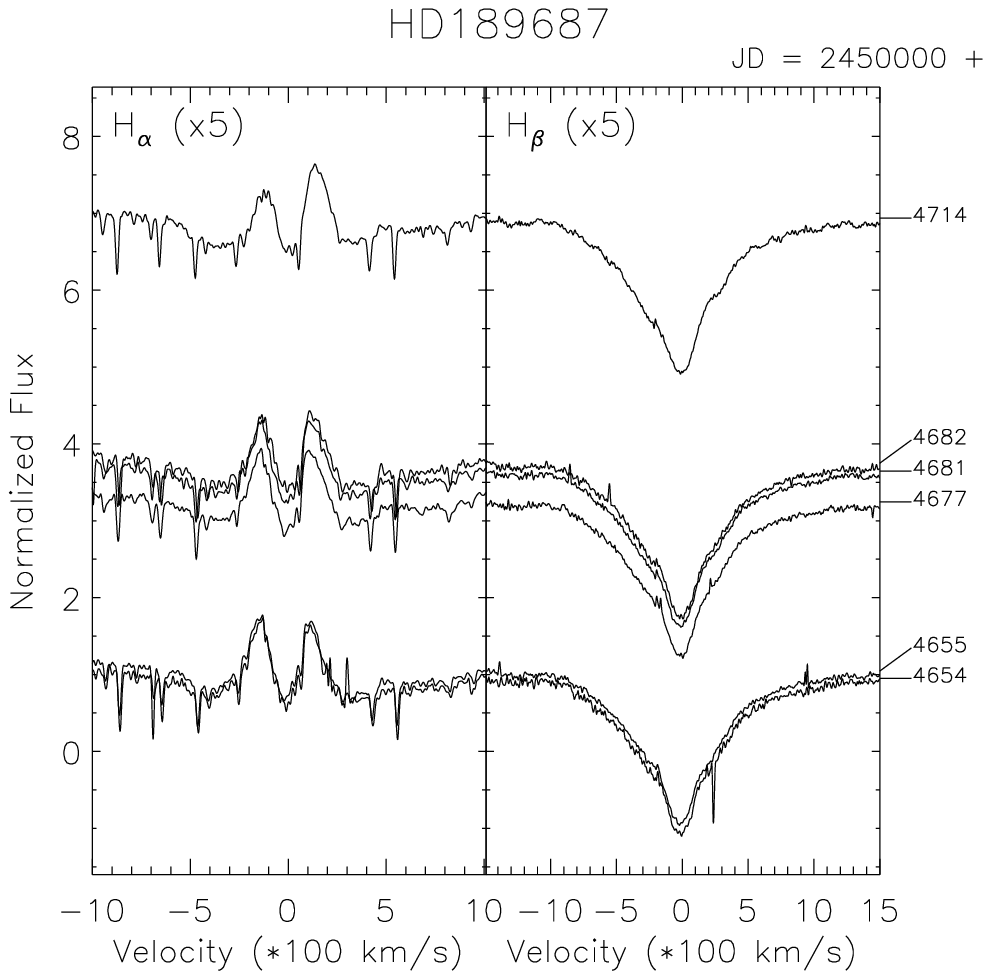}  
   \end{array}$
   \end{center}

      \caption{As in Fig.~\ref{variab1}
              }
         \label{variab4}
   \end{figure*}
   \begin{figure*}
   \begin{center}$
   \begin{array}{cc}
   \includegraphics[width=8cm]{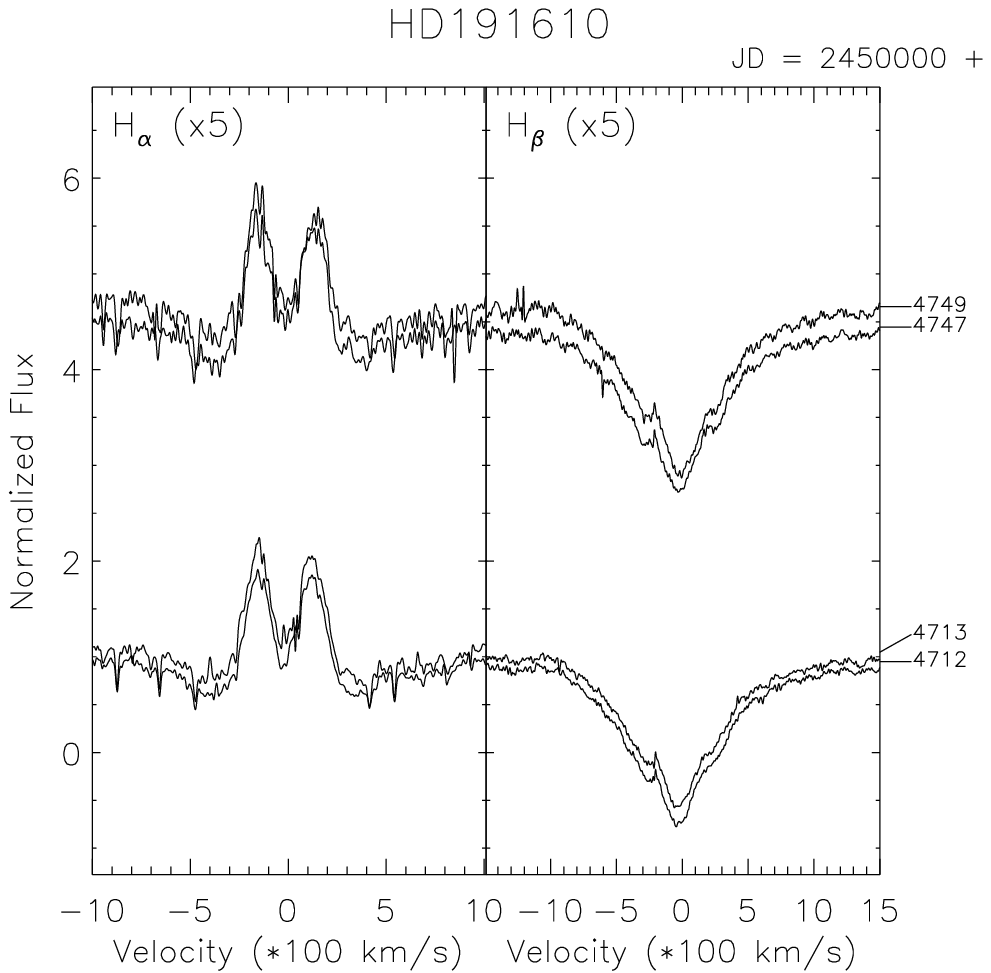} &
   \includegraphics[width=8cm]{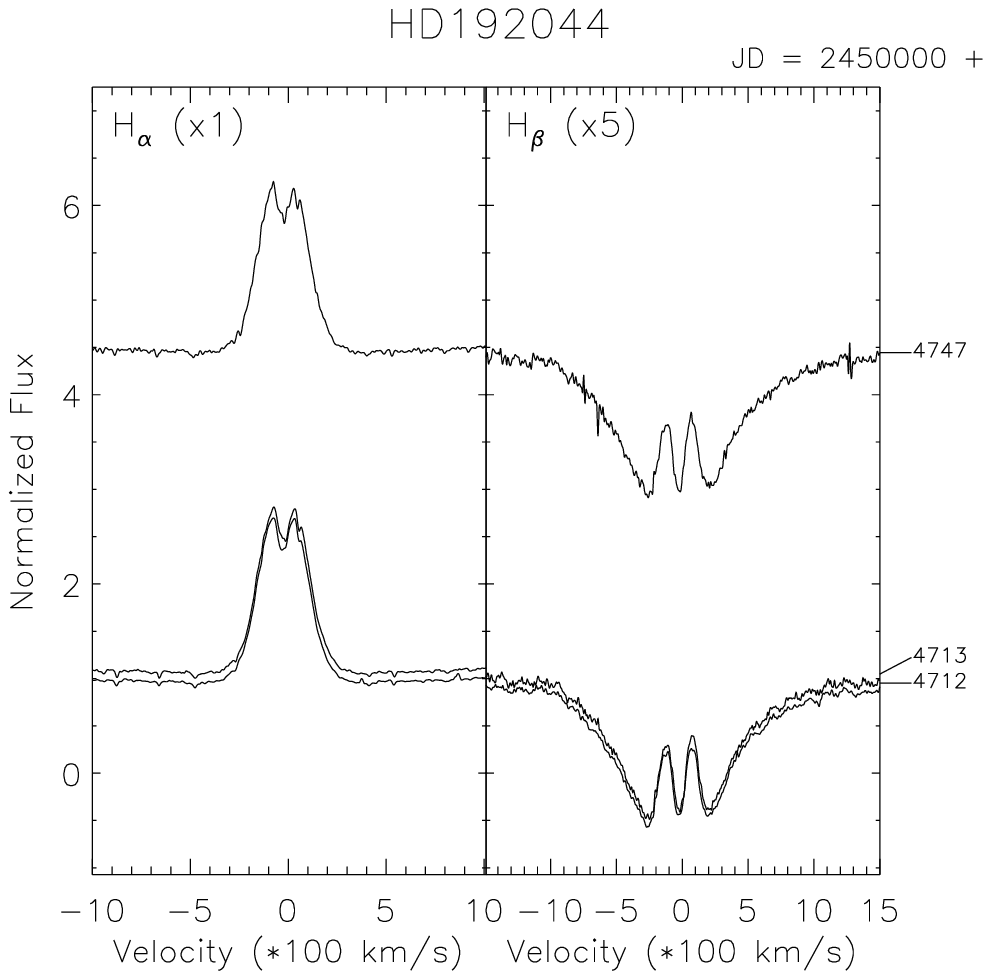} \\
   \includegraphics[width=8cm]{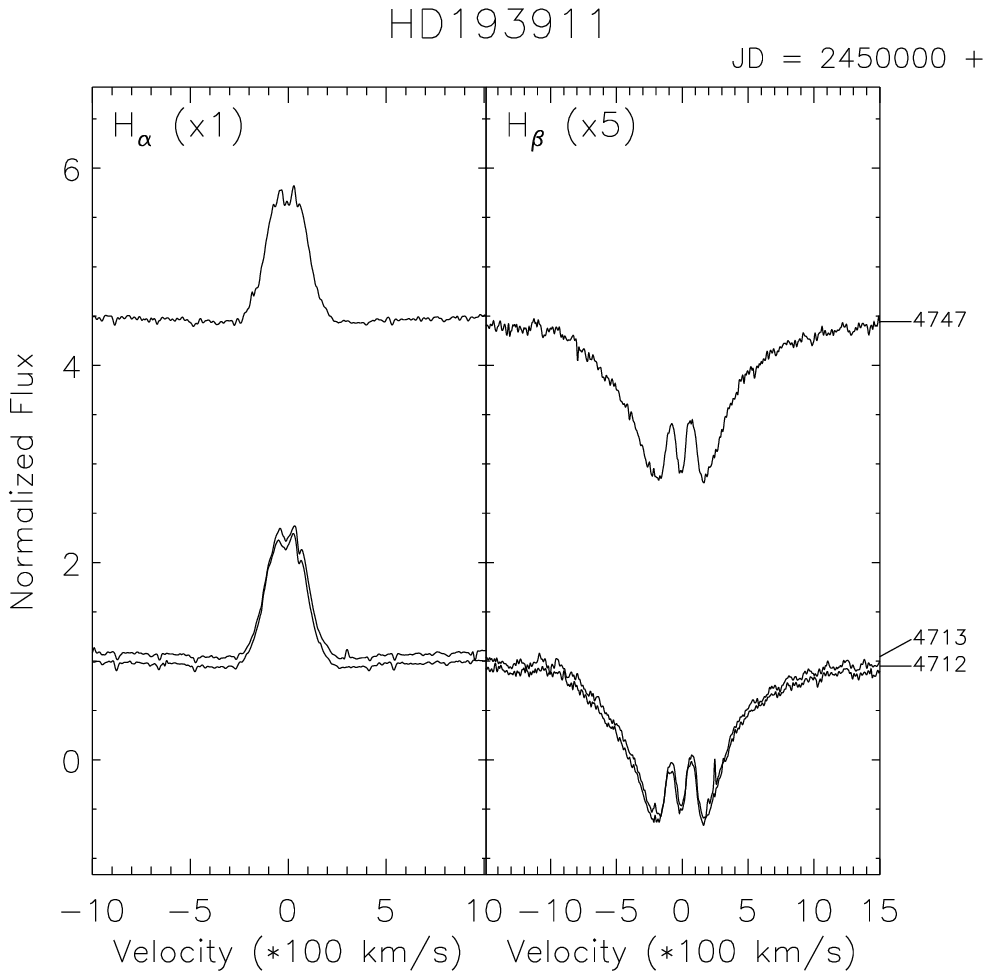} & 
   \includegraphics[width=8cm]{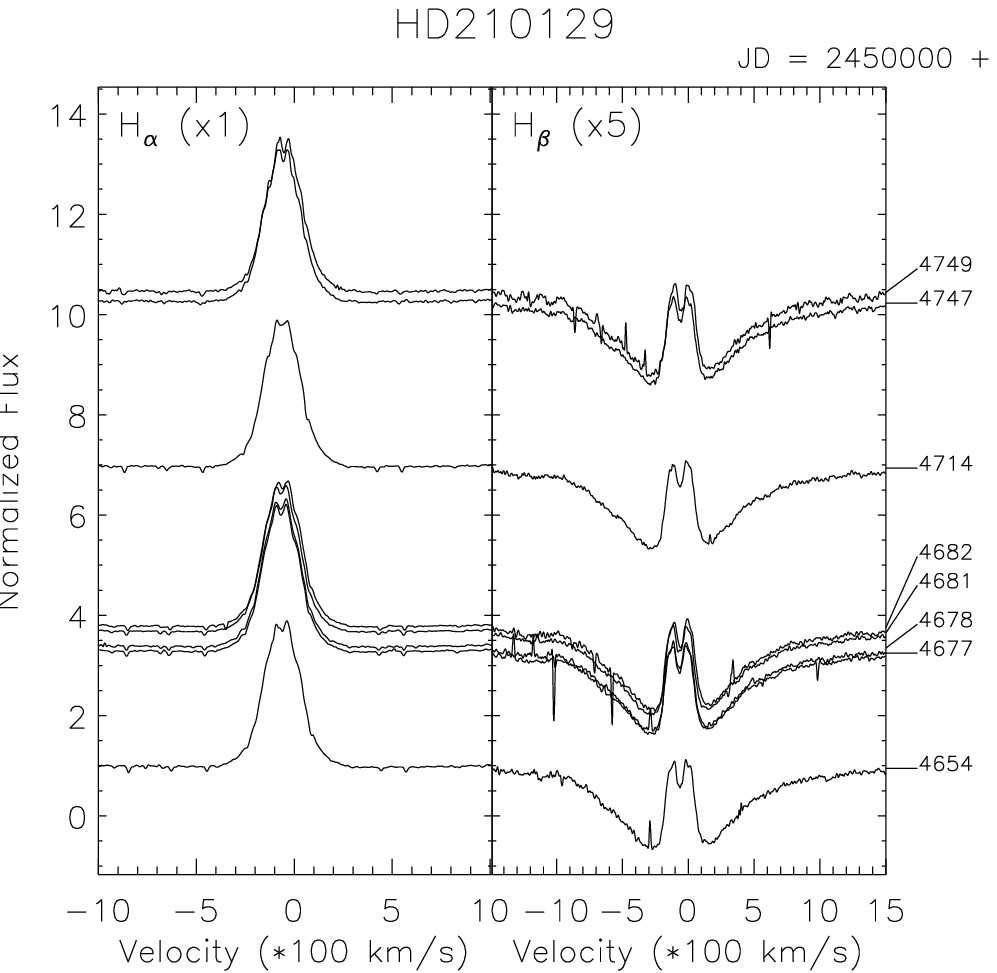} \\
   \includegraphics[width=8cm]{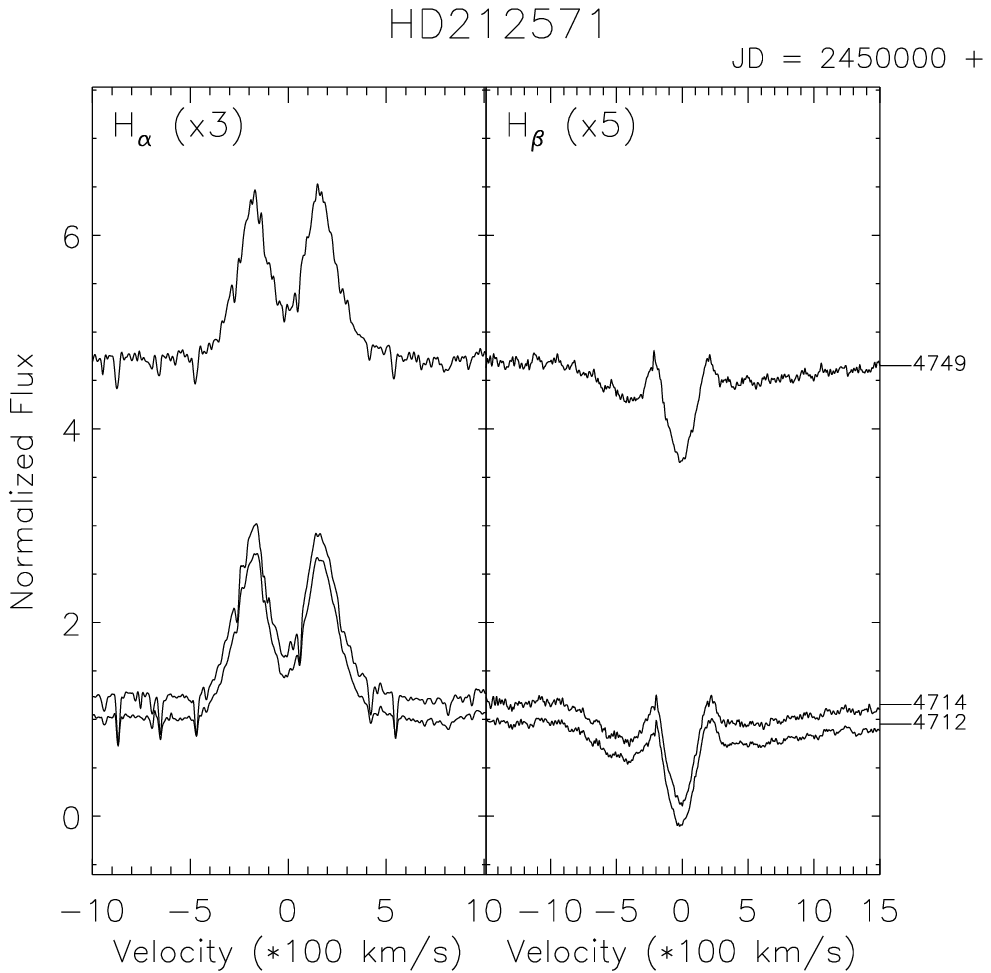} &
   \includegraphics[width=8cm]{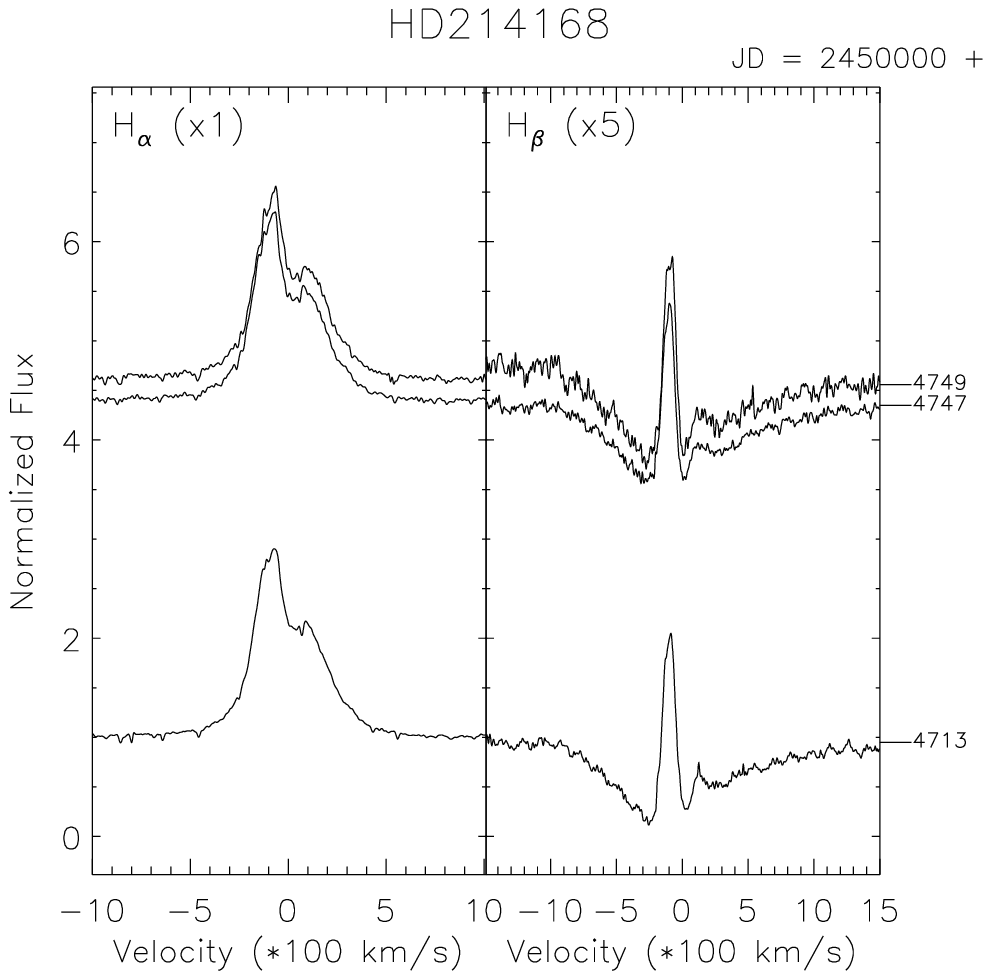} 
   \end{array}$
   \end{center}

      \caption{As in Fig.~\ref{variab1}
              }
         \label{variab5}
   \end{figure*}
   \begin{figure*}
   \begin{center}$
   \begin{array}{cc}
   \includegraphics[width=8cm]{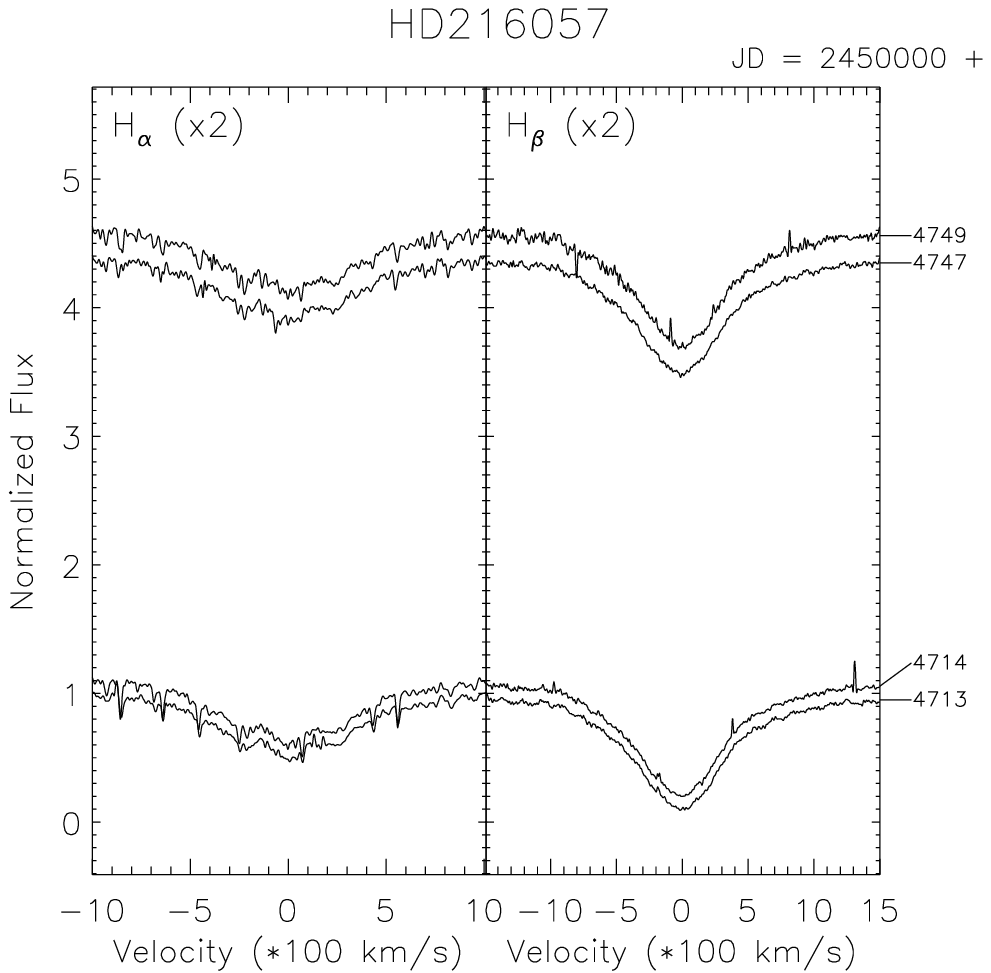} &
   \includegraphics[width=8cm]{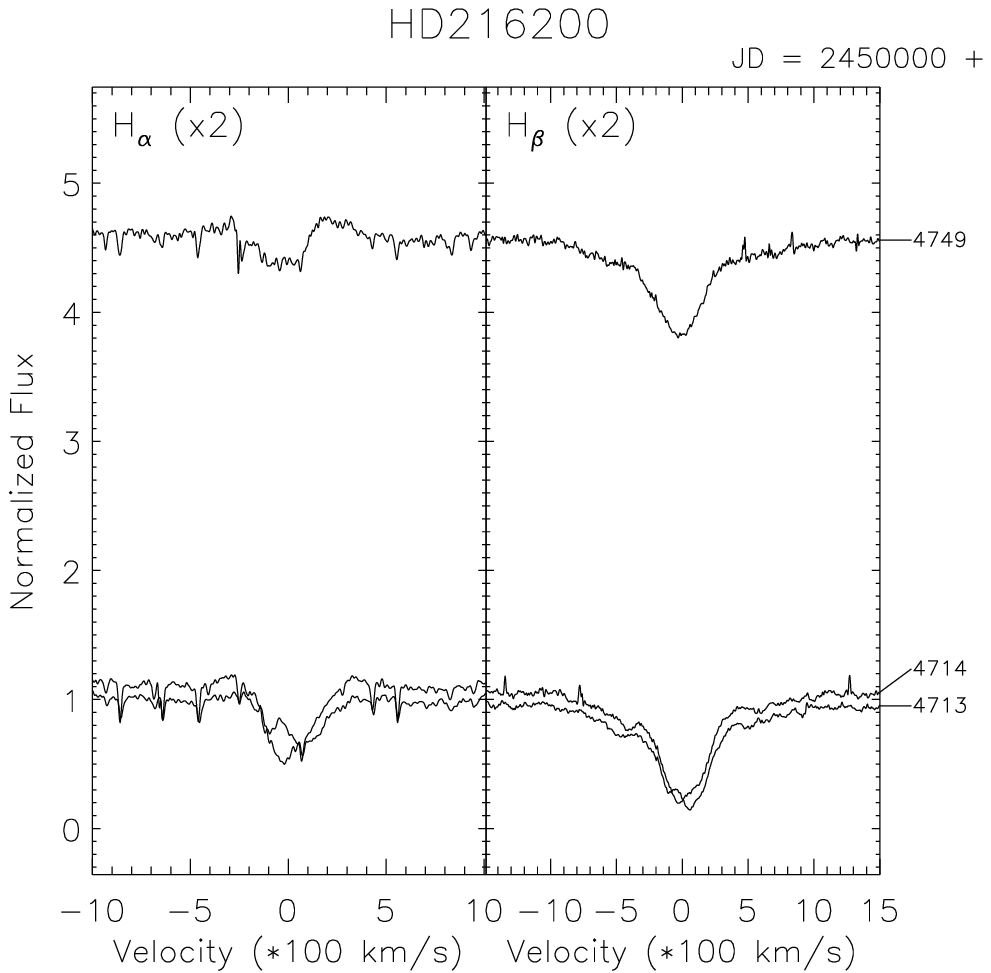} \\
   \includegraphics[width=8cm]{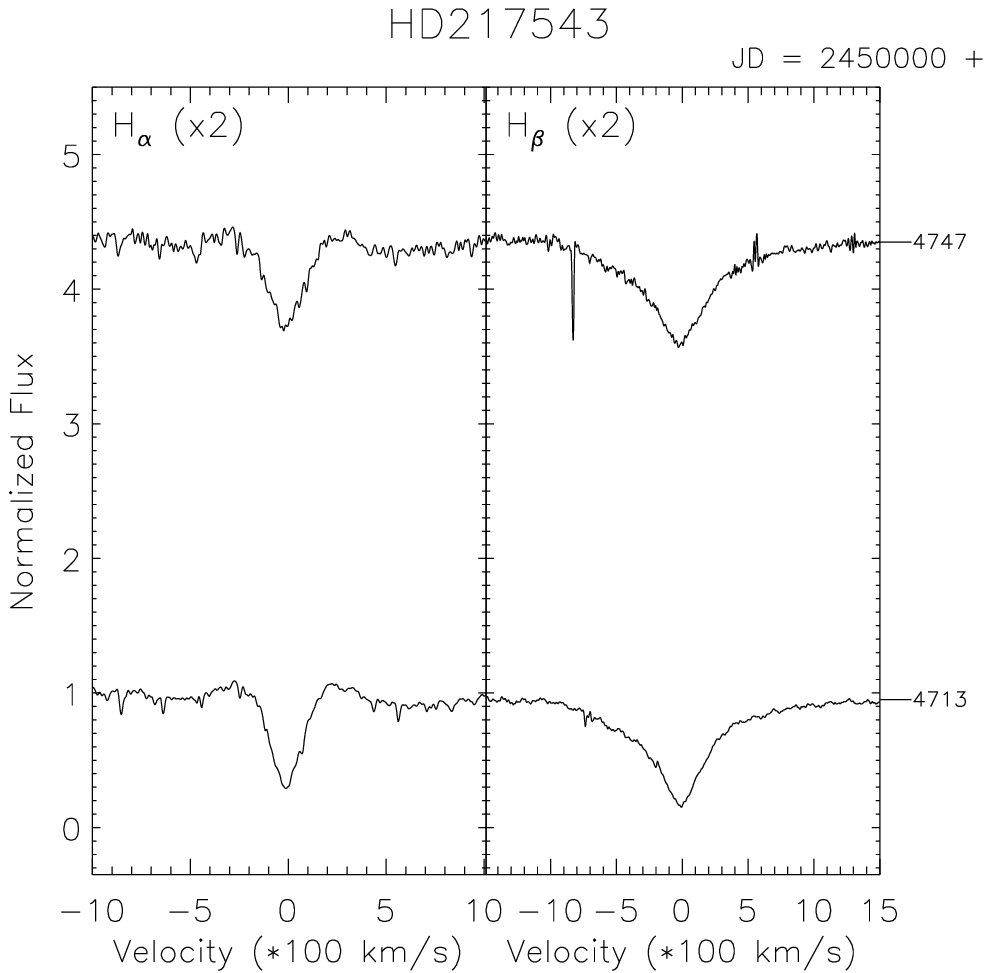} & 
   \includegraphics[width=8cm]{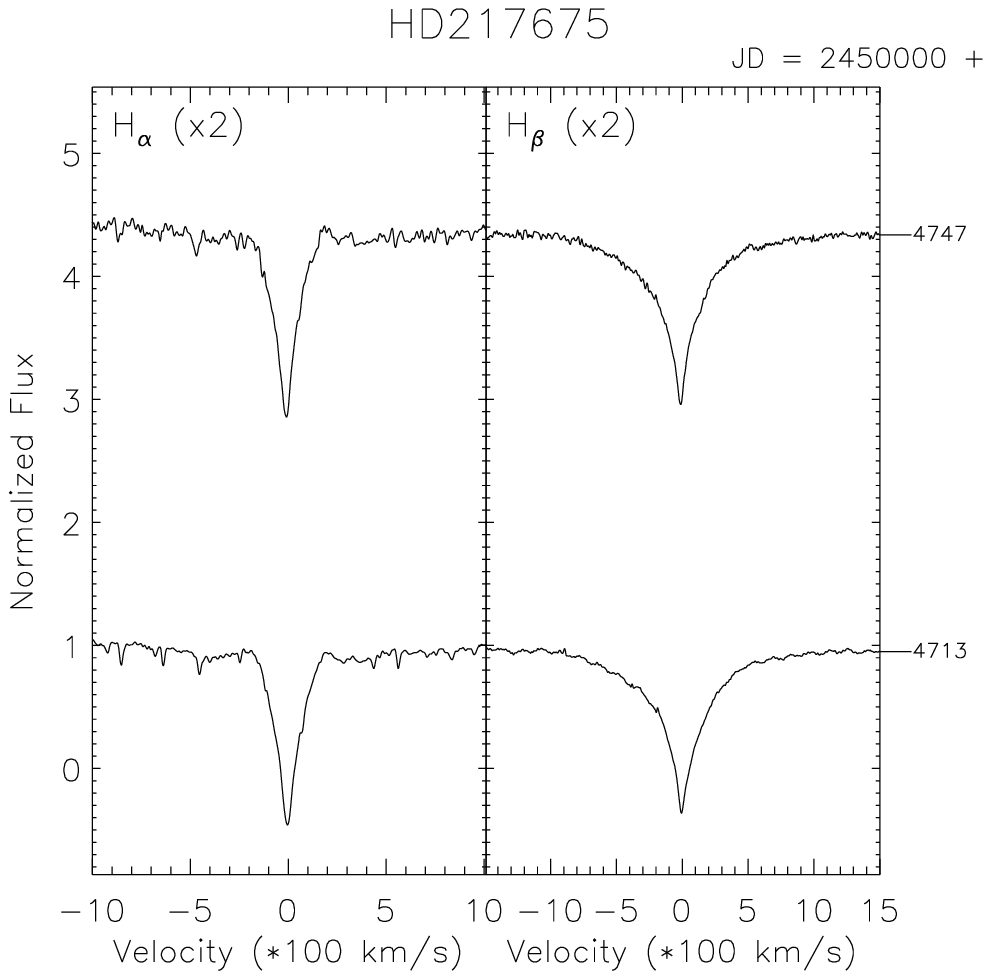} \\
   \includegraphics[width=8cm]{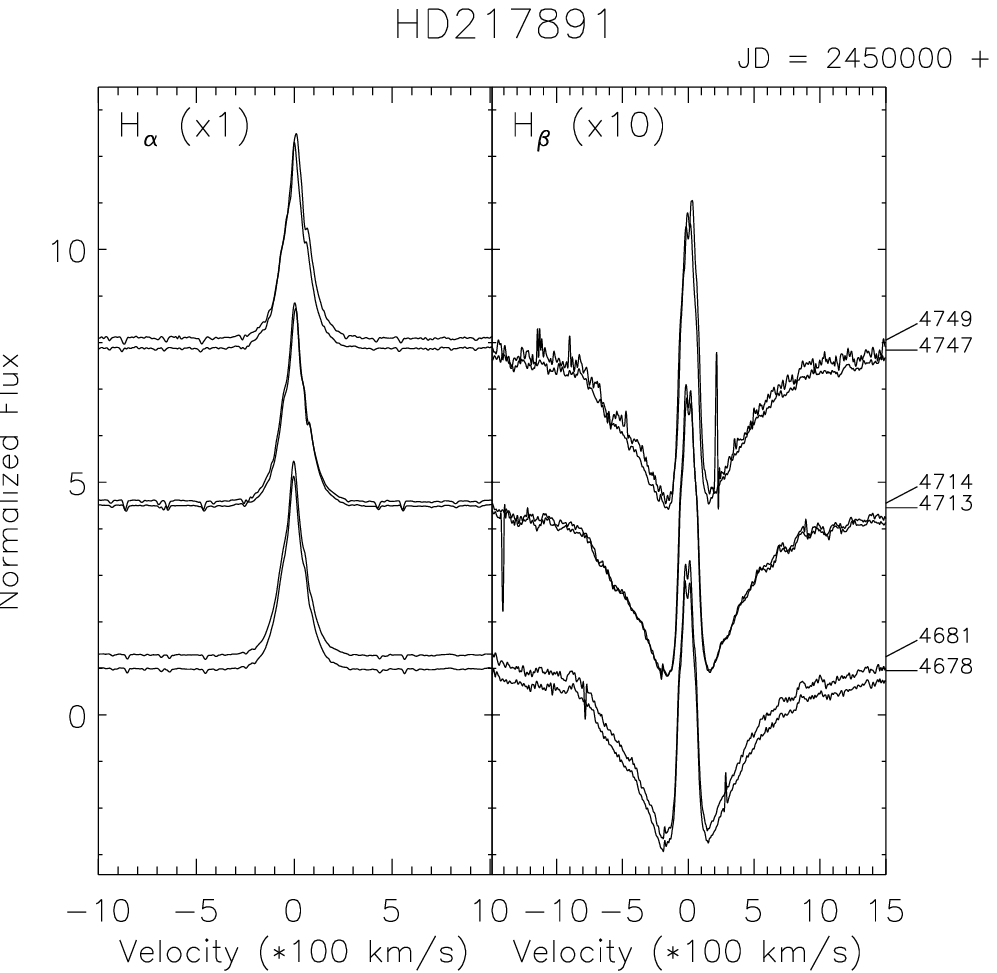} 
   \end{array}$
   \end{center}

      \caption{As in Fig.~\ref{variab1}
              }
         \label{variab6}
   \end{figure*}

\newpage

  \begin{figure*}
   \includegraphics[width=9cm]{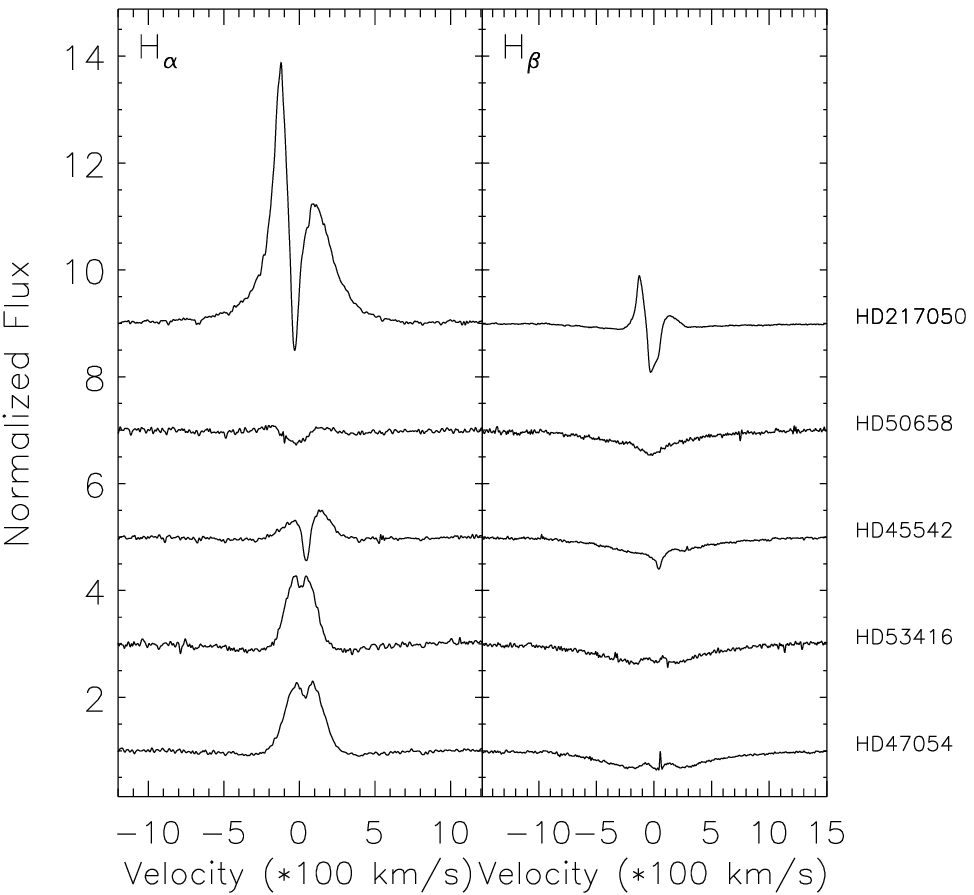}
   \includegraphics[width=9cm]{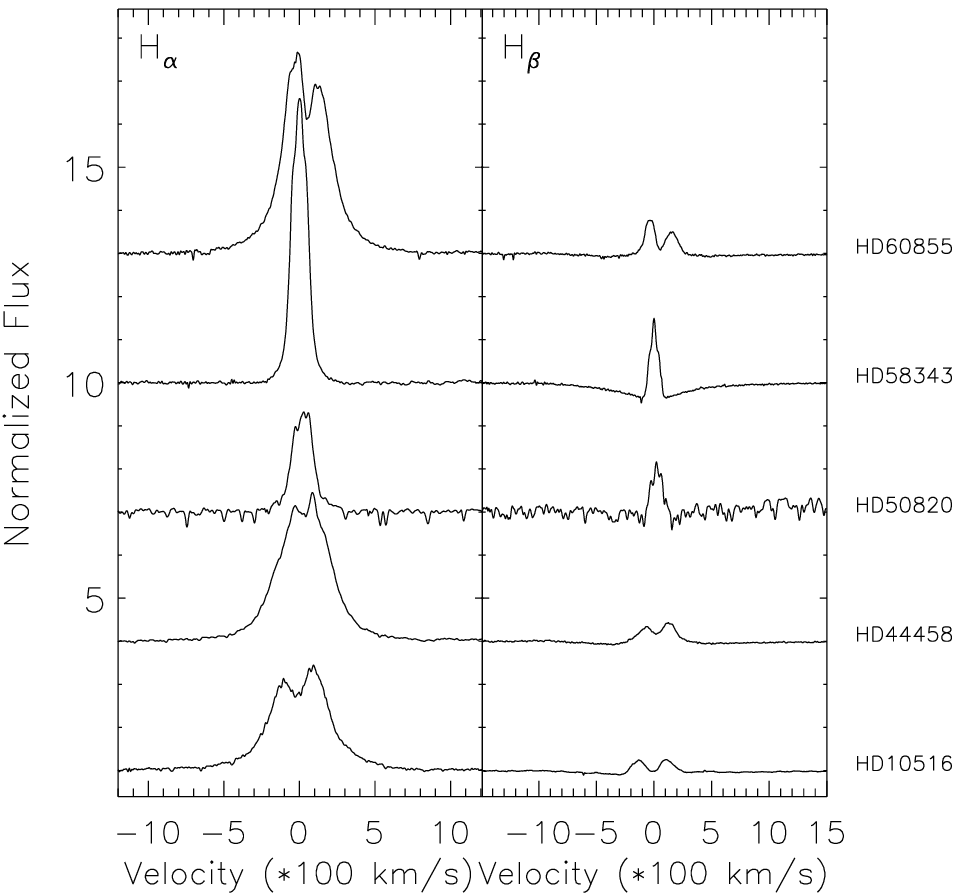}
   \includegraphics[width=9cm]{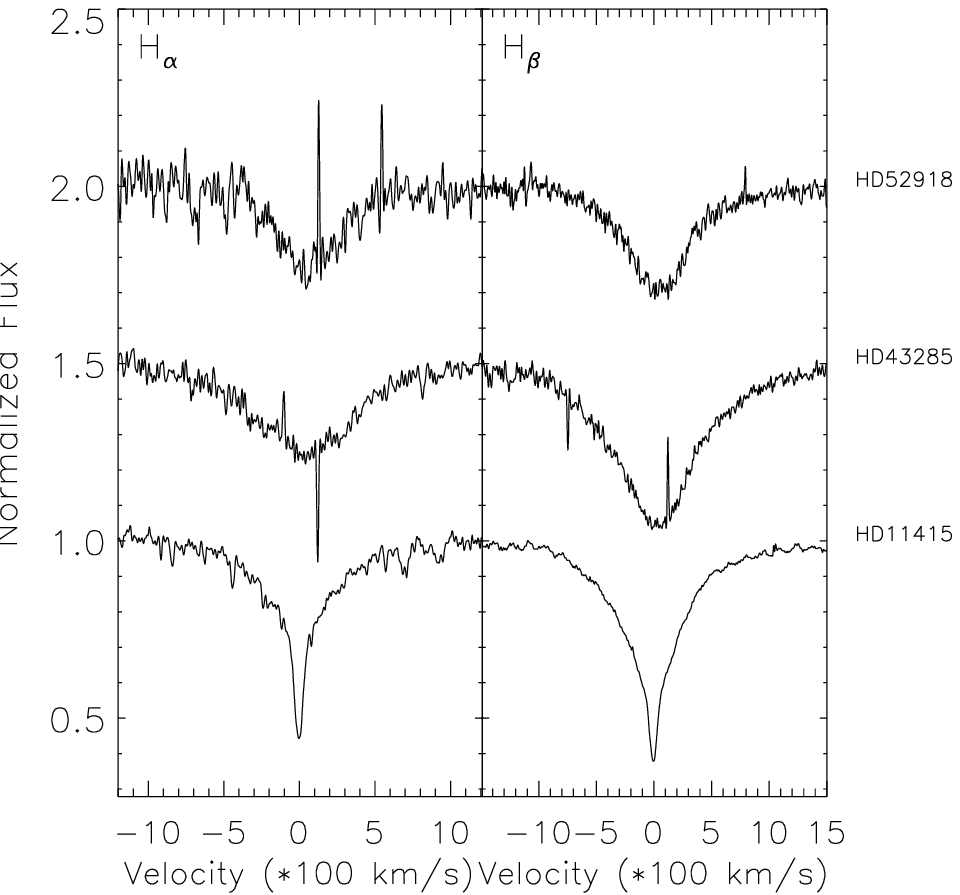}
      \caption{H$\alpha$ and H$\beta$ profile for stars for which we only have 
               one spectrum. In the boxes we display shell and class 1 stars, class 2 stars, and
               stars showing absorption, respectively.
                          }
         \label{one_shell}
   \end{figure*}

\end{document}